\documentclass[preprint,floatfix] {revtex4} 
\newcommand{\rvec}{\mathrm {\mathbf {r}}} 
\newcommand{\pvec}{\mathrm {\mathbf {p}}} 
\usepackage{graphicx}
\usepackage{subfigure}
\usepackage{xcolor}
\usepackage{amsmath}
\usepackage{enumerate}
\usepackage{tabularx}

\usepackage{color, soul}
\definecolor{darkblue}{rgb}{0,0,0.5}
\setulcolor{darkblue}

\begin{document}

\title{Information-entropic measures in free and confined hydrogen atom}

\author{Neetik Mukherjee}
\altaffiliation{Email: neetik.mukherjee@iiserkol.ac.in.}

\author{Amlan K.~Roy}
\altaffiliation{Corresponding author. Email: akroy@iiserkol.ac.in, akroy6k@gmail.com.}
\affiliation{Department of Chemical Sciences\\
Indian Institute of Science Education and Research (IISER) Kolkata, 
Mohanpur-741246, Nadia, WB, India}

\begin{abstract}
%%1234567890 %%1234567890 %%1234567890 %%1234567890 %%1234567890 %%1234567890 %%1234567890 %%1234567890 %%1234567890 %%1234567890

Shannon entropy ($S$), R{\'e}nyi entropy ($R$), Tsallis entropy ($T$), Fisher information ($I$) and Onicescu energy ($E$) have 
been explored extensively in both \emph{free} H atom (FHA) and \emph{confined} H atom (CHA). For a given quantum state, accurate 
results are presented by employing respective \emph{exact} analytical wave functions in $r$ space. The $p$-space wave functions are 
generated from respective Fourier transforms$-$for FHA these can be expressed analytically in terms of Gegenbauer polynomials, 
whereas in CHA these are computed numerically. \emph{Exact} mathematical expressions of $R_r^{\alpha}, R_p^{\beta}$, $T_r^{\alpha}, 
T_p^{\beta}, E_r, E_p$ are derived for \emph{circular} states of a FHA. Pilot calculations are done taking order of entropic moments 
($\alpha, \beta$) as $(\frac{3}{5}, 3)$ in $r$ and $p$ spaces. A detailed, systematic analysis is performed for both FHA and CHA 
with respect to state indices $n,l$, and with confinement radius ($r_c$) for the latter. In a CHA, at small $r_{c}$, kinetic energy 
increases, whereas $S_{\rvec}, R^{\alpha}_{\rvec}$ decrease with growth of $n$, signifying greater localization in high-lying states. 
At moderate $r_{c}$, there exists an interplay between two mutually opposing factors: (i) radial confinement (localization) and (ii) 
accumulation of radial nodes with growth of $n$ (delocalization). Most of these results are reported here for the first time, revealing 
many new interesting features. Comparison with literature results, wherever possible, offers excellent agreement. 

\vspace{3mm}

\vspace{3mm}

\end{abstract}
\maketitle

\section{introduction}
A quantum mechanical particle under extreme pressure displays many interesting and notable properties 
{\color{red}\cite{michels37,sabin2009,sen2014electronic}}. In the last few decades, quantum confinement has emerged as 
a very fascinating and relevant research area from both theoretical and experimental perspectives \cite{sabin2009,sarsa11,
sech11,katriel12,cabrera13,sen2014electronic}. Discovery and development of modern experimental techniques have given the 
required insight about responses of matter under confinement. Furthermore, advancement of nano-science and nano-technology 
has also stimulated extensive research activity to explore and study such systems. They have potential applications in a 
wide range of problems in physics and chemistry, namely, quantum wells, quantum wires, quantum dots, defects in solids, 
super-lattice structure, as well as nano-sized circuits such as quantum computer, etc. Besides, they have uses in cell-model 
of liquid, high-pressure physics, astrophysics \cite{pang11}, study of impurities in semiconductor materials, matrix isolated 
molecules, endohedral complexes of fullerenes, zeolites cages, helium droplets, nano-bubbles, \cite{sabin2009} etc. 

Extensive theoretical works have been published covering a broad variety of confining potentials. Two prototypical systems 
receiving maximum attention are harmonic oscillator (in 1D, 2D, 3D, D dimension) {\color{red}\cite{aquino97,campoy2002,montgomery07,roy14,
ghosal16}} and confined hydrogen atom (CHA) inside a spherical enclosure {\color{red}\cite{goldman92,aquino95,garza98,laughlin02,burrows06,
aquino07cha,baye08,ciftci09, sen2014electronic,roy15}}. A CHA within an impenetrable (as well as penetrable) cavity was studied 
quite vigorously leading 
to a host of attractive properties$-$both from physical and mathematical point of view. They offer many unique phenomena, 
especially relating to \emph{simultaneous, incidental and inter-dimensional} degeneracy \cite{montgomery07}. Effect of 
compression on energy levels of ground and excited states, as well as other properties like hyperfine splitting constant, dipole 
shielding factor, nuclear magnetic screening constant, pressure, static and dynamic polarizability, etc., were examined 
{\color{red}\cite{sabin2009,sen2014electronic,sen12}}. Numerous 
theoretical methods varying in complexity, sophistication were employed; a selected set includes perturbation theory, Pad\'e 
approximation, WKB method, Hypervirial theorem, power-series solution, super-symmetric quantum mechanics, Lie algebra, 
Lagrange-mesh method, asymptotic iteration method, generalized pseudo-spectral method, etc. \cite{goldman92,aquino95,garza98,
laughlin02,burrows06,aquino07cha,baye08,ciftci09,roy15} and references therein. \emph{Exact} solutions \cite{burrows06} 
are expressible in terms of Kummer M-function (confluent hypergeometric). 

In recent years, appreciable attention was paid to investigate various information measures, namely, Fisher information 
($I$), Shannon entropy ($S$), R{\'e}nyi entropy ($R$), Tsallis entropy ($T$), Onicescu energy ($E$) and several complexities 
in a multitude of physical and chemical systems including central potentials. The literature is quite vast. Here we restrict
ourselves to a few references pertaining to H atom. Some of these for an unconfined \emph{free} H atom (FHA) are: $I_r, I_p, I$ 
in 3D \cite{romera05}, in D-dimension \cite{romera06,dehesa06,dehesa07}, upper bounds of $S, R$ \cite{sanchez2011}, $S$ in 3D 
\cite{toranzo16a}, in D-dimension \cite{yanez94}, $R, T$ in 3D \cite{toranzo16a}, in $D$-dimension \cite{toranzo16}. 
Relativistic effects on the information measures of FHA are also examined \cite{katriel10}. A lucid review on information
theory of D-dimensional FHA is provided in \cite{dehesa10}. However, in CHA such studies are quite scarce, \emph{viz.}, $S$ 
{\color{red}\cite{sen05,jiao17}}, bounds of $I, S, R, T$ \cite{patil07}, $I, S$ in case of soft spherically CHA \cite{aquino13}. 

In a quantum system, $S$ and $I$ quantify the information content in different and complimentary way. Former refers to the 
expectation value of logarithmic probability density function and is a global measure of spread of density. On the other 
hand, $I$ is a gradient functional of density and in position ($r$) space, it quantifies the oscillatory nature and narrowness of 
density. In recent years, $S$ is examined in a number of occasions, such as, P\"oschl-Teller \cite{sun2013quantum}, 
Rosen-Morse \cite{sun2013quantum1}, pseudo-harmonic \cite{yahya2015}, squared tangent well \cite{dong2014quantum}, 
hyperbolic \cite{valencia2015quantum}, position-dependent mass Schr\"odinger equation \cite{chinphysb,yanez2014quantum}, 
infinite circular well \cite{song2015shannon}, hyperbolic double-well (DW) potential \cite{sun2015shannon}, etc. In parallel, $I$ 
is found to be a useful tool to analyze various atomic and molecular systems \cite{liu07,nagy07,szabo08}. Analytical expressions 
for $I$ are obtained for generalized central potentials in both $r$ and momentum ($p$) space \cite{romera05}. $E$ 
is quantified as the second-order entropic moment \cite{onicescu66}. It becomes minimum for equilibrium and hence often termed as 
disequilibrium. Recently, some of these measures have been found to be quite effective and useful to explain the oscillation and 
localization-delocalization behavior of a particle in symmetric and asymmetric DW potential \cite{neetik15,neetik16}, as well as 
in a confined 1D quantum harmonic oscillator \cite{ghosal16}. 

It is well known that, $R^{\alpha}, T^{\alpha}$, the so-called information generating functionals, are closely connected to
entropic moments (discussed later), and completely characterize density $\rho(\rvec)$. They are expressed in terms of 
expectation values of density, in following conventional forms,
\begin{equation}
\begin{aligned}
R^{\alpha}[\rho (\rvec)] & =  \frac{1}{(1-\alpha)} \ \mathrm{ln} \ \langle \rho(\rvec)^{(\alpha-1)}\rangle,       \\
T^{\alpha}[\rho (\rvec)] & =  \frac{1}{(\alpha-1)} \left[ 1-\langle \rho(\rvec)^{(\alpha-1)}\rangle \right]   \ \ \ \ 0 < \alpha < \infty, \ \ \alpha \neq 1. 
\end{aligned}
\end{equation}
Untitled Folder
They actually quantify the spatial delocalization of single-particle density of a system in various complimentary ways. Arguably, 
these are the most appropriate uncertainty measures, as they do not make any reference to some specific point of the 
corresponding Hilbert space. Moreover, these are closely related to energetic and experimentally measurable quantities 
\cite{gonzalez03,sen12} of a system. {\color{red} In case of $R$ and $S$, some lower bound is available, which does not
depend on quantum number. But, for $I$ both upper and lower bounds have been established, which strictly change with quantum numbers 
{\color{red}\cite{bbi06,bbi75,romera05}}}.  

It is interesting to note that, $S$, $E$ (disequilibrium) are two particular cases of $R^{\alpha}, T^{\alpha}$ 
\cite{sen12,bbi06}. Former measures total extent of density whereas $E$ quantifies separation of density with respect 
to equilibrium. They are related to $R^{\alpha}, T^{\alpha}$ in following way \cite{toranzo16}, 
\begin{equation}
\begin{aligned}
S[\rho] & =   -\int \rho(\rvec) \ \mathrm{ln} \  \rho(\rvec) d\rvec = \lim_{\alpha \rightarrow 1}R^{\alpha}[\rho] = 
\lim_{\alpha \rightarrow 1}T^{\alpha}[\rho],    \\
E & =  \langle \rho \rangle = \exp \left( R^{2}[\rho] \right), \ \ \ \ \ \langle \rho \rangle = \int \rho^2(\rvec) \ d\rvec.  
\end{aligned}
\end{equation} 
Lately, R{\'e}nyi entropy has been successfully employed to investigate and predict various quantum properties and phenomena like 
entanglement, communication protocol, correlation de-coherence, measurement, localization properties of Rydberg states, 
molecular reactivity, multi-fractal thermodynamics, production of multi-particle in high-energy collision, disordered systems, 
spin system, quantum-classical correspondence, localization in phase space \citep{varga03,renner05, levay05,verstraete06,
bialas06,salcedo09,liu15}, etc. Likewise, $T$ has been also been studied, albeit with rather lesser intensity. It has been 
implicated for non-extensive thermo-statistics \cite{tsallis04, naudts11} and exploited quite extensively in the field of 
image processing, power-signal analysis, gravitation \cite{plastino99, chen14}, etc.

{\color{red} These information measure may be used in FHA, CHA to understand diffused nature of orbitals. All these are statistical 
quantities and are directly related to single-particle density. Also, results of one quantity compliments 
the inferences of others. In FHA, they help to grasp the spreading of orbitals at higher states. Whereas, in CHA, 
$R, T, S, E$ qualitatively explain the effect of confinement on an arbitrary $n,~l,~m$ state. As found in later section, with increase 
in $R_{\rvec}, T_{\rvec}, S_{\rvec}$ and decrease in $E_{\rvec}$, system gets delocalized and vice versa.} 

The present communication has several objectives. Our primary motivation is to undertake a detailed analysis of $S, I, R, T, E$ 
in a FHA in a systematic fashion for an arbitrary state in both spaces. To put things in proper perspective, it is worth 
mentioning the scattered results that are available in literature for a FHA. The \emph{exact} mathematical form of $I$ for an 
\emph{arbitrary} state of FHA was given in both $r$ and $p$ space \cite{romera05} in terms of four expectation values 
$\langle r^2 \rangle, \langle r^{-2} \rangle, \langle p^2 \rangle, \langle p^{-2} \rangle$, and eventually in terms of the 
related quantum numbers. Likewise, an \emph{exact} analytical formula for $S$ in \emph{ground state} of a D-dimensional FHA was 
derived long times ago \cite{yanez94} in both $r$ and $p$ space. Later, similar analytical expressions of $S$ for \emph{circular 
or node-less} states of a D-dimensional FHA was offered in 2010 \cite{dehesa10} in both spaces. However, such a closed-form 
expression of $S$ is as yet lacking for a \emph{general} state. {\color{red} Very recently, in \cite{jiao17}, accurate radial Shannon 
entropies in $r,~p$ spaces with $n\leq 10$ are computed numerically. Moreover, a generalized form of angular shannon entropy was also
derived.} Recently, $R$ and $T$ were studied for Rydberg hydrogenic states within a 
strong Laguerre asymptotic approximation \cite{toranzo16a,toranzo16}. But, their \emph{exact} solutions are as yet unknown; moreover 
these were reported only in $r$ space and mostly for $l=0$ states. And to the best of our knowledge, $E$ has not yet been 
explored at all whatsoever. Thus there is some gap in the understanding of information-entropic measures in this system. This work makes an 
attempt to fill this void and embarks on an elaborate analysis of all these quantities in \emph{both} spaces for a \emph{general} 
state having principal and azimuthal quantum numbers $n,l$, while keeping magnetic quantum number $m=0$. For free system, all 
these can be calculated from \emph{exact} analytical wave functions in $r,p$ space. It is found that, for node-less states, expressions 
of $S, R, T, E$ are accessible in closed form in a FHA, as the required radial polynomial reduces to unity. But for all other, 
$n,l$, they need to be computed numerically, as presence of nodes in such wave functions leads to difficult 
polynomials. Next, we proceed for a parallel analysis for a CHA at varying $r_c$, taking \emph{exact} wave function in $r$ 
space. However such expressions are unavailable in $p$ space, and hence numerical Fourier transforms need to be carried out. 
Unlike the case of FHA, all measures in a CHA have to be obtained numerically--in both $r,p$ spaces. Note that such studies in 
CHA are very rare. {\color{red} Apart from the work of \cite{jiao17} for $S$ (as mentioned above), it was studied in the context of soft and hard 
confinement in lowest state \cite{aquino13,sen05}. Further in \cite{jiao17}, variation of $S$ in $s, p, d$ orbitals ($n \leq 7$) was 
followed with $r_c$.} Thus it is very desirable to probe these with respect to state indices and $r_c$. Throughout the article, comparison 
with existing 
literature results are made wherever possible. Organization of our article is as follows. Section~II gives essential components 
of methodology; then Sec.~III gives a detailed discussion on the results of above-mentioned 
quantities for FHA and CHA, while we conclude with a few remarks in Sec.~IV. 

\section{Methodology} 
Without any loss of generality, the time-independent non-relativistic wave function for a hydrogenic system, in $r$ space
can be written as ($\rvec = \{ r, \Omega \}$), 
\begin{equation} 
\Psi_{n,l,m} (\rvec) = \psi_{n, l}(r)  \ Y_{l,m} (\Omega), 
\end{equation}
with $r$ and $\Omega$ denoting radial distance and solid angle respectively. Here $\psi_{n,l}(r)$ corresponds to radial part 
and $Y_{l,m}(\Omega)$ the spherical harmonics of atomic state, determined by quantum numbers $(n,l,m)$. In what follows, atomic 
units employed unless otherwise mentioned and $\rvec, \pvec$ subscripts denote quantities in full $r$ and $p$ spaces (including
angular part) respectively.

The relevant radial Schr\"odinger equation under the influence of confinement is, 
\begin{equation}
	\left[-\frac{1}{2} \ \frac{d^2}{dr^2} + \frac{l (l+1)} {2r^2} + v(r) +v_c (r) \right] \psi_{n,l}(r)=
	{\color{red}\mathcal{E}_{n,l}}\ 
\psi_{n,l}(r),
\end{equation}
where $v(r)=-Z/r \ (Z=1$ for H atom). Our desired confinement inside an impenetrable spherical cage is accomplished by 
invoking the following potential: $v_c(r) = +\infty$ for $r > r_c$, and 0 for $r \leq r_c$, where $r_c$ signifies radius of 
the box. This equation needs to be solved under Dirichlet boundary condition, $\psi_{n,l} (0)=\psi_{n,l}(r_c)=0$.

\begingroup           %%Table 1
\squeezetable
\begin{table}
\caption{The co-efficients $a_k$ and $b_j$ for even-$l$ $p$-space wave functions in FHA. See text for details.} 
\centering
\begin{ruledtabular}
\begin{tabular}{l|ccccc|cccc}
$l$ & $b_{0}$ & $b_{2}$ & $b_{4}$ & $b_{6}$ & $b_{8}$ & $a_{1}$ & $a_{3}$ & $a_{5}$ & $a_{7}$   \\
\hline
$0$ & $\frac{1}{\sqrt{\pi}}$    & --         & --        & --     & --     & --     & --      & --       & --              \\
$2$ & $\frac{1}{\sqrt{\pi}}$    & $-$$\frac{3}{\sqrt{\pi}}$      & --        & --         & --        & $\frac{3}{\sqrt{\pi}}$     
    & --         & --       & --              \\
$4$ & $\frac{1}{\sqrt{\pi}}$  & $-$$\frac{105}{\sqrt{\pi}}$    & $\frac{315}{\sqrt{\pi}}$      & --         & --        
    & $\frac{30}{\sqrt{\pi}}$    & $-$$\frac{315}{\sqrt{\pi}}$    & --       & --              \\
$6$ & $\frac{1}{\sqrt{\pi}}$   & $-$$\frac{210}{\sqrt{\pi}}$    & $\frac{4725}{\sqrt{\pi}}$    & $-$$\frac{10395}{\sqrt{\pi}}$  
    & --  & $\frac{21}{\sqrt{\pi}}$    & $\frac{-1260}{\sqrt{\pi}}$   & $\frac{10395}{\sqrt{\pi}}$   & --         \\
$8$ & $\frac{1}{\sqrt{\pi}}$  & $-$$\frac{630}{\sqrt{\pi}}$  & $\frac{51975}{\sqrt{\pi}}$  & $-$$\frac{945945}{\sqrt{\pi}}$ 
    & $\frac{2027025}{\sqrt{\pi}}$ & $\frac{36}{\sqrt{\pi}}$   & $-$$\frac{6930}{\sqrt{\pi}}$ & $\frac{270270}{\sqrt{\pi}}$ 
    & $-$$\frac{2027025}{\sqrt{\pi}}$    \\
\end{tabular}
\end{ruledtabular}
\end{table}
\endgroup

Angular part has following common form in both $r$ and $p$ spaces ($P_{l}^{m} (\cos \theta)$ signifies usual associated 
Legendre polynomial), 
\begin{equation}
Y_{l,m} ({\Omega}) =\Theta_{l,m}(\theta) \ \Phi_m (\phi) = (-1)^{m} \sqrt{\frac{2l+1}{4\pi}\frac{(l-m)!}{(l+m)!}} 
\ P_{l}^{m}(\cos \theta)\ e^{-im \phi}.  
\end{equation}
The exact generalized radial wave function for a CHA can be expressed \cite{burrows06} as, 
\begin{equation}
\psi_{n, l}(r)= N_{n, l}\left(2r\sqrt{-2{\color{red}\mathcal{E}_{n,l}}}\right)^{l} \ _{1}F_{1}
\left[\left(l+1-\frac{1}{\sqrt{-2{\color{red}\mathcal{E}_{n,l}}}}\right),(2l+2),2r\sqrt{-2{\color{red}\mathcal{E}_{n,l}}}\right] 
e^{-r\sqrt{-2{\color{red}\mathcal{E}_{n,l}}}},
\end{equation}
where $N_{n, l}$ denotes normalization constant and ${\color{red}\mathcal{E}_{n,l}}$ corresponds to energy of a given state 
characterized by $n,l$ 
quantum numbers, whereas $_1F_1\left[a,b,r\right]$ represents confluent hypergeometric function. In case of FHA 
($r_c \rightarrow \infty$), the first-order hypergeometric function reduces to associated Laguerre polynomial with 
${\color{red}\mathcal{E}_{n,l}}=-\frac{Z^2}{2n^2}$ ($Z$ denotes atomic number); so the radial function simplifies to commonly used form, as given below, 
\begin{equation}
\psi_{n,l}(r)= \frac{2}{n^2}\left[\frac{(n-l-1)!}{(n+l)!}\right]^{\frac{1}{2}}\left[\frac{2Z}{n}r\right]^{l} 
e^{-\frac{Z}{n}r} \ L_{(n-l-1)}^{(2l+1)} \left(\frac{2Z}{n}r\right).  
\end{equation}
Thus allowed energies at a specific $r_c$ can be obtained by finding the zeros of $_{1}F_{1}$, 
\begin{equation}
_{1}F_{1}\left[\left(l+1-\frac{1}{\sqrt{-2{\color{red}\mathcal{E}_{n,l}}}}\right),(2l+2),2r_{c}
\sqrt{-2{\color{red}\mathcal{E}_{n,l}}}\right]=0. 
\end{equation}
For a particular $l$, first root corresponds to energy of the lowest-$n$ state $(n_{lowest}=l+1)$ with successive roots
identifying excited states. It is instructive to note that, in order to construct the exact wave function of CHA for a specific 
state, one needs to supply energy eigenvalue of that state. In our present calculation, ${\color{red}\mathcal{E}_{n,l}}$ of CHA, 
computed by means of 
the generalized pseudo-spectral (GPS) method is employed, because for a number of central potentials as in current 
situation, this has produced highly accurate eigenvalues and eigenfunctions, in both free and confinement situations; see e.g., 
references \cite{roy04, sen06, roy13, roy14, roy15} and therein for some of the developments. Note that in this communication, our 
objective is not precise calculation of energy; rather we are interested in the information measures in CHA, for which GPS 
energies are sufficiently accurate to obtain correct eigenfunctions. 

\begingroup           %%Table 2
\squeezetable
\begin{table}
\caption{The co-efficients $a_k$ and $b_j$ for odd-$l$ $p$-space wave functions in FHA. See text for details.} 
\centering
\begin{ruledtabular}
\begin{tabular}{l|ccccc|ccccc}
$l$ & $a_{0}$ & $a_{2}$ & $a_{4}$ & $a_{6}$ & $a_{8}$ & $b_{1}$ & $b_{3}$ & $b_{5}$ & $b_{7}$ & $b_{9}$   \\
\hline 
$1$ & $\frac{1}{\sqrt{\pi}}$   & --     & --        & --            & --          & $-$$\frac{1}{\sqrt{\pi}}$     & --       
    & --           & --         & --              \\
$3$ & $\frac{1}{\sqrt{\pi}}$  & $-$$\frac{15}{\sqrt{\pi}}$    & --        & --            & --       & $-$$\frac{3!}{\sqrt{\pi}}$ 
    & $\frac{15}{\sqrt{\pi}}$    & --           & --         & --              \\
$5$ & $\frac{1}{\sqrt{\pi}}$  & $-$$\frac{105}{\sqrt{\pi}}$    & $\frac{945}{\sqrt{\pi}}$   & --            & --          
    & $-$$\frac{15}{\sqrt{\pi}}$    & $\frac{420}{\sqrt{\pi}}$  & $-$$\frac{945}{\sqrt{\pi}}$      & --     & --          \\
$7$ & $\frac{1}{\sqrt{\pi}}$ & $-$$\frac{378}{\sqrt{\pi}}$   & $\frac{17325}{\sqrt{\pi}}$ & $-$$\frac{135135}{\sqrt{\pi}}$      
    & --  & $-$$\frac{28}{\sqrt{\pi}}$  & $\frac{3150}{\sqrt{\pi}}$ & $-$$\frac{2370}{\sqrt{\pi}}$  
    & $\frac{135135}{\sqrt{\pi}}$ & --          \\
$9$ & $\frac{1}{\sqrt{\pi}}$   & $-$$\frac{990}{\sqrt{\pi}}$ & $\frac{135135}{\sqrt{\pi}}$  & $-$$\frac{4729725}{\sqrt{\pi}}$ 
    & $\frac{34459425}{\sqrt{\pi}}$  & $-$$\frac{45}{\sqrt{\pi}}$  & $\frac{13860}{\sqrt{\pi}}$  & $-$$\frac{945945}{\sqrt{\pi}}$ 
    & $\frac{16216200}{\sqrt{\pi}}$ &  $-$$\frac{34459425}{\sqrt{\pi}}$  \\
\end{tabular}
\end{ruledtabular}
\end{table}
\endgroup

The $p$-space wave function ($\pvec = \{ p, \Omega \}$) for a particle in a central potential is obtained from respective 
Fourier transform of its $r$-space counterpart, and as such, is given below,
\begin{equation}
\begin{aligned}
\psi_{n,l}(p) & = & \frac{1}{(2\pi)^{\frac{3}{2}}} \  \int_0^\infty \int_0^\pi \int_0^{2\pi} \psi_{n,l}(r) \ \Theta(\theta) 
 \Phi(\phi) \ e^{ipr \cos \theta}  r^2 \sin \theta \ \mathrm{d}r \mathrm{d} \theta \mathrm{d} \phi,  \\
      & = & \frac{1}{2\pi} \sqrt{\frac{2l+1}{2}} \int_0^\infty \int_0^\pi \psi_{n,l} (r) \  P_{l}^{0}(\cos \theta) \ 
e^{ipr \cos \theta} \ r^2 \sin \theta  \ \mathrm{d}r \mathrm{d} \theta.  
\end{aligned}
\end{equation}
Note that $\psi(p)$ is not normalized; thus needs to be normalized. Integrating over $\theta$ and $\phi$ variables, {\color{red}Eq.~(9)} 
can be further reduced to, 
\begin{equation}
\psi_{n,l}(p)=(-i)^{l} \int_0^\infty \  \frac{\psi_{n,l}(r)}{p} \ f(r,p)\mathrm{d}r.    
\end{equation}
Depending on $l$, this can be rewritten in following simplified form ($m'$ starts with 0),  
\begin{equation}
\begin{aligned}
f(r,p) & = & \sum_{k=2m^{\prime}+1}^{m^{\prime}<\frac{l}{2}} a_{k} \ \frac{\cos pr}{p^{k}r^{k-1}} +  
            \sum_{j=2m^{\prime}}^{m^{\prime}=\frac{l}{2}} b_{j} \ \frac{\sin pr}{p^{j}r^{j-1}}, \ \ \ \ \mathrm{for} \ 
            \mathrm{even} \ l,   \\
f(r,p) & = & \sum_{k=2m^{\prime}}^{m^{\prime}=\frac{l-1}{2}} a_{k} \ \frac{\cos pr}{p^{k}r^{k-1}} +  
\sum_{j=2m^{\prime}+1}^{m^{\prime}=\frac{l-1}{2}} b_{j} \ \frac{\sin pr}{p^{j}r^{j-1}}, \ \ \ \ \mathrm{for} \ \mathrm{odd} \ l.
\end{aligned} 
\end{equation}
The coefficients $a_{k}$, $b_{j}$ of even-$l$ and odd-$l$ states are collected in Tables~I and II respectively. For a FHA, one can 
achieve the following analytical expression for wave function {\color{red}\cite{sanudo08a}},
\begin{equation}
\psi_{n,l}(p)=n^{2}\left[\frac{2}{\pi}\frac{(n-l-1)!}{(n+l)!}\right]^\frac{1}{2} 2^{(2l+2)} \ l! \ 
\frac{n^l}{ \{[\frac{np}{Z}]^2+1 \}^{l+2}} \left(\frac{p}{Z}\right)^l
C_{n-l-1}^{l+1} \left(\frac{[\frac{np}{Z}]^2-1}{[\frac{np}{Z}]^2+1}\right), 
\end{equation}
where $C_{\zeta}^{\eta}(t)$ signifies the Gegenbauer polynomial. 

\begin{figure}                         %%%Fig. 1, FHA
\begin{minipage}[c]{0.4\textwidth}\centering
\includegraphics[scale=0.75]{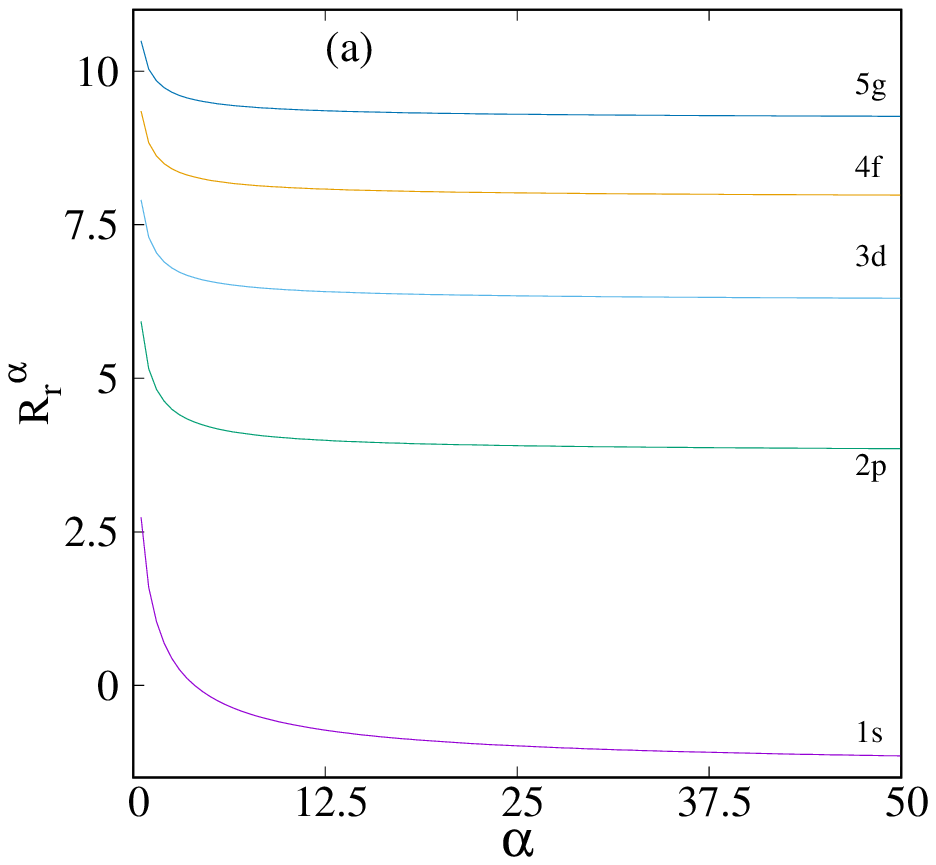}
\end{minipage}%
\hspace{0.1in}
\begin{minipage}[c]{0.5\textwidth}\centering
\includegraphics[scale=0.75]{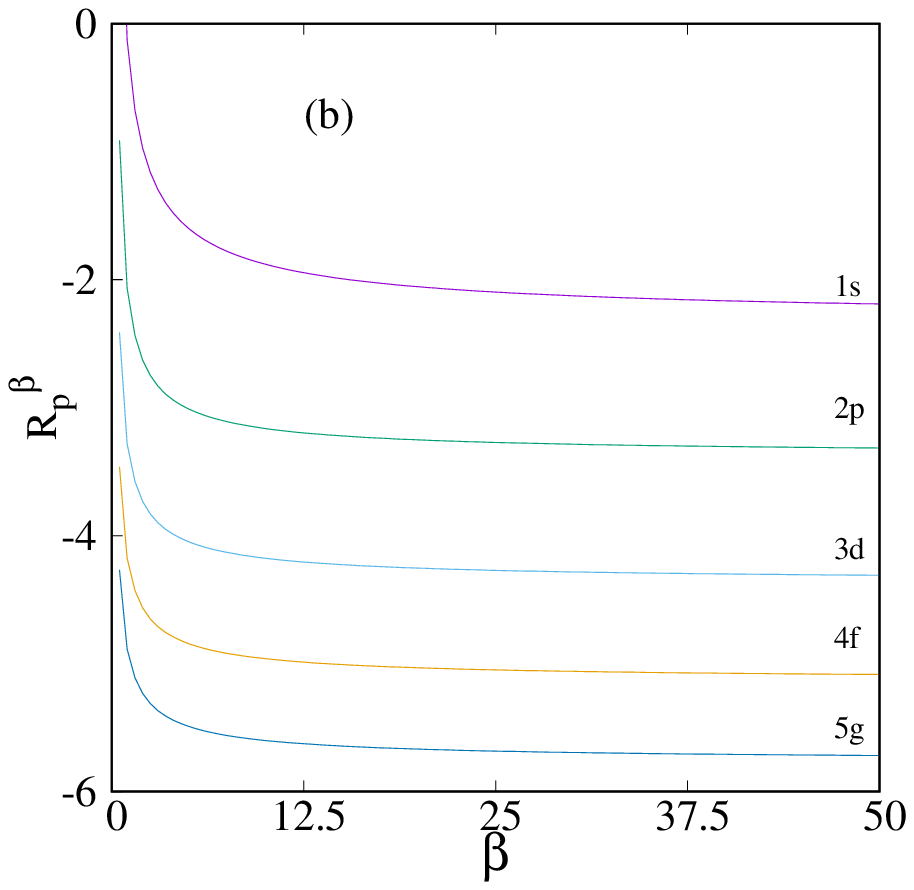}
\end{minipage}%
\caption{Variation of $R_{r}^{\alpha}$ and $R_{p}^{\beta}$ with respect to $\alpha$ and $\beta$ for $1s,~2p,~3d,~4f,~5g$ states 
of FHA, in panels (a), (b) respectively. More details can be found in the text.}
\end{figure}

It is known that $I_{\rvec}$, $I_{\pvec}$ for a single particle in a central potential can be written in terms of radial 
expectation values $\langle r^k \rangle $ and $ \langle p^k \rangle, (k = -2,2)$ \cite{romera05},
\begin{equation}
\begin{aligned} 
I_{\rvec} & =  & \int_{{\mathcal{R}}^3} \left[\frac{|\nabla\rho(\rvec)|^2}{\rho(\rvec)}\right] \mathrm{d}\rvec  & =  
4\langle p^2\rangle - 2(2l+1)|m|\langle r^{-2}\rangle; \ \ \ \rho(\rvec) = |\psi_{n,l,m}(\rvec)|^2,   \\
I_{\pvec} & =  & \int_{{\mathcal{R}}^3} \left[\frac{|\nabla\Pi(\pvec)|^2}{\Pi(\pvec)}\right] \mathrm{d} \pvec  & = 
4\langle r^2\rangle - 2(2l+1)|m|\langle p^{-2}\rangle; \ \ \ \Pi(\pvec) = |\psi_{n,l,m} (\pvec)|^2.
\end{aligned} 
\end{equation}
Whereas the {\color{red} total position-momentum (PM) Fisher information is expressed as, $I=I_{\rvec}I_{\pvec}$}. It satisfies the following bound 
\cite{romera05}, 
\begin{equation} 
{\color{red}\frac{81}{\langle r^2\rangle \langle p^2\rangle} \leq I_{\rvec} I_{\pvec} \leq 16 \langle r^2\rangle \langle p^2\rangle}.
\end{equation}                            
Here $\rho(\rvec), \Pi(\pvec)$ signify $r$- and $p$-space densities, both being normalized to unity.
Next, $S_{\rvec}, S_{\pvec}$ and {\color{red} total PM Shannon entropy $S$} is defined in terms of expectation values of logarithmic probability 
density functions, which for a central potential further simplifies {\color{red}\cite{bbi75}} as below, 
\begin{equation}
\begin{aligned} 
S_{\rvec} & =  -\int_{{\mathcal{R}}^3} \rho(\rvec) \ \ln [\rho(\rvec)] \ \mathrm{d} \rvec   = 
2\pi \left(S_{r}+S_{(\theta,\phi)}\right),    
S_{\pvec}  =  -\int_{{\mathcal{R}}^3} \Pi(\pvec) \ \ln [\Pi(\pvec)] \ \mathrm{d} \pvec   = 
2\pi \left(S_{p}+S_{(\theta, \phi)}\right), \\ 
S & = 2\pi \left[S_{r}+S_{p}+2S_{(\theta, \phi)}\right] \ \ {\color{red}\geq 3(1+\ln \pi)}, 
\end{aligned} 
\end{equation}
where the quantities $S_r, S_p$ and $S_{\theta}$ are defined as {\color{red}\cite{bbi75}},   
\begin{equation}
\begin{aligned} 
S_{r} & =  -\int_0^\infty \rho(r) \ \ln [\rho(r)] r^2 \mathrm{d}r, \ \ \ \ \ \ \ \ \ \ \ \ 
S_{p}  =  -\int_{0}^\infty \Pi(p) \ln [\Pi(p)]  \ p^2 \mathrm{d}p, \\
\rho(r) & = |\psi_{n,l}(r)|^{2},  \ \ \ \ \ \ \ \ \ \ \ \ \ \ \ \ \ \ \ \ \ \ \ \ \ \ \ \  \Pi(p) = |\psi_{n,l}(p)|^{2}, \\
S_{(\theta, \phi)} & =   -\int_0^\pi \chi(\theta) \ \ln [\chi(\theta)] \sin \theta \mathrm{d} \theta, \ \ \ \ \ \
\chi(\theta)   =  |\Theta(\theta)|^2.  \\   
\end{aligned} 
\end{equation}

\begin{figure}                         %%%Fig. 2, FHA
\begin{minipage}[c]{0.3\textwidth}\centering
\includegraphics[scale=0.5]{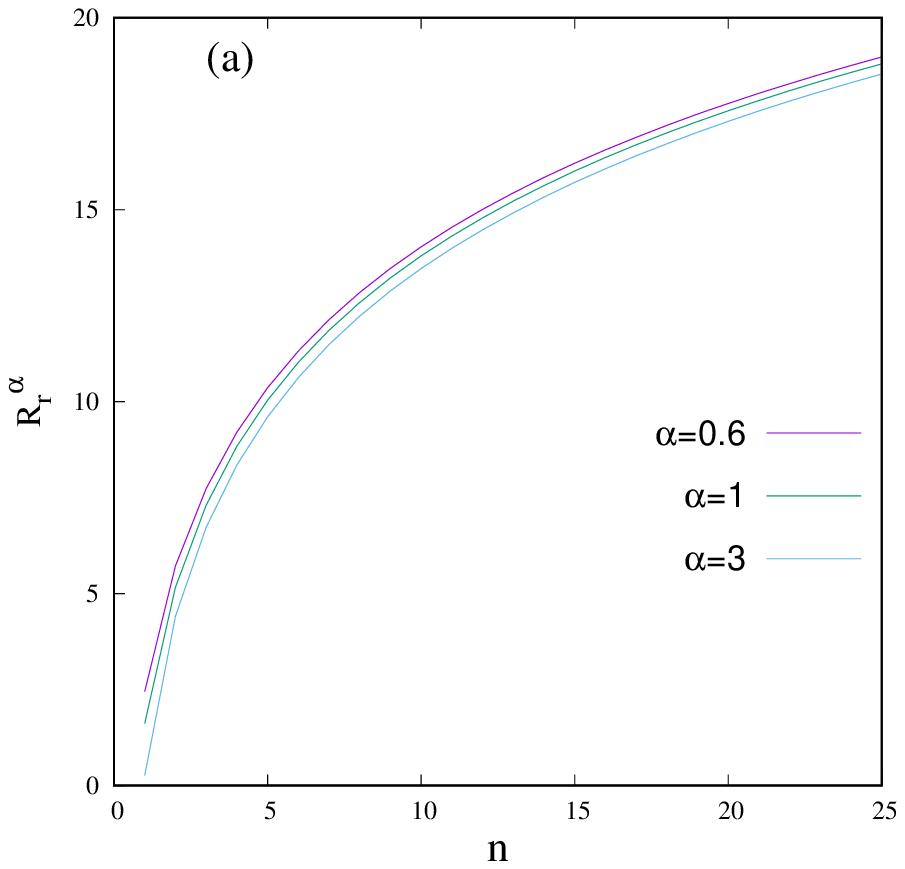}
\end{minipage}%
\hspace{0.02in}
\begin{minipage}[c]{0.3\textwidth}\centering
\includegraphics[scale=0.5]{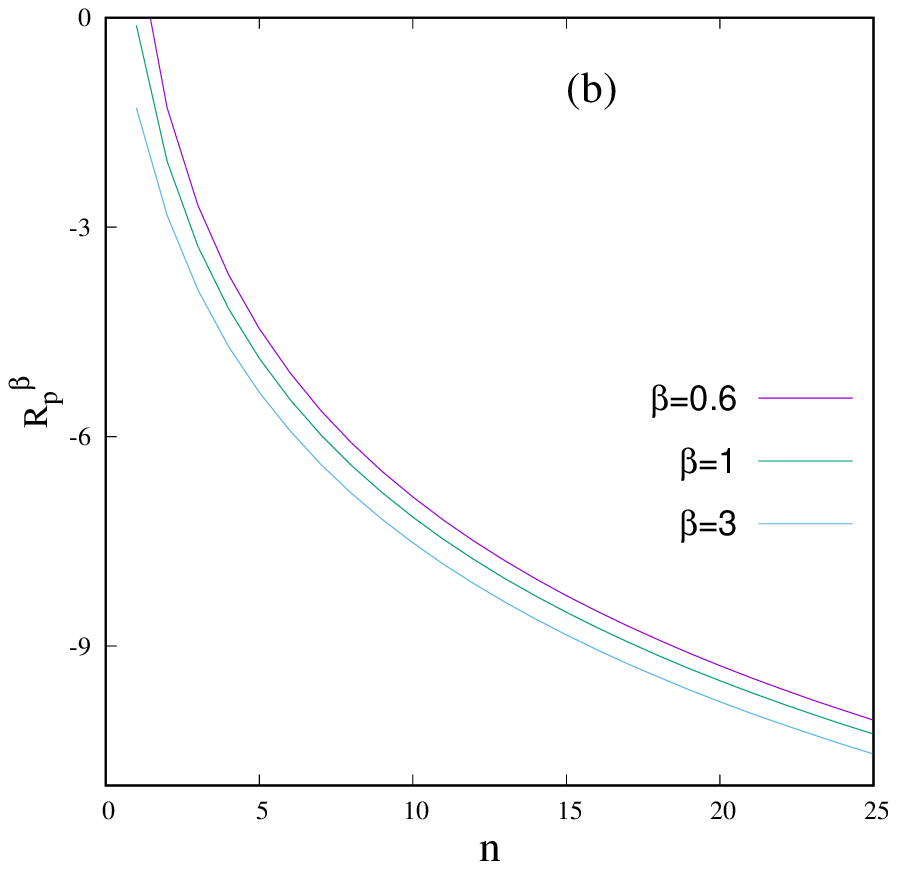}
\end{minipage}%
\hspace{0.02in}
\begin{minipage}[c]{0.3\textwidth}\centering
\includegraphics[scale=0.5]{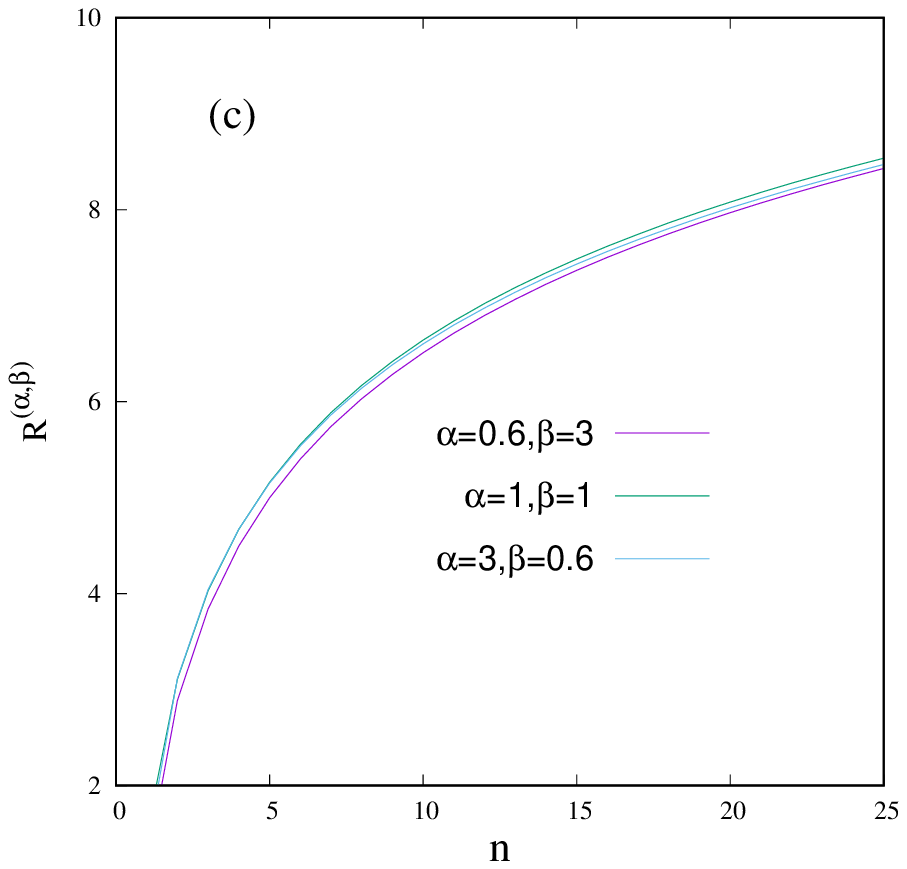}
\end{minipage}%
\caption{Plots of $R_{r}^{\alpha}$, $R_{p}^{\beta}$, {\color{red} radial PM R\'enyi entropy $R^{(\alpha, \beta)}$} against $n$ for some circular states of FHA at three 
specific sets of $\alpha, \beta$, $(\alpha, \beta)$ values, in panels (a), (b), (c) respectively. See text for details.}
\end{figure}

Similarly, R{\'e}nyi entropies of order $\lambda (\neq 1)$ are obtained by taking logarithm of $\lambda$-order entropic moment. 
In spherical polar coordinate these can be written in following simplified form by some straightforward mathematical manipulation, 
\begin{equation}
\begin{aligned} 
R_{\rvec}^{\lambda}  =  \frac{1}{1-\lambda} \ln \left(\int_{{\mathcal{R}}^3} \rho^{\lambda}(\rvec)\mathrm{d} \rvec \right)  = &
{\color{red}\frac{1}{(1-\lambda)} \ln \left(2\pi\int_0^\infty [\rho(r)]^{\lambda} r^2 \mathrm{d}r \int_0^\pi [\chi(\theta)]^{\lambda} \sin \theta \mathrm{d}\theta \right)} \\ 
 = & \frac{1}{(1-\lambda)}\left( \ln 2\pi + \ln [\omega^{\lambda}_r] + \ln [\omega^{\lambda}_{(\theta, \phi)}] \right),  \\
R_{\pvec}^{\lambda}  =  \frac{1}{1-\lambda} \ln \left[\int_{{\mathcal{R}}^3} \Pi^{\lambda}(\pvec)\mathrm{d} \pvec \right]  = &
{\color{red}\frac{1}{(1-\lambda)} \ln \left(2\pi \int_{0}^\infty [\Pi(p)]^{\lambda} p^2 \mathrm{d}p \int_0^\pi [\chi(\theta)]^{\lambda} \sin \theta \mathrm{d}\theta \right)} \\
 = & \frac{1}{(1-\lambda)}\left( \ln 2\pi + \ln [\omega^{\lambda}_p] + \ln [\omega^{\lambda}_{(\theta, \phi)}] \right).
\end{aligned} 
\end{equation}          
Here $\omega^{\lambda}_{\tau}$s are entropic moments in $\tau$ ($r$ or $p$ or $\theta$) space with order $\lambda$, having forms,
\begin{equation}
\omega^{\lambda}_r= \int_0^\infty [\rho(r)]^{\lambda} r^2 \mathrm{d}r, \ \ \  
\omega^{\lambda}_p= \int_{0}^\infty [\Pi(p)]^{\lambda} p^2 \mathrm{d}p, \ \ \  
\omega^{\lambda}_{(\theta, \phi)}= \int_0^\pi [\chi(\theta)]^{\lambda} \sin \theta \mathrm{d}\theta. 
\end{equation}          

\begingroup           %%Table 3, 1s-4f are all checked  
\squeezetable
\begin{table}
\caption{Angular contributions, $S_{(\theta, \phi)}, R^{\alpha}_{(\theta, \phi)}, R^{\beta}_{(\theta, \phi)}, 
T^{\alpha}_{(\theta, \phi)}, T^{\beta}_{(\theta,\phi)}, E_{(\theta, \phi)}$ in H atom ($m=0$), for the selected value of 
$\alpha=\frac{3}{5}, \beta=3$ for 10 lowest $l$ states. More details are available in text.}
\centering
\begin{ruledtabular}
\begin{tabular}{c|l|ll|ll|l}
$l$    &  $S_{(\theta, \phi)}^{\P}$  & $R^{\alpha}_{(\theta, \phi)}$  &  $R^{\beta}_{(\theta, \phi)}$   
       &  $T^{\alpha}_{(\theta, \phi)}$  &  $T^{\beta}_{(\theta, \phi)}$  &  $E_{(\theta, \phi)}$ \\ 
\hline 
0   &   2.531024246969     &   2.531024246969    &   2.531024246969   & 4.380562660576 & 0.4968337130111  &   0.0795774715459   \\
1   &   2.0990786249678    &   2.207799279060    &   1.856060888495   & 3.546081906570 & 0.4877871787573  &   0.1432394487826   \\
2   &   2.0411250061339    &   2.1880740866193   &   1.586098811200   & 3.498565470109 & 0.4790443043946  &   0.17052315331268  \\   
3   &   2.0206596227683    &   2.1838712989476   &   1.4135979721010  & 3.488489128929 & 0.4704107193889  &   0.18775831398309  \\   
4   &   2.0105368074094    &   2.1825847862425   &   1.2861478982321  & 3.485418316773 & 0.4618199815337  &   0.20037698056464  \\  
5   &   2.0045776990712    &   2.1821358741265   &   1.1848960592462  & 3.484336298921 & 0.4532499193936  &   0.21034302374067  \\
6   &   2.0006768495387    &   2.1819848295620   &   1.1008390899096  & 3.483972643154 & 0.4446913166609  &   0.21858446105644  \\
7   &   1.997934606130     &   2.1819528334935   &   1.028955122477   & 3.483896265386 & 0.4361397020166  &   0.22561345675926  \\
8   &   1.9959057777584    &   2.1819710153868   &   0.96615017473812 & 3.483938671551 & 0.4275926546791  &   0.23174282746972  \\
9   &   1.9943460712042    &   2.1820101260317   &   0.91038050803346 & 3.484031759027 & 0.4190487531790  &   0.23717779214936  \\
\end{tabular}
\end{ruledtabular}
\begin{tabbing}
{\color{red} $^{\P}$Literature results {\color{red}\cite{jiao17}} of $S_{(\theta, \phi)}$ for $l= 0-9$ and $m=0$ states are: 
2.5310242469692,~2.0990786249678,~2.0411250061339,}\\{\color{red} 
2.0206596227683,~2.0105368074095,~2.0045776990714,~2.0006768495387,~1.9979346061302,~1.9959057777583,} \\
{\color{red} ~1.9943460712038 respectively.} 
\end{tabbing}
\end{table}
\endgroup

If $\lambda$ corresponds to $\alpha$, $\beta$ in $r$, $p$ spaces respectively, then for R\'enyi and Tsallis entropies, they obey the 
condition $\frac{1}{\alpha}+\frac{1}{\beta}=2.$ Then one can define {\color{red} total PM R{\'e}nyi entropy as $R^{(\alpha,\beta)}$} {\color{red}\cite{bbi06,sen12}}, 
\begin{equation}
\begin{aligned}
R^{(\alpha,\beta)} & =  \frac{2-\alpha-\beta}{(1-\alpha)(1-\beta)} \ \ln 2\pi+ \frac{1}{(1-\alpha)}  
\left( \ln [\omega^{\alpha}_r]+ \ln [\omega^{\alpha}_{(\theta, \phi)}]\right)
 + \frac{1}{(1-\beta)}\left( \ln [\omega^{\beta}_p]+ \ln [\omega^{\beta}_{(\theta, \phi)}]\right)  \\
& {\color{red}\geq 3 \times \left[ -\frac{1}{2}\left(\frac{\ln \alpha}{1-\alpha}+\frac{\ln \beta}{1-\beta}\right)-
\ln \left(\frac{\Delta r \Delta p}{\hbar \pi}\right) \right].} 
\end{aligned} 
\end{equation}
$\Delta r$ and $\Delta p$ are standard deviation in position and momentum space respectively.

Finally, Tsallis entropy \cite{tsallis88} in $r, p$ spaces are expressed as below, 
\begin{equation}
\begin{aligned}
T_{\rvec}^{\alpha}= \left(\frac{1}{\alpha-1}\right)\left[1-\int_{{\mathcal{R}}^3} \rho^{\alpha}(\rvec)\mathrm{d} \rvec \right] 
    & = & \left(\frac{1}{\alpha-1}\right)\left[1-2\pi \omega^{\alpha}_r \omega^{\alpha}_{(\theta, \phi)}\right], \\
T_{\pvec}^{\beta}= \left(\frac{1}{\beta-1}\right)\left[1-\int_{{\mathcal{R}}^3} \Pi^{\beta}(\pvec)\mathrm{d} \pvec \right] 
    & = & \left(\frac{1}{\beta-1}\right)\left[1-2 \pi \omega^{\beta}_r \omega^{\beta}_{(\theta, \phi)}\right]. \\
\end{aligned}
\end{equation}             
The corresponding {\color{red} total PM Tsallis entropy is then given by, $T^{(\alpha,\beta)}=T_{\rvec}^{\alpha}T_{\pvec}^{\beta}$.}
       
{\color{red} By definition, $E$ refers to the 2nd order entropic moment \cite{sen12}; therefore} 
{\color{red}choice of $\alpha= \beta =2$} {\color{red} transforms {\color{red}Eq.~(18)} into the form,
\begin{equation}
E_{r}= \int_0^\infty [\rho(r)]^{2} r^2 \mathrm{d}r, \ \   
E_{p}= \int_{0}^\infty [\Pi(p)]^{2} p^2 \mathrm{d}p, \ \   
E_{\theta, \phi}= \int_0^\pi [\chi(\theta)]^{2} \sin \theta \mathrm{d}\theta, \ \
E=E_{r}E_{p}E_{\theta, \phi}^{2}. 
\end{equation}
where, $E$ is the total PM Onicescu energy. Note that, the restriction $\frac{1}{\alpha}+\frac{1}{\beta}=2$ holds for $R$ and $T$ only, 
and not $E$. Hence in our study of $R, T$, $\alpha=\frac{3}{5}$ and $\beta=3$ have been chosen.}

\begin{figure}                         %%%Fig. 3, FHA
\begin{minipage}[c]{0.35\textwidth}\centering
\includegraphics[scale=0.65]{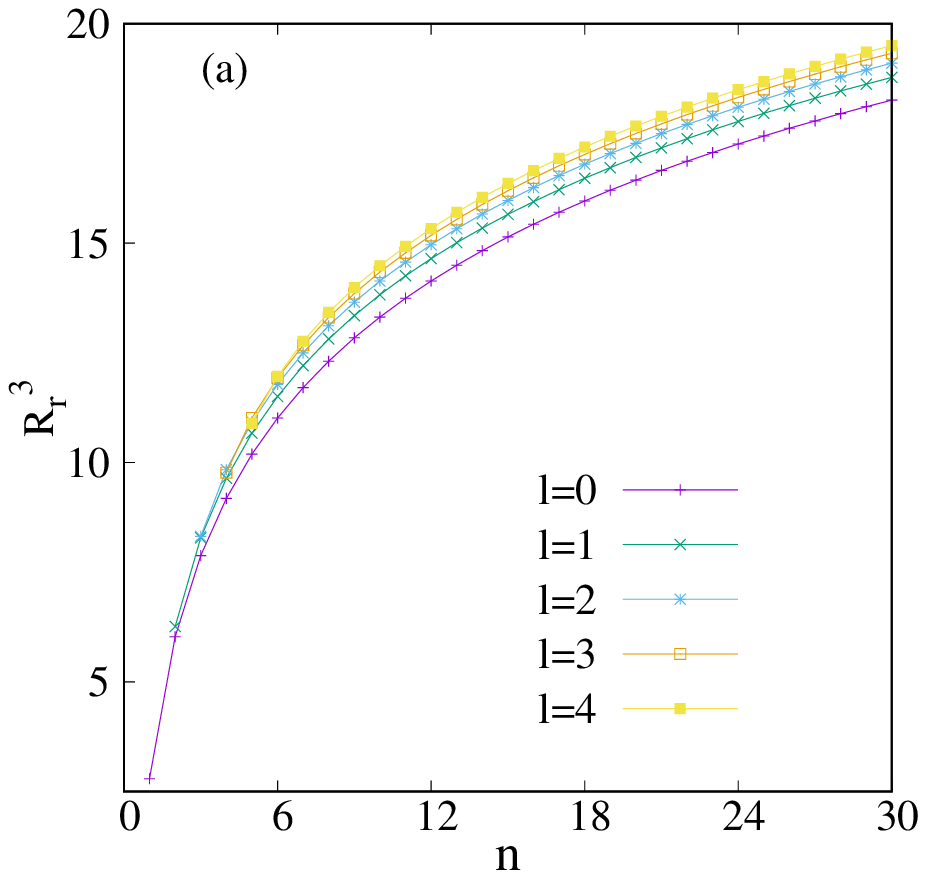}
\end{minipage}%
\hspace{0.5in}
\begin{minipage}[c]{0.35\textwidth}\centering
\includegraphics[scale=0.65]{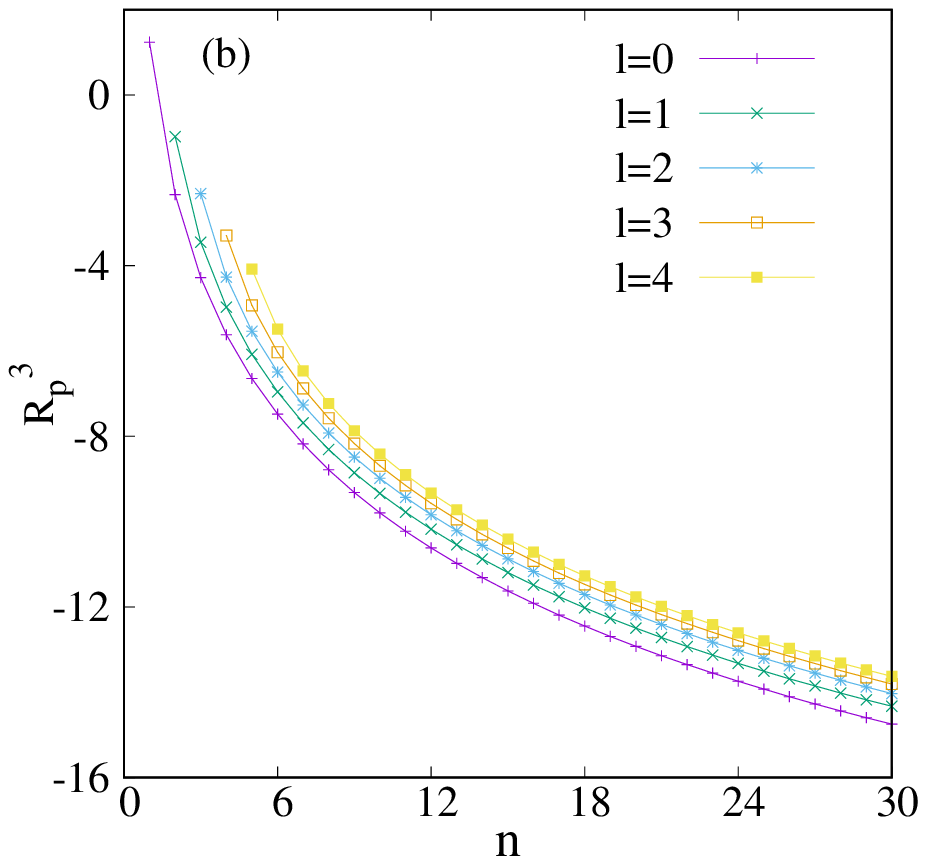}
\end{minipage}%
\caption{Variation of $R^{\alpha}_{r}, R^{\beta}_{p}$ with $n$, for lowest five $l$ (0-4) states, in panels (a)-(b), in a FHA. 
Both $\alpha, \beta$ are chosen as 3. For more details, consult text.}
\end{figure}

\section{Result and Discussion}
For ease of presentation, it would be appropriate to mention a few things at the outset. The \emph{net} information measures in 
$r$ and $p$ space of FHA and CHA may be branched into two separate contributions, \emph{viz.}, (i) a radial and (ii) an angular 
part. It is clear from {\color{red}Eqs.~(6) and (7)} that, general form of radial wave function changes from CHA to FHA. As mentioned earlier,
except the $p$ space of CHA, radial wave functions are available in closed analytical forms, in $r$ and $p$ spaces, both for FHA 
and CHA; and hence employed throughout all tables and figures in this section. As discussed later, it follows that in case of FHA, 
one can derive analytical expressions for all these quantities, for the special case of \emph{node-less} ($n-l=1$) states. Note 
that, angular portion of these measures remains invariant in $r$, $p$ spaces in both systems, and they also will not 
change with respect to boundary condition in $r_c$ in a CHA. Furthermore, they change with $l$, $m$ quantum numbers. In present 
calculation, we have chosen magnetic quantum number $m$ as 0, unless stated otherwise. Lastly, in case of FHA, only \emph{radial} 
parts of information measures are presented in tables and figures, whereas for CHA, these correspond to respective \emph{combined} 
(containing radial and angular) quantities.

\begin{figure}                    %%% Fig. 4, FHA
\begin{minipage}[c]{0.35\textwidth}\centering
\includegraphics[scale=0.65]{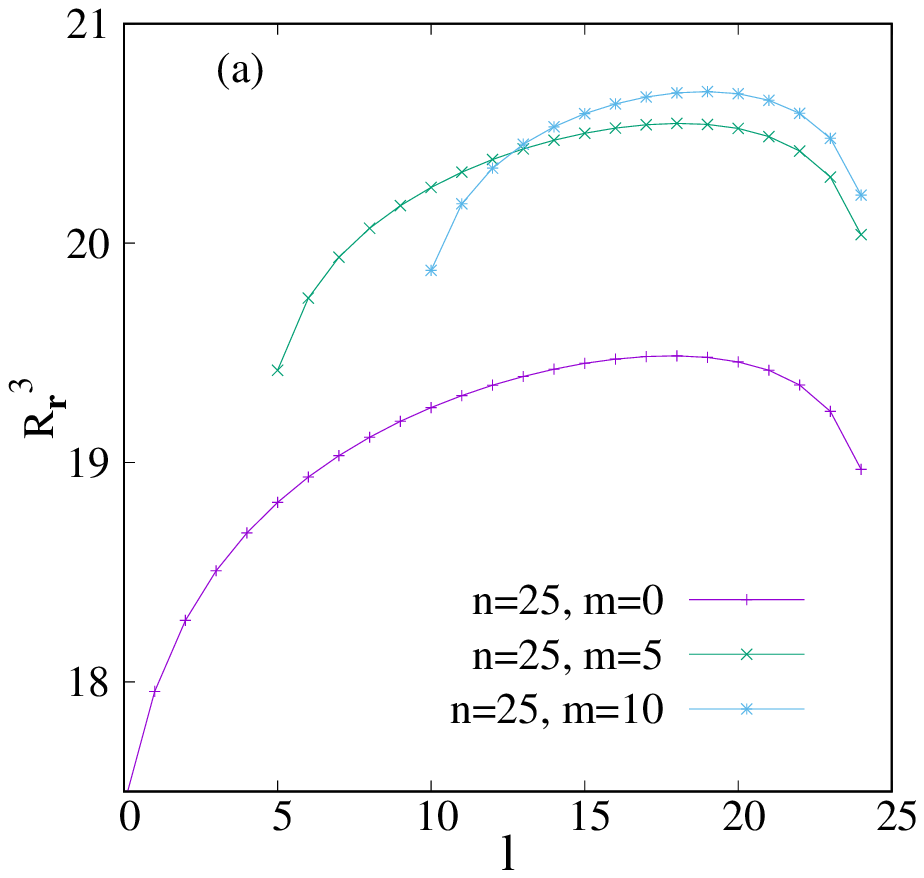}
\end{minipage}%
\hspace{0.5in}
\begin{minipage}[c]{0.35\textwidth}\centering
\includegraphics[scale=0.65]{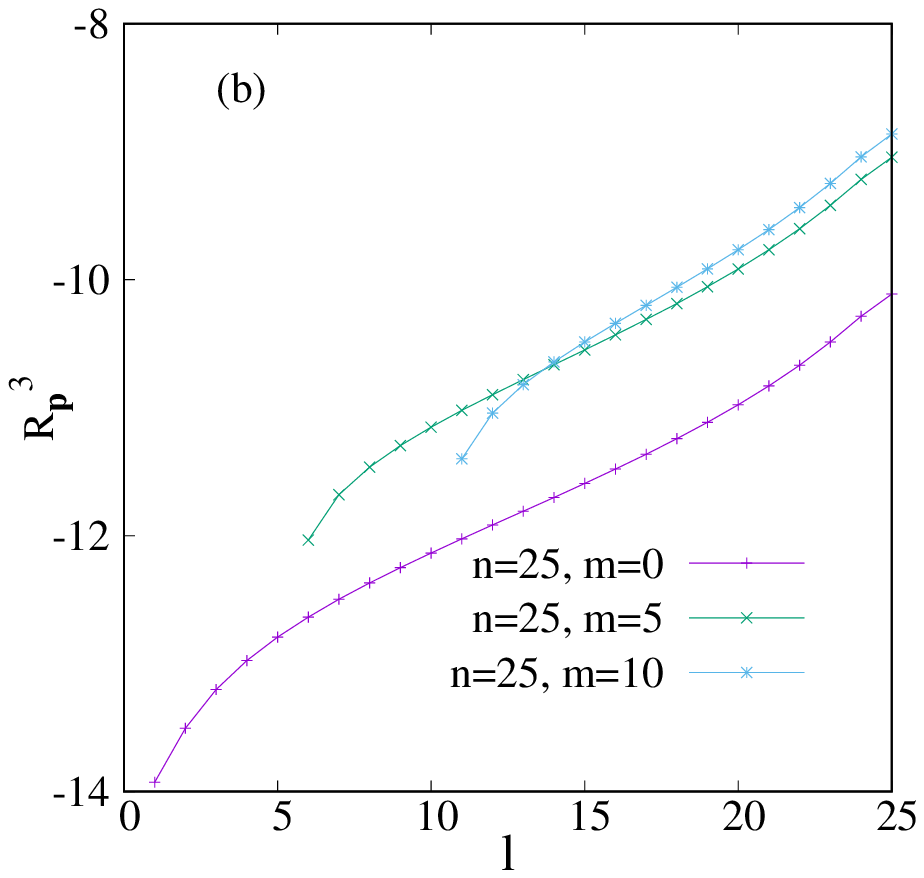}
\end{minipage}%                                                    
\caption{Changes in $R^{\alpha}_{\rvec}, R^{\beta}_{\pvec}$, in panels (a), (b), with $l$, for three particular pairs of 
$(n, m)$, namely, $(25, 0), (25, 5), (25, 10)$, in a FHA. Both $\alpha, \beta$ are kept fixed at 3. See text for more details.}
\end{figure}

\subsection{Free Hydrogen-like atom}
We begin by noting that, radial wave function for H-like atoms (in a.u.) in $r$ and $p$ spaces are given in 
{\color{red}Eqs.~(7) and (12)} respectively. It is well known \cite{romera05,romera06} that for a single particle in central potential, 
$I_{\rvec}, I_{\pvec}$ are amenable to simple closed-form expressions; former in terms of kinetic energy and radial expectation 
value $\langle r^{-2} \rangle$, while latter in terms of root-mean square radius and momentum expectation value $\langle p^{-2} 
\rangle$. For H-like atom, they may be further simplified in terms of $n,l,m$ state indices and $Z$,
\begin{equation}
I_{\rvec}=\frac{4Z^2}{n^2}\left[1-\frac{|m|}{n}\right], \ \ \ I_{\pvec}=\frac{2n^2}{Z^2} 
          \left[ \left(5n^2+1-3l(l+1)\right)-|m|(8n-6l-3) \right]. 
\end{equation}
In case of $m=0$, which we restrict ourselves here, the rightmost term vanishes, leading to, 
\begin{equation}
I_{\rvec}=\frac{4Z^2}{n^2} = 4 \langle p^2 \rangle, \ \ \ I_{\pvec}=\frac{2n^2}{Z^2} \left[ 5n^2+1-3l(l+1) \right] 
         = 4 \langle r^2 \rangle. 
\end{equation}
Thus, like energy, $I_{\rvec}$, solely depends on $n$, whereas, $I_{\pvec}$ on $l$, besides $n$. Thus, it one infers that,
$I_{\rvec}$ $I_{\pvec}$ show opposite behavior with $n$; former falls down whereas latter grows up. This happens because, 
as $n$ increases, kinetic energy lowers, whereas mean square root rises. Additionally, at a fixed $n$, 
$p$-space quantity diminishes with growth of $l$, implying its reduction as number of nodes goes down. In what follows, we 
give detailed tables of $R, T, S, E$ and not $I$. However systematic variations in figures cover all of them including $I$.

\begingroup           %%Table 4, 1s-4f are all checked  
\squeezetable
\begin{table}
\caption{$R_r^{\alpha}$ and $R_p^{\beta}$ for some selected states in FHA $(\alpha=\frac{3}{5}, \beta=3)$. See text for details.}
\centering
\begin{ruledtabular}
\begin{tabular}{cll|cll}   
State  &    $R_r^{\alpha}$    & $R_p^{\beta}$      & State      &  $R_r^{\alpha}$    &     $R_p^{\beta}$ \\ 
\hline 
$1s$   & 2.4448978171250  &  $-$1.29370300309     &   $10s$   &  15.050243          & $-$12.32498491             \\
$2s$   & 6.0819732        &  $-$4.8664081148      &   $10p$   &  15.035316          & $-$11.195758752            \\
$2p$   & 5.7179773964224  &  $-$2.8297112656580   &   $10d$   &  15.005147          & $-$10.57157435657          \\
$3s$   & 8.2848709        &  $-$6.809107089       &   $10f$   &  14.958899          & $-$10.10288080007          \\
$3p$   & 8.1262512        &  $-$5.3065985061863   &   $10g$   &  14.894965          & $-$9.700944996180          \\
$3d$   & 7.7379345080228  &  $-$3.8973824203957   &   $10h$   &  14.810529          & $-$9.32574529             \\
$4s$   & 9.8747416        &  $-$8.1489547505      &   $10i$   &  14.700625          & $-$8.95119472938          \\
$4p$   & 9.7849766        &  $-$6.8261191707376   &   $10k$   &  14.555826          & $-$8.55253099             \\
$4d$   & 9.5879299        &  $-$5.8530683362380   &   $10l$   &  14.355085          & $-$8.0952748              \\
$4f$   & 9.2078601178873  &  $-$4.7097820569485   &   $10m$   &  14.031044068431    & $-$7.521354              \\
\end{tabular}
\end{ruledtabular}
\end{table}
\endgroup

\subsubsection{Circular states}
In this subsection, some exact analytical results are given for the node-less ($n-l=1$) or so-called \emph{circular} states in a 
FHA in $r$, $p$ space. Note that for such states, the two respective polynomials $L_{2l+1}^{n-l-1}\left(\frac{2Z}{n}r\right)$ and 
$C_{n-l-1}^{l+1}\left(\frac{(\frac{np}{Z})^2-1}{(\frac{np}{Z})^2+1}\right)$ both reduce to unity. Hence radial components of 
wave functions in $r$, $p$ spaces simplify to,
\begin{equation}
\begin{aligned}
\psi_{n-l=1}(r) & =  \frac{2}{n^2} \left[\frac{1}{(n+l)!}\right]^{\frac{1}{2}}\left[\frac{2Z}{n}r\right]^{l}e^{-\frac{Z}{n}r}   \\
\psi_{n-l=1}(p) & =  n^{2} \left[\frac{2}{\pi}\frac{1}{(n+l)!}\right]^\frac{1}{2} 2^{(2l+2)} \ l! \ 
\frac{n^l}{ \{[\frac{np}{Z}]^2+1 \}^{l+2}} \left(\frac{p}{Z}\right)^l. 
\end{aligned} 
\end{equation}

At first, the radial entropic moments $\omega_{r}^{\alpha}$ and $\omega_{p}^{\beta}$ are calculated using wave functions in 
{\color{red}Eq.~(24)} and definition in {\color{red}Eq.~(18)}, leading to following forms,
\begin{equation}
\begin{aligned}
\omega_{r}^{\alpha} & =\left(\frac{2Z}{n}\right)^{(3\alpha-3)} 
\left \{ \frac{\Gamma (2l\alpha+3)}{\alpha^{(2l\alpha+3)} \ [\Gamma(2l+3)]^{\alpha}}\right \},  \\
\omega_{p}^{\beta} & =\left(\frac{n}{Z}\right)^{(3\beta-3)}  \ 
\left \{ \frac{\Gamma (2l+2)}{\Gamma\left(\frac{2l+3}{2}\right) \Gamma\left(\frac{2l+5}{2}\right)}\right \}^{\beta} \ 
\left \{ \frac{\Gamma\left(\frac{2l\beta+3}{2}\right) \Gamma\left(\frac{2l\beta+8\beta-3}{2}\right)}{\Gamma(2l\beta+4\beta)}\right \}.
\end{aligned}
\end{equation}
Now, routine mathematical manipulation leads to $R_{r}^{\alpha}$ and $R_{p}^{\beta}$ as,
\begin{equation}
\begin{aligned}
 R_{r}^{\alpha} & =   3 \ln\left[\frac{n}{2Z}\right]+\frac{(2l\alpha+3)}{(\alpha-1)} \ln \alpha +\frac{1}{(\alpha-1)}
\left[ \alpha \ln \{\Gamma (2l+3) \} - \ln \{\Gamma(2l\alpha+3) \} \right], \\
R_{p}^{\beta} &  =   3 \ln\left[\frac{Z}{2^{\frac{1}{3}} n}\right]+\frac{1}{(1-\beta)} \left[\beta \ln\left \{ \frac{\Gamma(2l+4)}
{\Gamma\left(\frac{2l+3}{2}\right) \Gamma\left(\frac{2l+5}{2}\right)} \right \}+ 
\ln\left \{ \frac{\Gamma\left(\frac{2l\beta+3}{2}\right) \Gamma\left(\frac{2l\beta+8\beta-3}{2}\right)}
{\Gamma(2\beta(l+2))}\right \} \right],  
\end{aligned}
\end{equation}
whereas the {\color{red} radial PM R\'enyi entropy} can be written as, {\color{red} $R^{(\alpha, \beta)}=R_{r}^{\alpha}+  R_{p}^{\beta}$}.  
\begingroup           %%Table 5, FHA
\squeezetable
\begin{table}
\caption{$T_r^{\alpha}$ and $T_p^{\beta}$ for some selected states in FHA $(\alpha=\frac{3}{5}, \beta=3)$. See text for details.}
\centering
\begin{ruledtabular}
\begin{tabular}{cll|cll}
State  &    $T_r^{\alpha}$    & $T_p^{\beta}$      & State      &  $T_r^{\alpha}$    &     $T_p^{\beta}$ \\ 
\hline 
$1s$    &    4.14755992475 & $-$6.1476195330   &     $10s$ & 1026.566912  &  $-$2.53697 $\times 10^{10} $ \\
$2s$    &   25.9765245     & $-$8430.486351    &     $10p$ & 1020.508405  &  $-$2.65144 $\times 10^{9}  $ \\
$2p$    &   22.11809373949 & $-$142.99143553   &     $10d$ & 1008.283591  &  $-$7.57534 $\times 10^{8}  $ \\
$3s$    &   66.2336586     & $-$4.104731$\times 10^{5}$  & $10f$ &  989.659422  &  $-$2.98007 $\times 10^{8}  $ \\
$3p$    &   62.0081804     & $-$20333.5038833    &   $10g$ &  964.605525  &  $-$1.33389 $\times 10^{8}  $ \\
$3d$    &   52.72769426026 & $-$1213.4292120388  &   $10h$ &  932.546877  &  $-$6.29880 $\times 10^{7}  $ \\
$4s$    &  127.32504795    & $-$5.984972 $\times 10^{6}$   & $10i$ &  893.652698  &  $-$2.97745 $\times 10^{7}  $ \\
$4p$    &  122.74685003    & $-$4.246794 $\times 10^{5}$   & $10k$ &  841.846208  &  $-$1.34213 $\times 10^{7}  $ \\
$4d$    &  113.56339555    & $-$6.0656449$\times 10^{4}$   & $10l$ &  776.747706  &  $-$5.37571 $\times 10^{7}  $     \\
$4f$    &   96.92810090882 & $-$6163.10389494375    & $10m$ &  682.0134947207 & $-$1.70583 $\times 10^{6} $  \\
\end{tabular}
\end{ruledtabular}
\end{table}
\endgroup 
                   
{\color{red}Equation~(26)} provides  $R_{r}^{\alpha}$, $R_{p}^{\beta}$ for arbitrary $\alpha$, $\beta$. In order to compute 
{\color{red} radial PM R\'enyi entropy $R^{(\alpha, \beta)}$,} the relation $(\frac{1}{\alpha}+\frac{1}{\beta}=2)$ should be satisfied between them. Figure~1 
graphically shows variations of $R_{r}^{\alpha}$, $R_{p}^{\beta}$ in panels (a), (b) with $\alpha$, $\beta$ 
respectively, for lowest five node-less states $1s, 2p, 3d, 4f, 5g$. It follows that with progression in $\alpha, \beta$, both 
$R_{r}^{\alpha}$, $R_{p}^{\beta}$ lessen for all of them. Further, $R_r^{\alpha}$'s, $R_p^{\beta}$'s appear to behave contrastingly
with upward changes in $n,l$; former assume progressively higher values, while latter go down. Starting from an initial 
value, all these fall quite sharply in lower $\alpha, \beta$ regions and tend to flatten as the latter two widen. Moreover, the 
extent of fall-off slows down as $n,l$ tend to grow. 

Now, Fig.~2 displays the plots of $R_{r}^{\alpha}, R_{p}^{\beta}$ and {\color{red} radial PM R\'enyi entropy $R^{(\alpha, \beta)}$,} with $n$ at three chosen 
sets of $\alpha \ (\frac{3}{5}, 1, 3)$, $\beta \ (\frac{3}{5}, 1, 3)$ and $(\alpha, \beta)$, namely $(\frac{3}{5}, 3), (1,1),
(3, \frac{3}{5})$ in three panels (a)-(c). Recall that, the above $R$'s corresponding 
to set $\alpha=\beta=1$ represent $S_{r}, S_{p}$, {\color{red} radial PM Shannon entropy $S$} respectively. While range of $n$ remains fixed in all three plots, same 
for $y$ axis differs in all cases. For all $\alpha$, $R_{r}^{\alpha}$'s rise with $n$--more sharply at
smaller $n$'s, See text for details. and rate of progress taking a dive with $n$. At a given $n$, $R^{\alpha}_r$ tends to diminish continuously 
as order of moment enhances. On the other hand, $R_{p}^{\beta}$ in (b) shows a complimentary behavior to (a), steadily 
falling as $n$ grows. A combined effect of these two produces the plot in panel (c), quite similar in qualitative 
nature as in (a), with visible differences in the values in $y$ axis. Since $\alpha, \beta$ obey the relation $\frac{1}{\alpha}+
\frac{1}{\beta}=2$, when $\alpha > 1$, $\beta < 1$ and vice versa. Evidently, in both situations, corresponding changes 
in $R_{r}^{\alpha}, R_{p}^{\beta}$, $R^{(\alpha, \beta)}$ maintain similar trend. The above reasoning may be interpreted in terms
of radial probability distribution getting more diffused with $n$. 
{\color{red} It is appropriate to mention here that, the bounds provided in Eqs.~(14), (15) and (19) are applicable to total PM Fisher 
information, total PM Shannon entropy and total PM R\'enyi entropy. Thus, in case of FHA, the \emph{radial} PM R\'enyi, PM Shannon entropy and 
PM Onicescu energies, reported here, are not subject to such bounds.}  

\begingroup           %%Table 6, 1s-4f are all checked  
\squeezetable
\begin{table}
\caption{$S_r, S_p$ for some selected states in FHA. See text for details.}
\centering
\begin{ruledtabular}
\begin{tabular}{cll|cll}
State  &    $S_{r}^{\ddag}$          & $S_{p}^{\ddag}$               & State         &  $S_r^{\S}$              &     $S_p^{\S}$ \\ 
\hline 
$1s^\dagger$   & 1.6137056388801     &  $-$0.1091619058    &   $10s$   &  14.83421801  & $-$8.5831     \\
$2s$   & 5.579905117         &  $-$3.288603        &   $10p$   &   14.81546079         & $-$8.40306    \\
$2p$   & 5.1658184934843     &  $-$2.056657825     &   $10d$   &   14.779519706        & $-$8.22058    \\
$3s$   & 7.895456983         &  $-$4.71928         &   $10f$   &   14.726933588        & $-$8.03408    \\
$3p$   & 7.706768439         &  $-$3.988042        &   $10g$   &   14.657200818        & $-$7.84526    \\
$3d$   & 7.3045091959407     &  $-$3.273842250     &   $10h$   &   14.568453746        & $-$7.65744    \\
$4s$   & 9.543883432         &  $-$5.67677         &   $10i$   &   14.456574316        & $-$7.47587    \\
$4p$   & 9.434788623         &  $-$5.162422        &   $10k$   &   14.312835432        & $-$7.30982    \\
$4d$   & 9.220979188         &  $-$4.635591        &   $10l$   &   14.116240399        & $-$7.1796     \\
$4f$   & 8.8401955766914     &  $-$4.169046134     &   $10m$   &   13.794498337697     & $-$7.1533     \\
\end{tabular}
\end{ruledtabular}
\begin{tabbing}
$^{\dagger}$Literature results \cite{aquino13} for $S_{\rvec}=S_r+S_{(\theta, \phi)}$ and $S_{\pvec}=S_p+S_{(\theta, \phi)}$ are 
4.1447 and 2.4219 respectively. Present values are \\ 4.14472988585 and 2.42186234117. \\
[0pt \color{red}] $^{\ddag}$Literature results {\color{red}\cite{jiao17}} of $S_{r},~S_{p}$ for $1s$-$4f$ states are: 
(1.6137056388,$-$0.1091619058),~(5.5799051176,~$-$3.2886034474), \\
~(5.1658184934,~$-$2.0566578254),~(7.8954569837,~$-$4.7192844860),~(7.7067684395,~$-$3.9880420674),~(7.3045091959,~$-$3.2738422502), \\
~(9.5438834322,~$-$5.6767751478),~(9.4347886234,~$-$5.1624221872),~(9.2209791882,~$-$4.6355912037),~(8.8401955766,~$-$4.1690461340)  \\
respectively. \\ 
$^{\S}$Literature results {\color{red}\cite{jiao17}} of $S_{r},~S_{p}$ for $10s$-$10m$ states are: (14.834218018,~-8.583082598), ~(14.815460797,~$-$8.403065247),  \\
~(14.779519706,~$-$8.220588961),~(14.726933588,~$-$8.034081080), ~(14.657200818,~$-$7.845266303),~(14.568453746,~$-$7.657443887),  \\
~(14.456574316,~$-$7.475870737), ~(14.312835432,~$-$7.309826844),~(14.116240399,~$-$7.179685623),~(13.794498337,~$-$7.153386777) \\
 respectively.
\end{tabbing}
\end{table}
\endgroup

Next we proceed for $T_{r}^{\alpha}$ and $T_{p}^{\beta}$, which are obtained from {\color{red}Eq.~(20)} as,
\begin{equation}
\begin{aligned}
T_{r}^{\alpha} & =\frac{1}{\alpha-1}\left[1-\left(\frac{2Z}{n}\right)^{(3\alpha-3)} 
\left \{ \frac{\Gamma (2l\alpha+3)}{\alpha^{(2l\alpha+3)}[\Gamma(2l+3)]^{\alpha}}\right \} \right], \\
T_{p}^{\beta} & =\frac{1}{\beta-1}\left[1-2^{(\beta-1)}\left(\frac{n}{Z}\right)^{(3\beta-3)} 
\left \{  \frac{\Gamma (2(l+2))}{\Gamma\left(\frac{2l+3}{2}\right)
\Gamma\left(\frac{2l+5}{2}\right)}\right \}^{\beta} \left \{ \frac{\Gamma\left(\frac{2l\beta+3}{2}\right)
\Gamma\left(\frac{2l\beta+8\beta-3}{2}\right)}{\Gamma(2\beta l+4\beta)}\right \} \right].
\end{aligned}
\end{equation}
Evidently, $T_{r}^{\alpha}, T_{p}^{\beta}$ show analogous behavior as $R_{r}^{\alpha}, R_{p}^{\beta}$ in {\color{red}Eq.~(26)}, namely,
$T_{r}^{\alpha}$ grows up, whereas $T_{p}^{\beta}$ reduces with successive upward changes in $n$.  
A characteristic feature of $R$, $T$ is that, when $(\alpha, \beta) \rightarrow 1$ we have $(R_{r}^{\alpha}, T_{r}^{\alpha}) 
\rightarrow S_{r}$ and $(R_{p}^{\beta}, T_{p}^{\beta}) \rightarrow S_{p}$. Therefore we employ {\color{red}Eqs.~(26), (27)} separately (both 
lead to same result obviously) in this limit to secure following expressions for $S_r, S_p$ in node-less states, \emph{viz.,} 
\begin{equation}
\begin{aligned}
S_{r} & =3 \ln\left[\frac{n}{2Z}\right]+(2l+3)+ \ln[\Gamma(2l+3)]-2l\left[\sum_{k=1}^{2l+2}\left(\frac{1}{k}\right)-C\right], \\
S_{p} & = \ln\left[ \frac{Z^3}{2n^3} \ \frac{\Gamma \left(\frac{2l+3}{2}\right) \Gamma\left(\frac{2l+5}{2}\right)}
{\Gamma(2l+4)}\right]- l \ \frac{\Gamma'\left(\frac{2l+3}{2}\right)}{\Gamma\left(\frac{2l+3}{2}\right)} 
 -(l+4)\frac{\Gamma'\left(\frac{2l+5}{2}\right)}{\Gamma\left(\frac{2l+5}{2}\right)}    
  +(2l+4) \frac{\Gamma'\left(2l+2\right)}{\Gamma\left(2l+2\right)}. 
\end{aligned}
\end{equation}
Here $\frac{\Gamma' (t)}{\Gamma(t)}$ refers to the Poly-Gamma function with order $0$, and Euler Constant 
$C$ equals to $0.57721~56649~01532~86060~651209 \cdots $. The {\color{red} radial PM Shannon entropy $S$} is then gathered as sum of individual components 
$S_r$ and $S_p$. 

\begingroup           %%Table 7, 1s-4f are all checked  
\squeezetable
\begin{table}
\caption{$E_r$ and $E_p$ for some selected states in FHA. See text for details.}
\centering
\begin{ruledtabular}
\begin{tabular}{cll|cll}
State  &    $E_r$          & $E_p$               & State         &  $E_r$              &     $E_p$ \\ 
\hline 
$1s$   & 0.5               &    2.626056561016    &   $10s$   &  0.000000943317      &  93503.5290      \\
$2s$   & 0.009765625       &   96.1295856275048   &   $10p$   &  0.000000841581      &  39072.5385      \\
$2p$   & 0.001398822737580 &   13.793428401297597 &   $10d$   &  0.000000801822      &  23397.786166    \\
$3s$   & 0.000964506172839 &  599.4570931556239   &   $10f$   &  0.000000785289      &  15697.1858628 \\
$3p$   & 0.000884130658436 &  160.07008401467374  &   $10g$   &  0.000000784339      &  11036.824894    \\
$3d$   & 0.001012731481481 &   41.63970776113258  &   $10h$   &  0.000000797955      &   7885.727210    \\
$4s$   & 0.000185489654541 & 2072.833978828845    &   $10i$   &  0.000000829206      &   5609.7326559   \\
$4p$   & 0.000166416168212 &  676.412752272371    &   $10k$   &  0.000000887467      &   3899.8392585   \\
$4d$   & 0.000170230865478 &  277.6680799158443   &   $10l$   &  0.000000999887      &   2595.614180    \\
$4f$   & 0.000204563140869 &   95.71851114591887  &   $10m$   &  0.000001285853      &   1659.760152    \\
\end{tabular}
\end{ruledtabular}
\end{table}
\endgroup

Additionally, we have derived $S_{r},S_{p}$ for such states using an alternate method applying the definition in {\color{red}Eq.~(16)}.  
It turns out that the analytical expression of $S_r$ obtained from this route is completely identical to that in {\color{red}Eq.~(28)}. However for $S_p$,
one obtains the following expression from {\color{red}Eq.~(16)}, 
\begin{equation}
{\color{red} \color{red} S_{p} = \ln\left[\frac{Z^{3}\Gamma \left(\frac{2l+3}{2}\right) 
\Gamma\left(\frac{2l+5}{2}\right)}{2n^{3}\Gamma(2l+4)}\right]+  \left[\frac{\sqrt \pi l \ \Gamma (2l+3)}{2^{(2l+2)}\Gamma(l+2)\Gamma\left(\frac{2l+5}{2}\right)}\right]
 +(4l+8)\left[\frac{\Gamma(2l+4)}{\Gamma\left(\frac{2l+5}{2}\right)\Gamma\left(\frac{2l+3}{2}\right)}\right]I_{p}^{l}.} 
\end{equation}
The integration in last line is defined as given below, 
\begin{equation}
I_{p}^{l}=\int_{0}^{\infty}\frac{p^{(2l+2)}}{(1+p^2)^{(2l+4)}} \ \ln(1+p^2) \ \mathrm{d}p,  
\end{equation} 
which can be computed for a specific $l$ numerically quite easily.

{\color{red} Table~S1}, however, demonstrates that the two expressions of $S_{p}$, in {\color{red}Eqs.~(28) and (29)}, although apparently different, actually produce 
virtually identical numerical results. Here we offer lowest eleven node-less states for the 
purpose of illustration, but this has been found to be generally valid for other states as well. One can surmise that $S_p$ 
falls down steadily as $n$ rises. Further, the graphs $\alpha=1$ and $\beta=1$ in panels (a), (b) of Fig.~2
endorse that, for circular states, $S_{r}$ rises and $S_{p}$ diminishes with $n$ advancing forward.                 

From the foregoing analysis, it is realized that, $R_{r}^{\alpha}~,T_{r}^{\alpha}$ and $S_{r}$ gain with $n$. Conversely, 
$R_{p}^{\beta}~,T_{p}^{\beta}$, $S_{p}$ assume reverse trend with $n$. Because, in circular states, the $r$-space density 
gets more diffused as $n,l$ progress, without changing number of nodes. Finally, we move on to $E$ in such states
in FHA, keeping in mind that, $\alpha=\beta=2$ in {\color{red}Eq.~(25)} leads to radial Onicescu energy in position ($E_{r}$) and momentum 
($E_{p}$) spaces respectively. After some straightforward algebraic manipulation, one gets, 
\begin{equation}
\begin{aligned}
E_{r} & =\left(\frac{2Z}{n}\right)^{3}\left[\frac{\Gamma(4l+3)}{2^{(4l+3)} \ \Gamma(2l+3)^{2}}\right],  \\
E_{p} & =2\left(\frac{n}{Z}\right)^{3}\left[\frac{\Gamma (2(l+2))}{\Gamma\left(\frac{2l+3}{2}\right)
\Gamma\left(\frac{2l+5}{2}\right)}\right]^{2} \left[\frac{\Gamma\left(\frac{4l+3}{2}\right)
\Gamma\left(\frac{4l+13}{2}\right)}{\Gamma(4l+8)}\right].
\end{aligned}
\end{equation}
One discerns from {\color{red}Eq.~(31)} that, $E_{r}$ declines whereas $E_{p}$ grows as $n$ rises. This is in accordance with our previous 
conclusion that, delocalization escalates, as $n$ becomes larger.    

\begin{figure}                         %%%Fig. 5, FHA
\begin{minipage}[c]{0.3\textwidth}\centering
\includegraphics[scale=0.55]{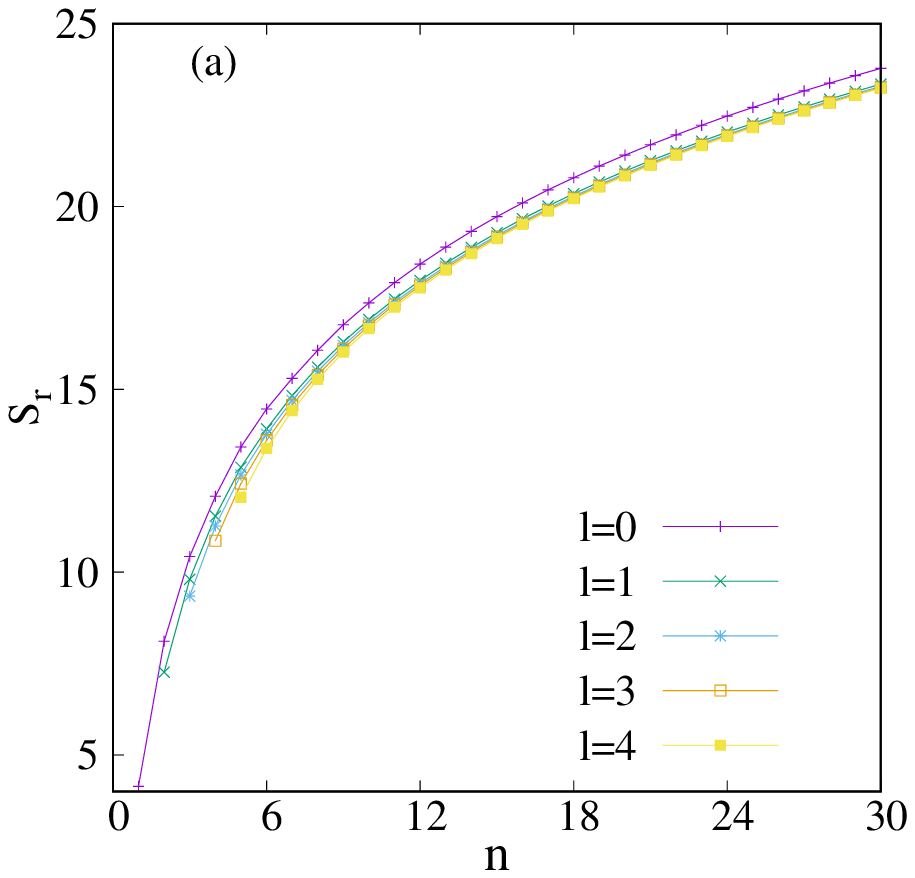}
\end{minipage}%
\hspace{0.15in}
\begin{minipage}[c]{0.3\textwidth}\centering
\includegraphics[scale=0.55]{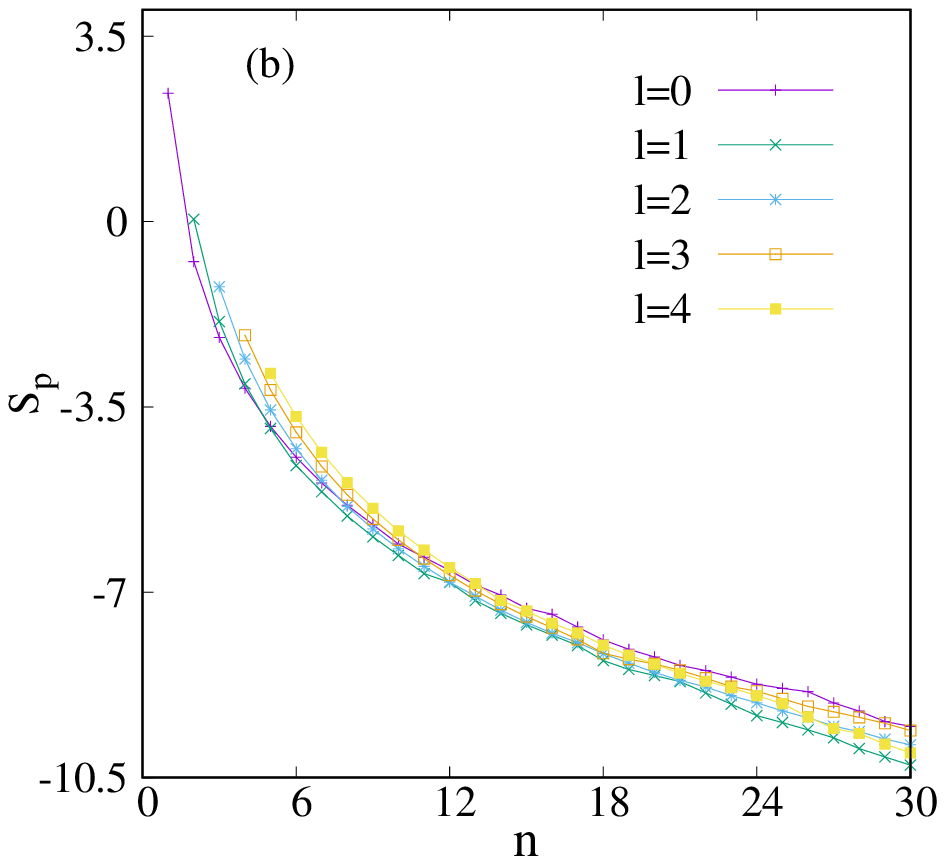}
\end{minipage}%
\hspace{0.15in}
\begin{minipage}[c]{0.3\textwidth}\centering 
\includegraphics[scale=0.55]{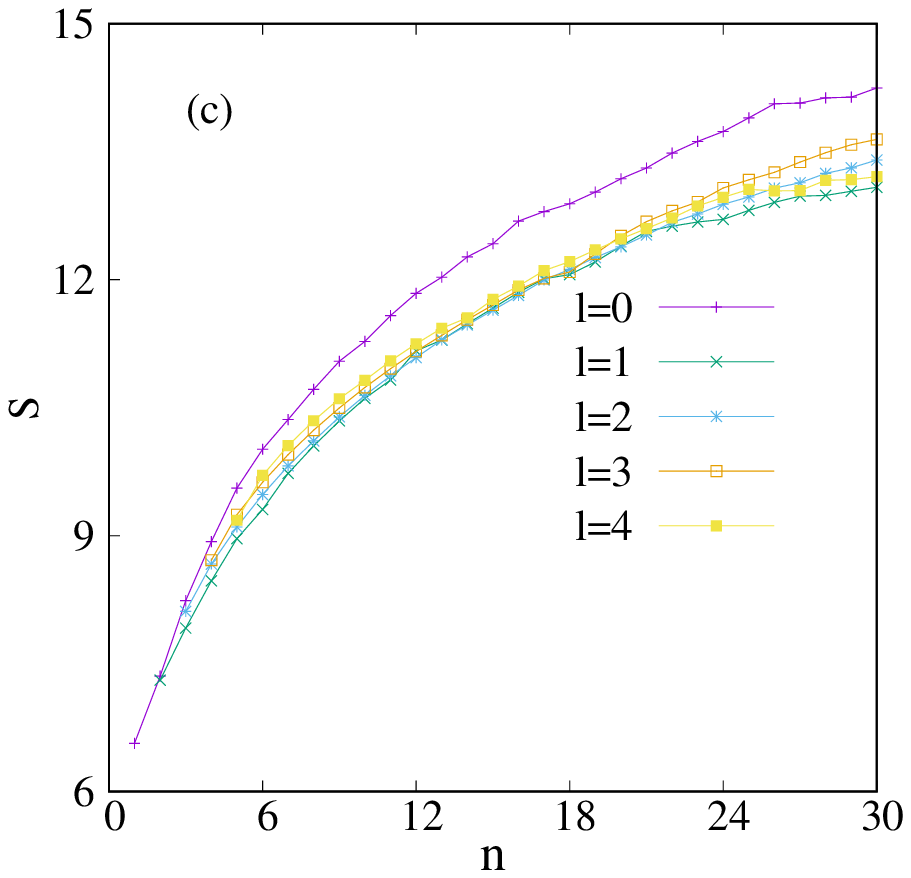} 
\end{minipage}%
\caption{Changes in $S_{r}, S_{p}$, {\color{red}radial PM Shannon entropy $S$} with $n$ at five lowest $l$ (0-4) in FHA. For more details, see text.}
\end{figure}

\subsubsection{States with arbitrary $n,l$}
This subsection is now devoted to a discussion of $S, R, T, E$ for an \emph{arbitrary} state (not necessarily \emph{circular}) in a FHA. 
Several attempts were made to derive analytical results for $S, R, T, E$ in such states. The main problem remains rooted in 
integrating the occurring polynomials with certain power ($\alpha$ or $\beta$) for $R, T$ or in logarithmic form for $S$. 
These functions are exactly integrable when $\alpha, \beta$ are integers (as they assume finite form). In same token, $E$ 
($\alpha=\beta=2$) for FHA can be written analytically for integer $\alpha, \beta$ for all states-though for larger $n,l$, 
number of contributing terms in polynomial appearing in wave function populates considerably, making it rather cumbersome. 
However, for fractional $\alpha, \beta$, these ($R, T$) lead to infinite series; hence become quite intractable. Some recent works 
\cite{toranzo16,toranzo16a} have reported a few analytical expressions of $R$ with particular approximations in terms of Airy and 
Bessel functions, for $l=0$ states of a D-dimensional H-like atom in $r$ space.

\begin{figure}             %%% Fig. 6, FHA
\begin{minipage}[c]{0.3\textwidth}\centering
\includegraphics[scale=0.55]{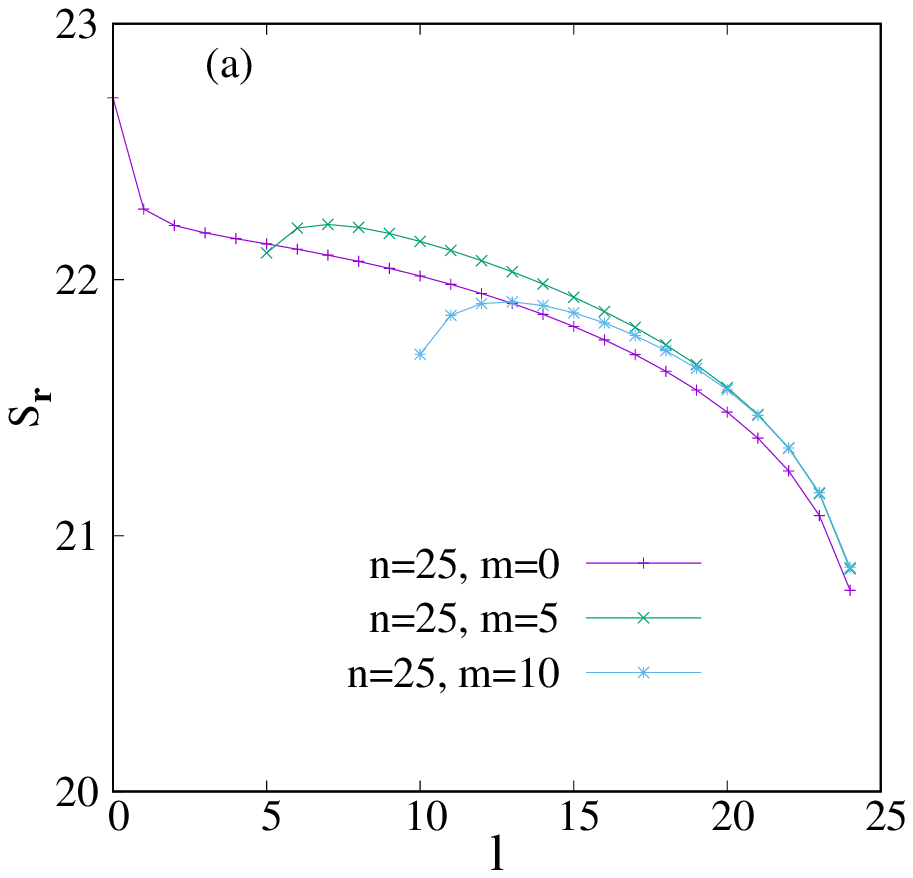}
\end{minipage}%
\hspace{0.2in}
\begin{minipage}[c]{0.3\textwidth}\centering
\includegraphics[scale=0.55]{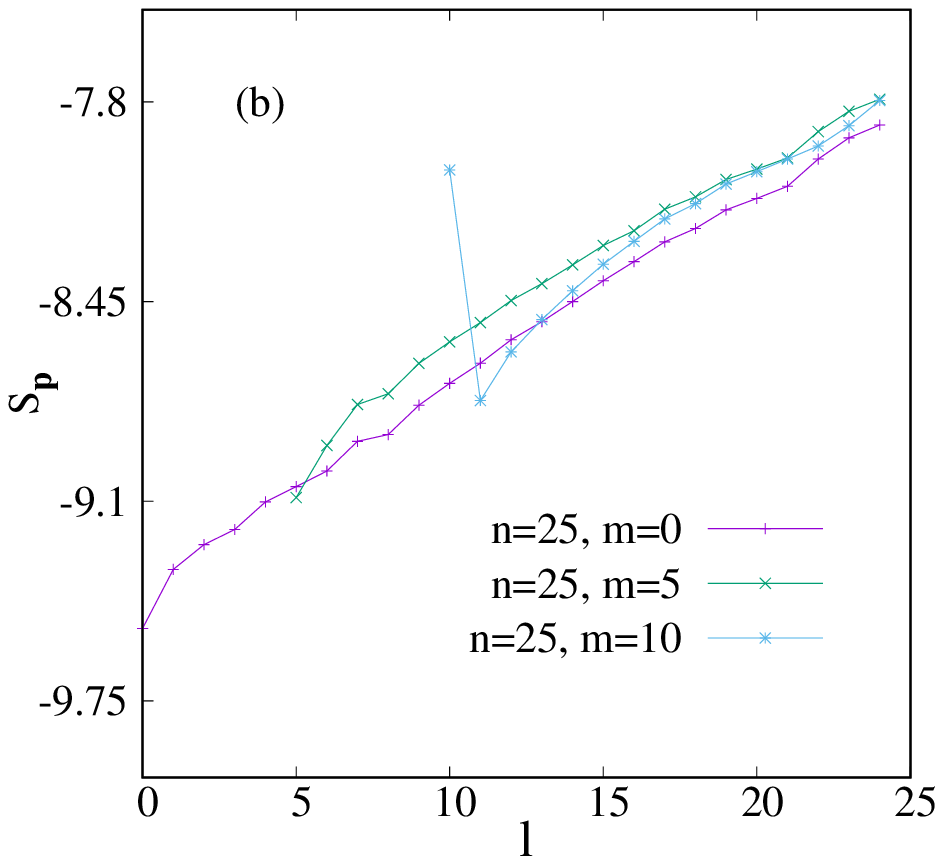}
\end{minipage}%
\hspace{0.2in}
\begin{minipage}[c]{0.3\textwidth}\centering
\includegraphics[scale=0.55]{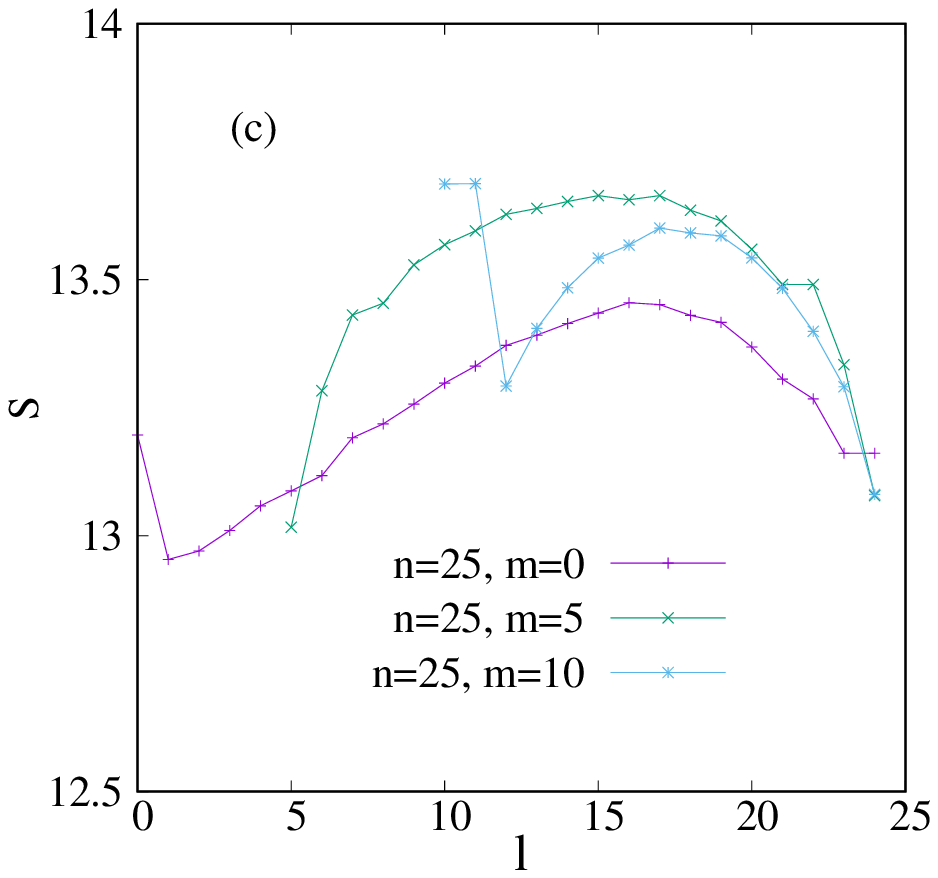}
\end{minipage}%
\caption{Variation of $S_r, S_p$ and {\color{red} radial PM Shannon entropy $S$} respectively, in panels (a), (b), (c), with $l$, for three specific pairs of $(n, m)$, 
namely, $(25, 0), (25, 5), (25, 10)$, in a FHA. More details are available in the text.}
\end{figure}

Before going to a detailed analysis, at first in {\color{red} Table~III} we present the angular parts of concerned information quantities, 
\emph{viz.}, $S_{(\theta, \phi)}, R^{\alpha}_{(\theta, \phi)}, R^{\beta}_{(\theta, \phi)}, T^{\alpha}_{(\theta, \phi)}, 
T^{\beta}_{(\theta, \phi)}, E_{(\theta, \phi)}$ in columns 2-7 in a H atom. These are offered for 10 lowest states, keeping 
$m$ fixed at 0, with specific set of $\alpha, \beta$ corresponding to $\frac{3}{5}$, 3 respectively. It may be noted that
all future calculations employ same $\alpha, \beta$. {\color{red} So far such results are only known for $S_{\theta, \phi}$ 
{\color{red}\cite{jiao17}}, where the wave functions were expanded in cut-off STOs and L\"owdin's canonical othogonalization method
was used. Our computed $S_{\theta, \phi}$ values are in complete consistence with these results (given below the {\color{red}Table~III}). But,
no reported results are available for $ R^{\alpha}_{(\theta, \phi)}, R^{\beta}_{(\theta, \phi)}, T^{\alpha}_{(\theta, \phi)}, 
T^{\beta}_{(\theta, \phi)}, E_{(\theta, \phi)}$, to the best of our knowledge.} Their deviations with respect to numerical parameters 
were carefully checked and are reported here up to the extent to which convergence was attained. These would be applicable to both 
FHA and CHA, as we are interested only in radial confinement. It is seen that, $S_{(\theta, \phi)}$  as well as $p$-space components 
of $R, T$, i.e., $R^{\beta}_{(\theta, \phi)}$ $T^{\beta}_{(\theta, \phi)}$ gradually fall off as $l$ grows, while an opposite behavior 
is recorded for $E_{(\theta, \phi)}$; in both occasions the extent is lowered as we go down the table. It is interesting to observe that, both 
$R^{\alpha}_{(\theta,\phi)}$, $T^{\alpha}_{(\theta,\phi)}$ initially decline, then attain minima at $l=7$ and again tend to 
ascend thereafter. 

Let us now focus on the nature of radial parts of such information quantities. Figure~3 presents rise, fall of $R_{r}^{\alpha}$, 
$R_{p}^{\beta}$ respectively with changes of $n$ in panels (a), (b) for FHA in case of lowest five $l$ states. This time, we 
have chosen $\alpha=\beta=3$. Both quantities assume larger value with $l$ for a given $n$. This may occur due to spreading of state 
function with $n,l$. Also, separation amongst $l$ widens as $n$ goes up. Further, Fig.~4 depicts changes in $R_{\rvec}^{\alpha}$, 
$R_{\pvec}^{\beta}$ with $l$ keeping $n$ fixed (25) at three different $m$ namely, $0,~5,~10$, for same $\alpha, \beta$ of previous 
figure. Note that, unlike other figures of FHA, here the graphs include angular contributions. In each of these three $m$, one finds
a hump in (a) segment, all passing through a maximum. Whereas right side reveals that $R_{\pvec}^{\beta}$ climbs up with $l$. It may 
be noted that, for a fixed $n$, number of radial nodes lowers as $l$ increments. On the other hand, if $n$ is sufficiently large 
(which is the case here), radial orbitals get comparably (and considerably as well) extended amongst all available $l$. Now, from 
discussion of {\color{red} Table~III}, we know that, with $l$, angular contribution initially grows up, then attains a maximum and in the 
end falls down. Thus it is not quite straightforward to explain such graphs. Rather, these calculations may be used
to grasp the distribution pattern of total probability in high-$n,l$ states. 

\begin{figure}                         %%%Fig. 7, CHA
\begin{minipage}[c]{0.35\textwidth}\centering
\includegraphics[scale=0.55]{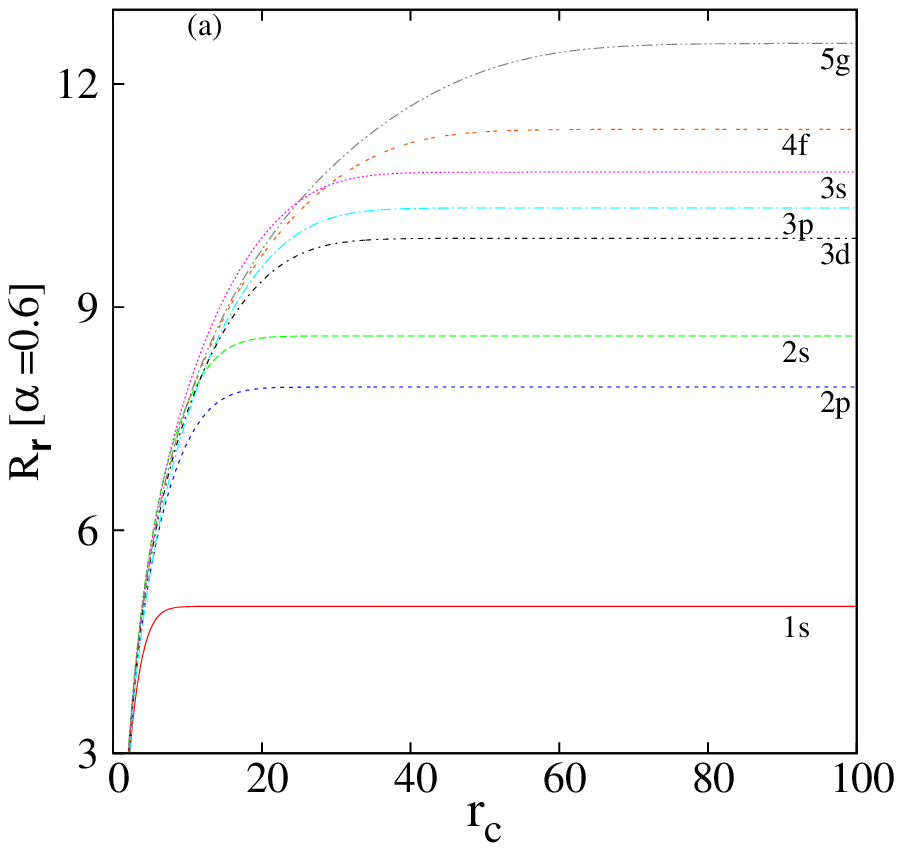}
\end{minipage}%
\begin{minipage}[c]{0.35\textwidth}\centering
\includegraphics[scale=0.55]{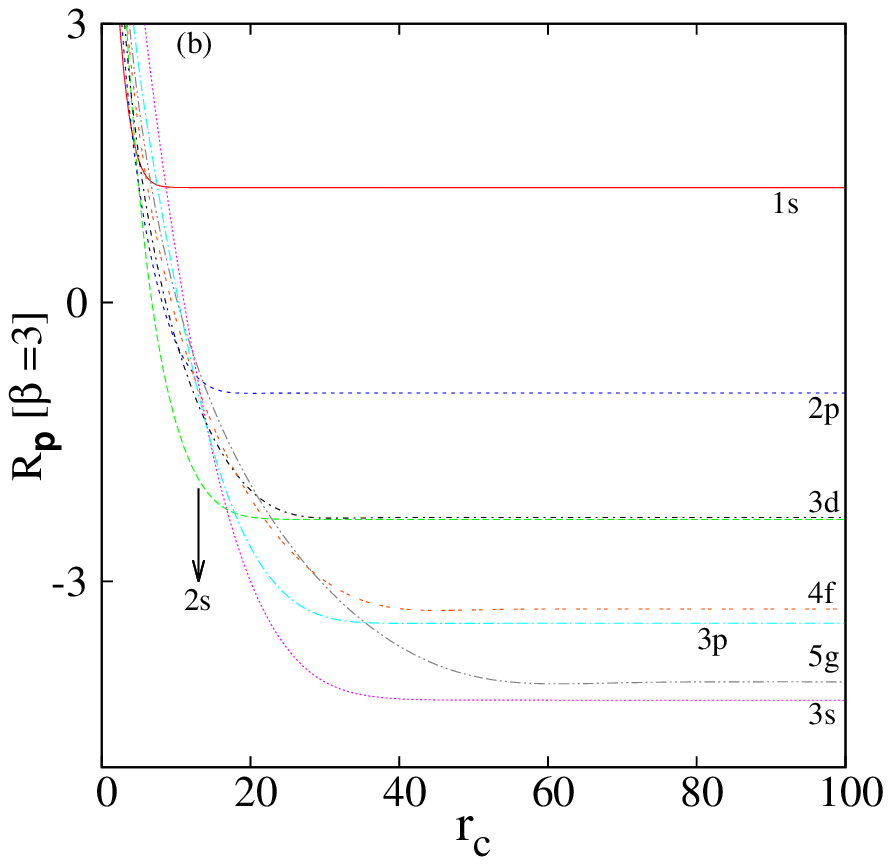}
\end{minipage}%
\begin{minipage}[c]{0.35\textwidth}\centering
\includegraphics[scale=0.55]{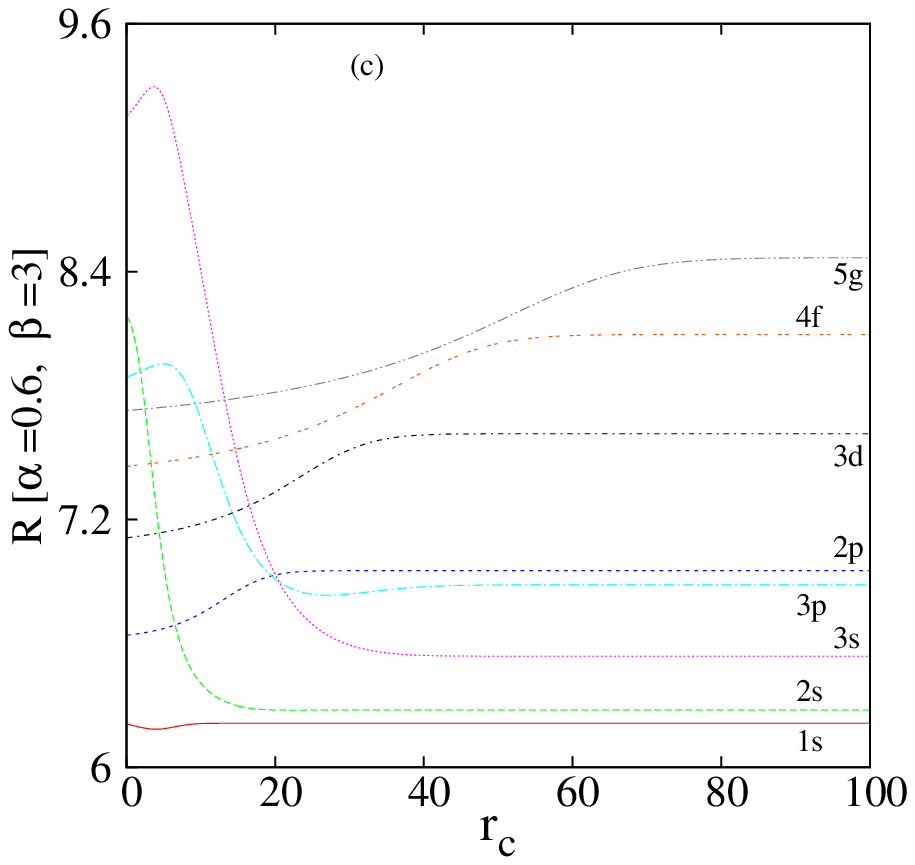}
\end{minipage}%
\caption{Plot of $R_{\rvec}^{\alpha}$, $R_{\pvec}^{\beta}$, {\color{red} total PM R\'enyi entropy $R^{(\alpha, \beta)}$} against $r_c$ for some low-lying states 
of CHA, having $\alpha=\frac{3}{5}, \beta=3$, in panels (a), (b), (c) respectively. {\color{red} $R^{(\alpha, \beta)}$'s for all these states obey the 
lower bound condition given in Eq.~(19).} More details can be found in text.}
\end{figure}

Now we turn to the main results of FHA. {\color{red}Table~IV} provides our numerically estimated values for $R_r^{\alpha}, R_p^{\beta}$
in left portion, for $1s$-$4f$ states, while right side correspond to those for all ten $l$ states belonging to $n=10$. Left, 
right side of {\color{red}Table~V} display similar entries for $T_r^{\alpha}$, $T_p^{\beta}$, for above same states, while results for  
$S_r, S_p$ and $E_r, E_p$ are tabulated in {\color{red}Tables~VI and VII}. {\color{red} Except for the case of $S_{r}, S_{p}$ in {\color{red}Table~VI}, 
all these are offered here for first time and we hope they would be useful for future referencing and stimulate further 
research in this direction. Literature result for $S_r, S_p$ is based on a variational calculation {\color{red}\cite{jiao17,aquino13}}, 
which shows reasonable agreement with our present finding.} Note that reference data of \cite{aquino13} correspond to $S_{\rvec /\pvec}$ 
rather than $S_{r/p}$; therefore need to be adjusted for their appropriate angular contributions $S_{(\theta, \phi)}$. At this point, 
it may be recalled that, a lowering and raising in $R,~T,~S$ reflects global extension and concentration of density distribution 
in corresponding spaces. It is noticed from {\color{red} Tables~IV-VI} that, for all $n$ considered, $R,T,S$ diminish and  
progress in $r$, $p$ spaces as $l$ ascends. On the other hand, {\color{red} Table~VII} suggests that, for a specific $l$, $E_r$ and 
$E_p$ show contrasting behavior (fall, rise respectively) with $n$, which could be attributed to a gain in number 
of radial nodes, leading to a delocalization of electron. Interestingly, as $l$ changes for a given $n$, $E_r$ at first drops
down, attains a minimum before surging further. Conversely, for same reason, $E_p$ descends with growth of $l$. 
This time, number of radial nodes dips, but with $l$ going up, there is a spread in probability distribution. Due 
of these two contrasting factors, one encounters inflection points in $E_r$. 
 
Figure~5 now registers nature of changes in $S_{r}$ (a), $S_{p}$ (b), {\color{red} radial PM Shannon entropy $S$} (c) with $n$ at first five $l$ ($0-4$). Evidently, 
$S_r$, $S$ tend to grow with $n$, whereas $S_{p}$ shows reverse effect. This is in keeping with that, as $n,l$ are raised, radial 
orbitals get extended in space. It is worth pointing out that unlike $R$, $S_r$ and $S$ slump with $l$. But, $S_{p}$ does 
not permit any straightforward motif.   

\begingroup           %%Table 8, R for 1s-2s in CHA, all checked 
\squeezetable
\begin{table}
\caption{$R_{\rvec}^{\alpha}, R_{\pvec}^{\beta}$ and {\color{red} total PM R\'enyi entropy $R^{(\alpha, \beta)}= (R_{\rvec}^{\alpha}+R_{\pvec}^{\beta})$} for lowest 
two $s$ states in a CHA, at various $r_c$, for $\alpha=\frac{3}{5}, \beta=3$ respectively. See text for more details.}
\centering
\begin{ruledtabular}
\begin{tabular}{llllllll}
\multicolumn{4}{c}{$1s$}    &      \multicolumn{4}{c}{$2s$}    \\
\cline{1-4} \cline{5-8}
$r_c$  &    $R_{\rvec}^{\alpha}$     & $R_{\pvec}^{\beta}$  &  $R^{(\alpha, \beta)}$  &  
$r_c$  &    $R_{\rvec}^{\alpha}$     & $R_{\pvec}^{\beta}$  &  $R^{(\alpha, \beta)}$  \\
\hline 
0.1   & $-$6.0449530234201  & 12.2544945    &  6.2095414   &  0.1   &  $-$6.0652785667052  &   14.2461812    & 8.18090263      \\
0.2   & $-$3.9740686542021  & 10.1826733    &  6.20860464  &  0.2   &  $-$3.9857010841026  &   12.1605425    & 8.17484143      \\
0.3   & $-$2.7665379461615  &  8.97420833   &  6.20767038  &  0.3   &  $-$2.7690858567973  &   10.9370670    & 8.16798114       \\
0.5   & $-$1.2527639276520  &  7.4585759    &  6.20581197  &  0.5   &  $-$1.2359050275123  &    9.3875148    & 8.15160977      \\
0.6   & $-$0.7156642633591  &  6.9205535    &  6.2048892   &  0.6   &  $-$0.6884504327452  &    8.83042433   & 8.1419738      \\
0.8   &    0.1265545289041  &  6.0765062    &  6.2030607   &  0.8   &     0.1758653404988  &    7.9436427    & 8.1195080      \\
1.0   &    0.7735958787514  &  5.4276644    &  6.20126027  &  1.0   &     0.8469685980263  &    7.2454746    & 8.0924431       \\
1.5   &    1.9263580259098  &  4.27057585   &  6.19693387  &  3.0   &     4.1824208766826  &    3.3944866    & 7.5769074      \\
2.5   &    3.2916372871390  &  2.89792262   &  6.18955990  &  5.0   &     5.7661541562104  &    1.2155313    & 6.9816854      \\
3.0   &    3.7310884276653  &  2.45579844   &  6.1868868   &  7.5   &     6.9655961664353  & $-$0.3916768    & 6.5739193     \\
4.0   &    4.3257559261586  &  1.85876674   &  6.1845226   &  10.0  &     7.7053107207956  & $-$1.2984101    & 6.4069006      \\
5.0   &    4.6620663954973  &  1.5246585    &  6.1867248   &  12.0  &     8.0874246952222  & $-$1.7443702    & 6.3430544      \\
7.5   &    4.9391522549392  &  1.2635057    &  6.20265795  &  15.0  &     8.4139254912730  & $-$2.1168056    & 6.2971198      \\
10.0  &    4.9726811434694  &  1.23872097   &  6.21140211  &  20.0  &     8.5857873091459  & $-$2.3081634    & 6.27762390      \\
20.0  &    4.9759220330329  &  1.23732124   &  6.21324327  &  30.0  &     8.6127521696429  & $-$2.33524516   & 6.27750700     \\
40.0  &    4.9759220625078  &  1.23732124   &  6.21324330  &  40.0  &     8.6129969633475  & $-$2.33538378   & 6.27761318    \\
\end{tabular}
\end{ruledtabular}
\end{table}
\endgroup

Next Fig.~6 depicts modulation of $S_{\rvec}, S_{\pvec}$, {\color{red} radial PM Shannon entropy $S$} in left, middle, right segments (a)-(c), with changes 
in $l$, keeping $n$ fixed (25) at three distinct $m$, namely, 0, 5, 10. Unlike $R^{3}_{\pvec}$, of Fig.~4, $S_{\rvec}$ for $m=0$ 
changes inversely with $l$. But, for the other two $m$, like $R^{3}_{\rvec}$ of Fig.~4, $S_{\rvec}$'s advance with $l$, attain some 
plateau and then decline. On the other hand, $S_{\pvec}$'s for first two $m$, follow the same trend as $R^{3}_{\pvec}$ did. In both 
cases, $S_{\pvec}$'s steadily upturn with $l$. Only, for $m=10$, it deviates from $R^{3}_{\pvec}$ and shows a downward trend reaching 
a minimum, and then climbing up again. For $S$, one encounters a plateau for $m=0,~5$, whereas, for $m=10$, first there 
appears a maximum followed by a minimum and lastly a plateau. Like the $R$ plots of Figs.~3, 4, in this case also, 
while $n$ variations are rather direct and straightforward, same for $l$ are relatively intricate. However, like Fig.~4, this result 
may also be useful to understand transmission of total probability distribution of atomic orbitals. 

\subsection{The confined Hydrogen atom}
In this subsection now, all information measures of previous subsection are presented, in case of a CHA, in order to help 
uncover the impact of impenetrable spherical cage on these. The radial boundary now changes from infinity to a 
finite region without affecting angular boundary conditions. Thus, angular information contributions, as produced in {\color{red} Table~III}, 
remain invariant to a change in potential from FHA to CHA. It is expected that, a progressively larger $r_{c}$ should lead to a
delocalization in the system in such a fashion that, when $r_{c} \rightarrow \infty$, it should evolve to FHA. Whereas, when 
$r_{c} \rightarrow 0$, influence of confinement is maximum. Thus, it will be convenient to pursue our calculation by choosing some 
specific $r_{c}$ values starting from $0.1$ to $100$. This parametric increase in $r_{c}$ reveals manifestation of the system 
from maximum confinement to a free system. 

\begin{figure}                                            %%%Fig. 8, CHA
\begin{minipage}[c]{0.30\textwidth}\centering
\includegraphics[scale=0.40]{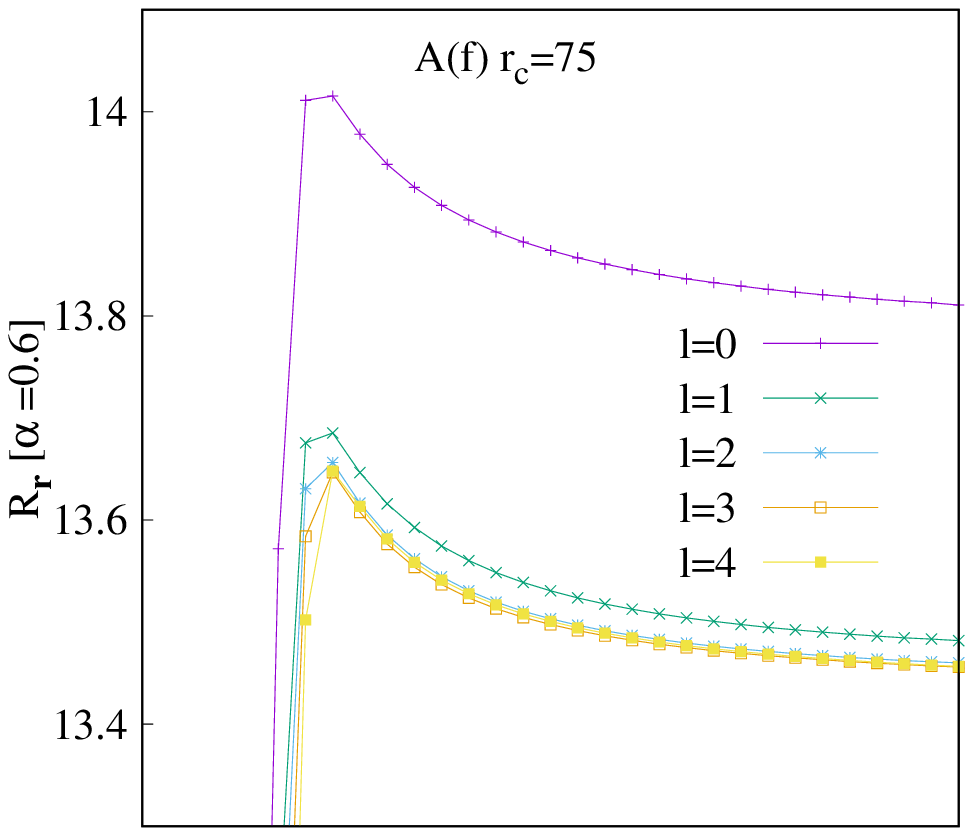}
\end{minipage}%
\begin{minipage}[c]{0.30\textwidth}\centering
\includegraphics[scale=0.40]{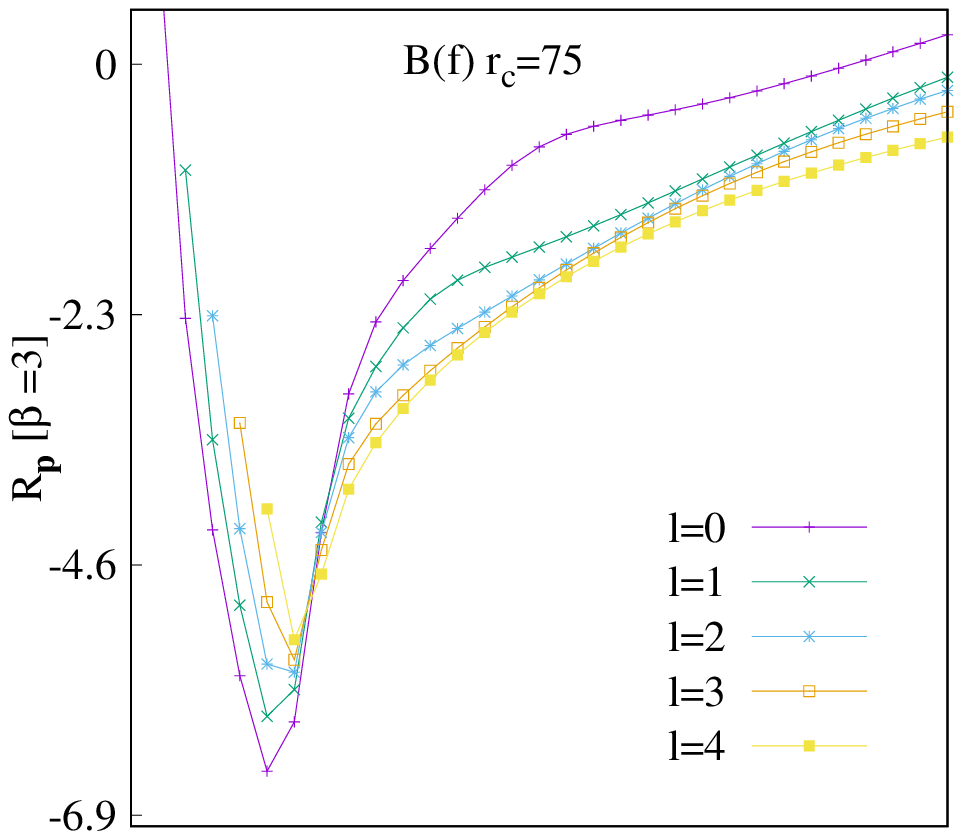}
\end{minipage}%
\begin{minipage}[c]{0.30\textwidth}\centering
\includegraphics[scale=0.40]{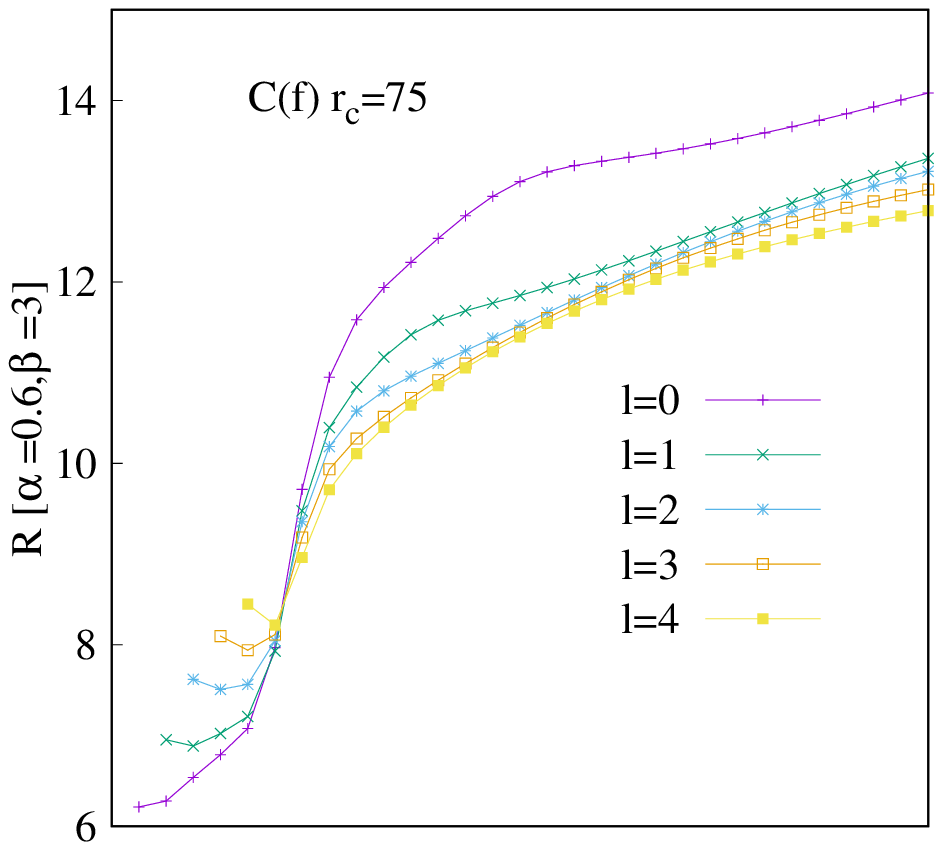}
\end{minipage}%
\hspace{0.2in}
\begin{minipage}[c]{0.30\textwidth}\centering
\includegraphics[scale=0.40]{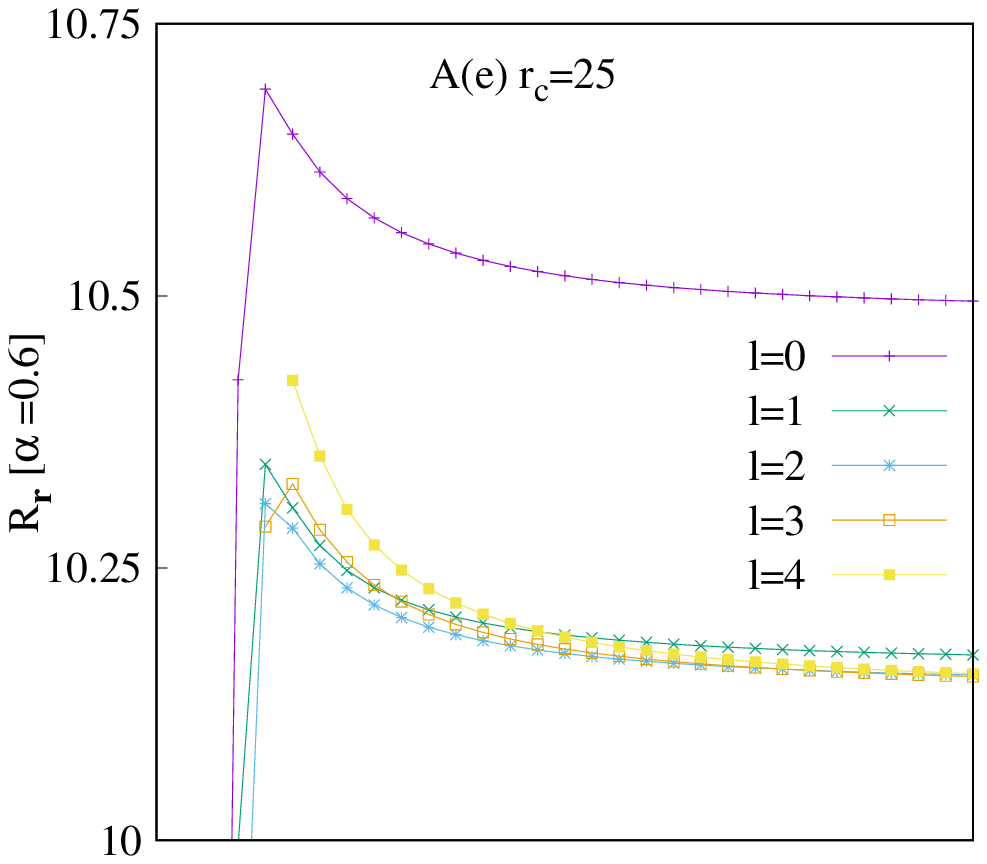}
\end{minipage}%2
\begin{minipage}[c]{0.30\textwidth}\centering
\includegraphics[scale=0.40]{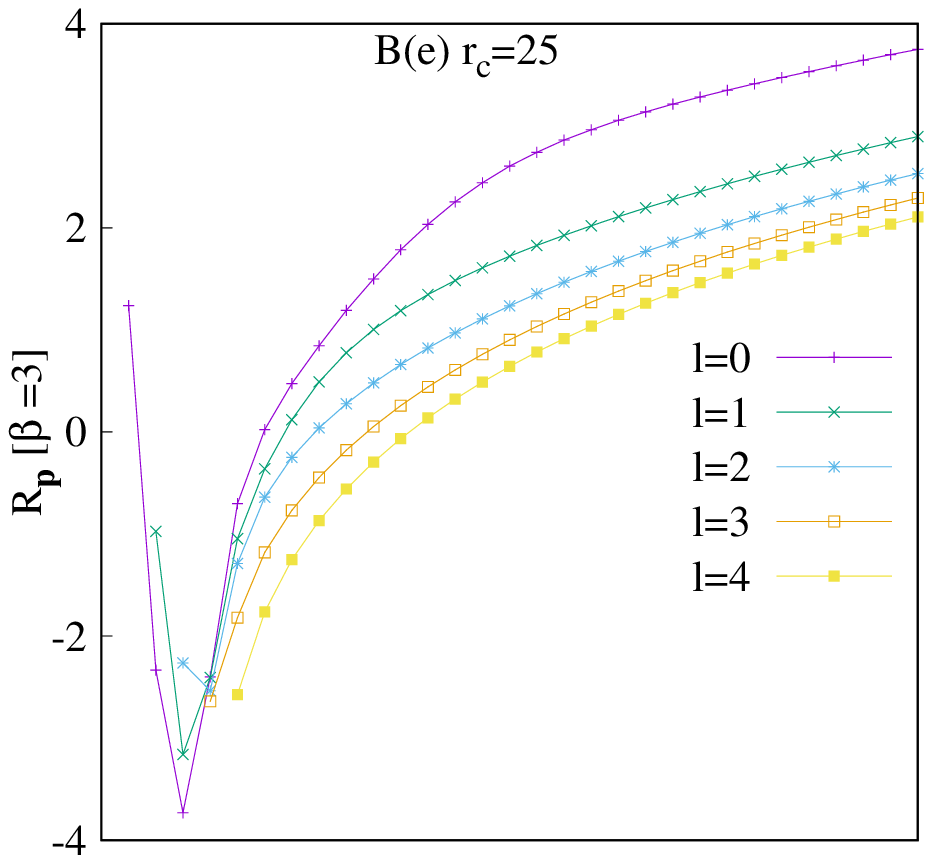}
\end{minipage}%
\begin{minipage}[c]{0.30\textwidth}\centering
\includegraphics[scale=0.40]{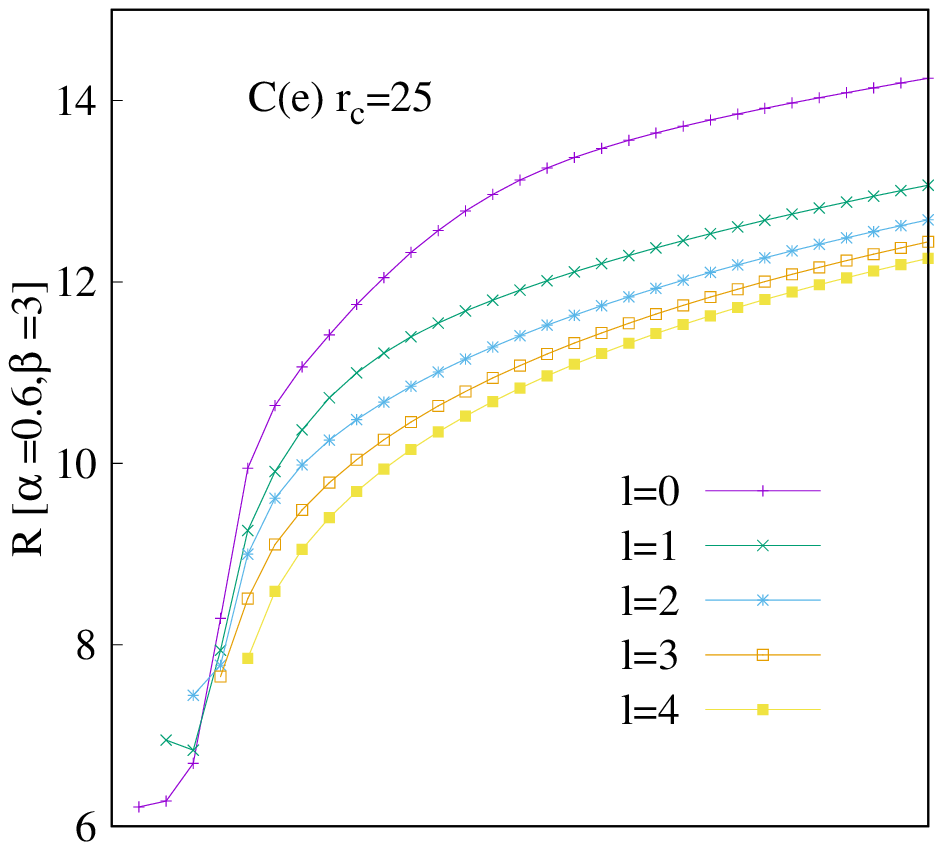}
\end{minipage}%
\hspace{0.2in}
\begin{minipage}[c]{0.30\textwidth}\centering
\includegraphics[scale=0.40]{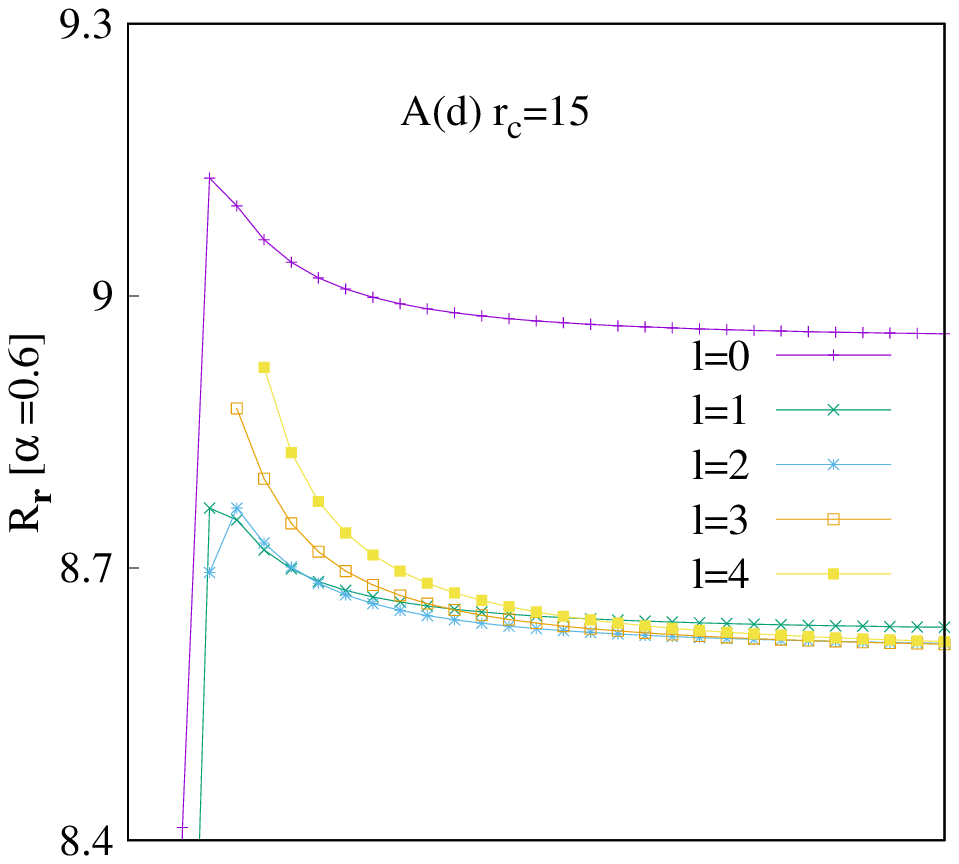}
\end{minipage}%
\begin{minipage}[c]{0.30\textwidth}\centering
\includegraphics[scale=0.40]{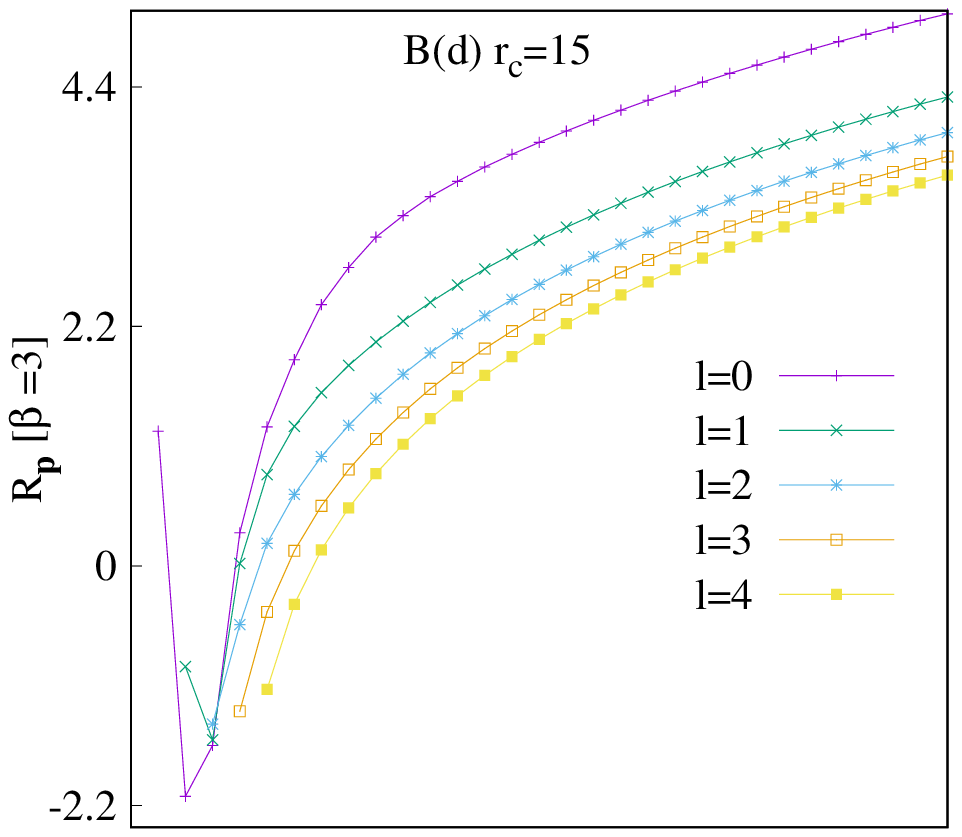}
\end{minipage}%
\begin{minipage}[c]{0.30\textwidth}\centering
\includegraphics[scale=0.40]{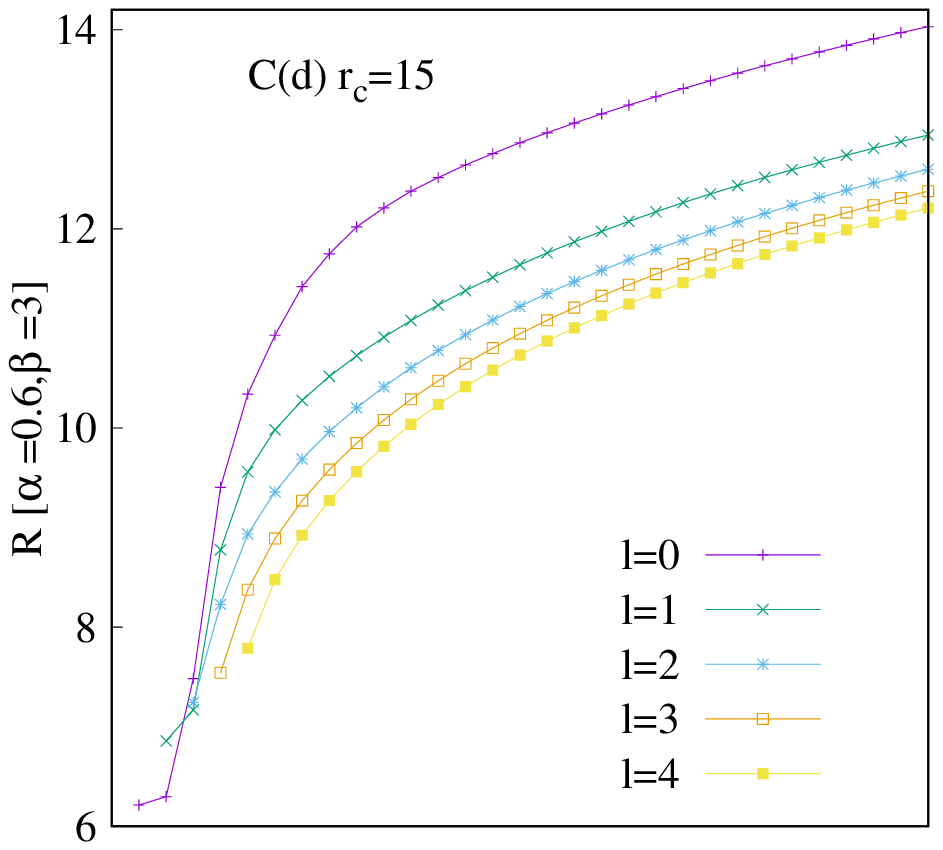}
\end{minipage}%
\hspace{0.2in}
\begin{minipage}[c]{0.30\textwidth}\centering
\includegraphics[scale=0.40]{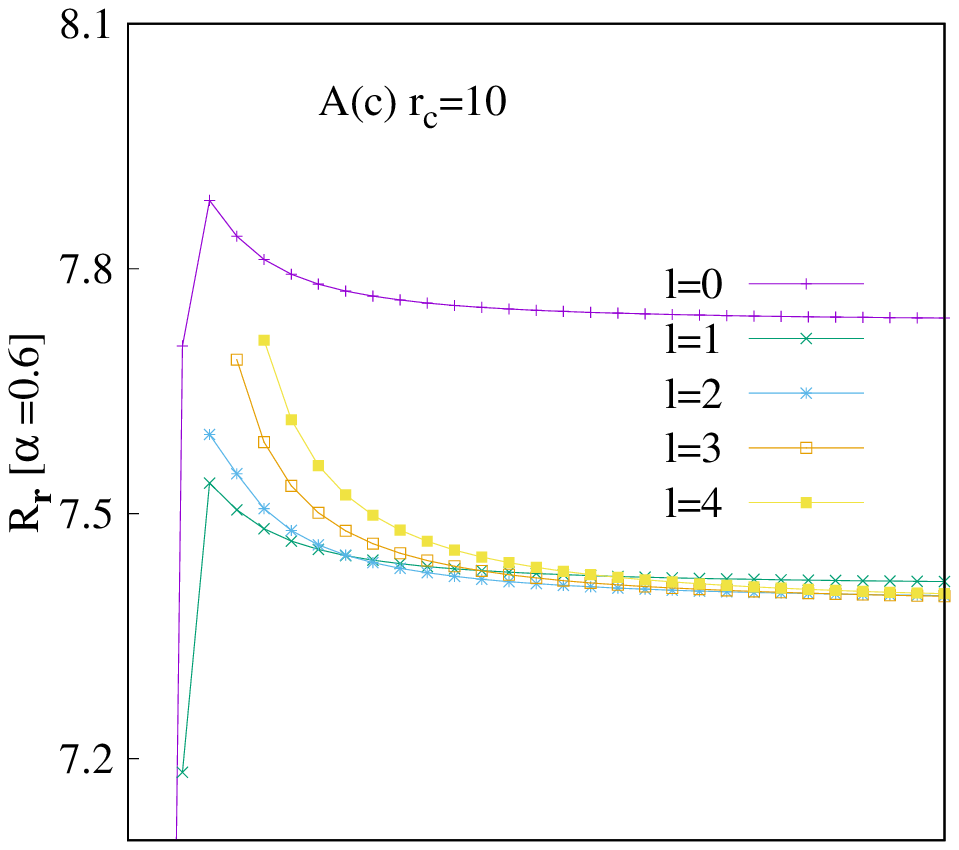}
\end{minipage}%
\begin{minipage}[c]{0.30\textwidth}\centering
\includegraphics[scale=0.40]{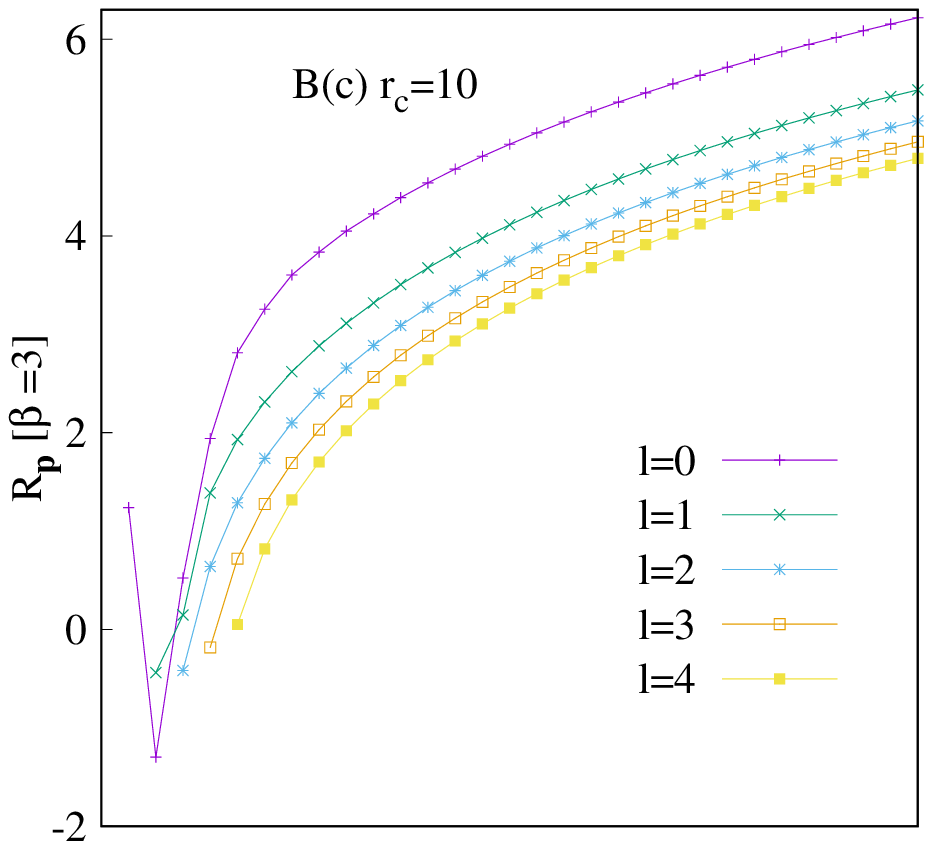}
\end{minipage}%
\begin{minipage}[c]{0.30\textwidth}\centering
\includegraphics[scale=0.40]{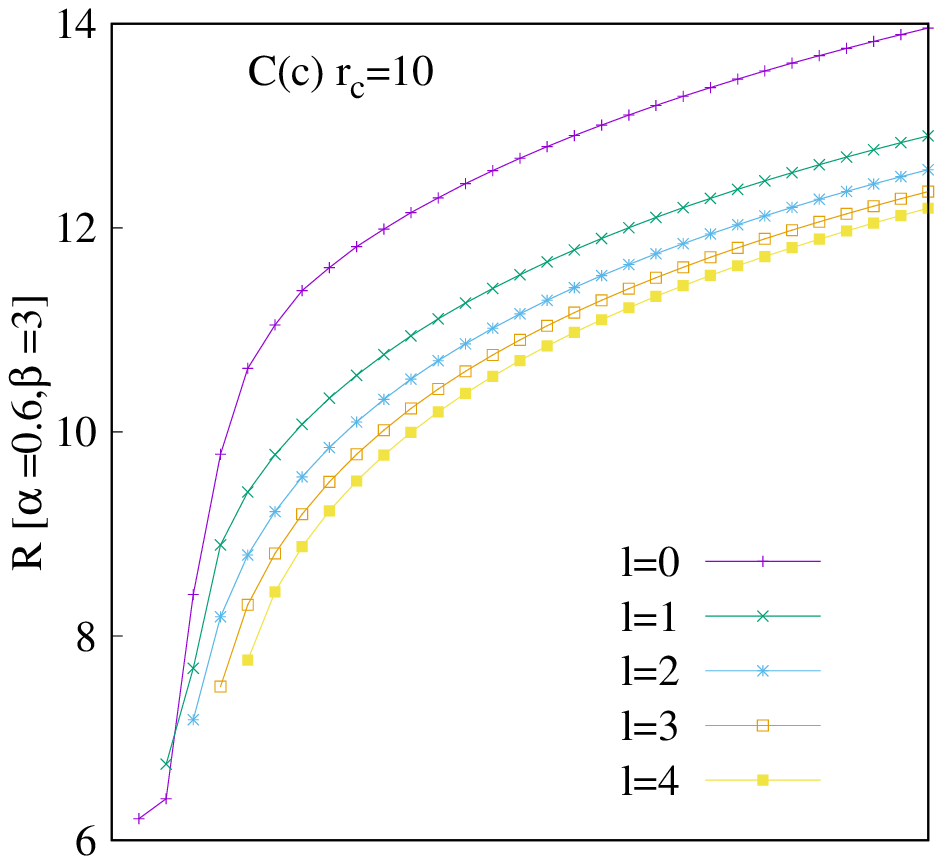}
\end{minipage}%
\hspace{0.2in}
\begin{minipage}[c]{0.30\textwidth}\centering
\includegraphics[scale=0.40]{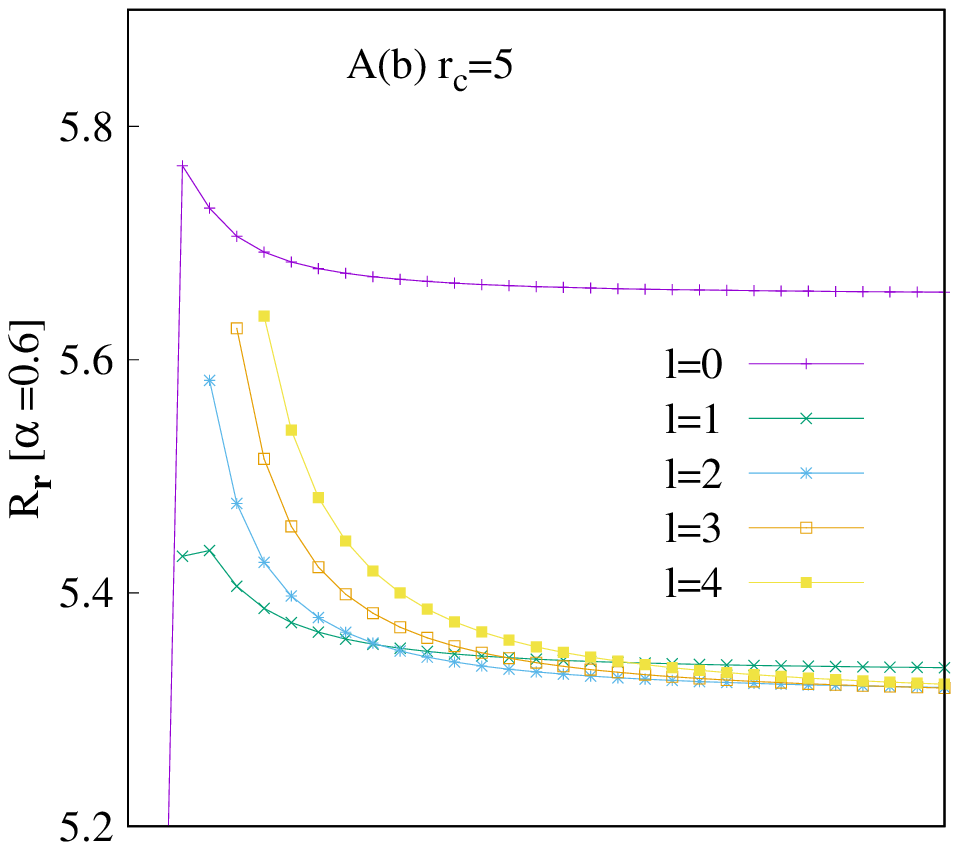}
\end{minipage}%
\begin{minipage}[c]{0.30\textwidth}\centering
\includegraphics[scale=0.40]{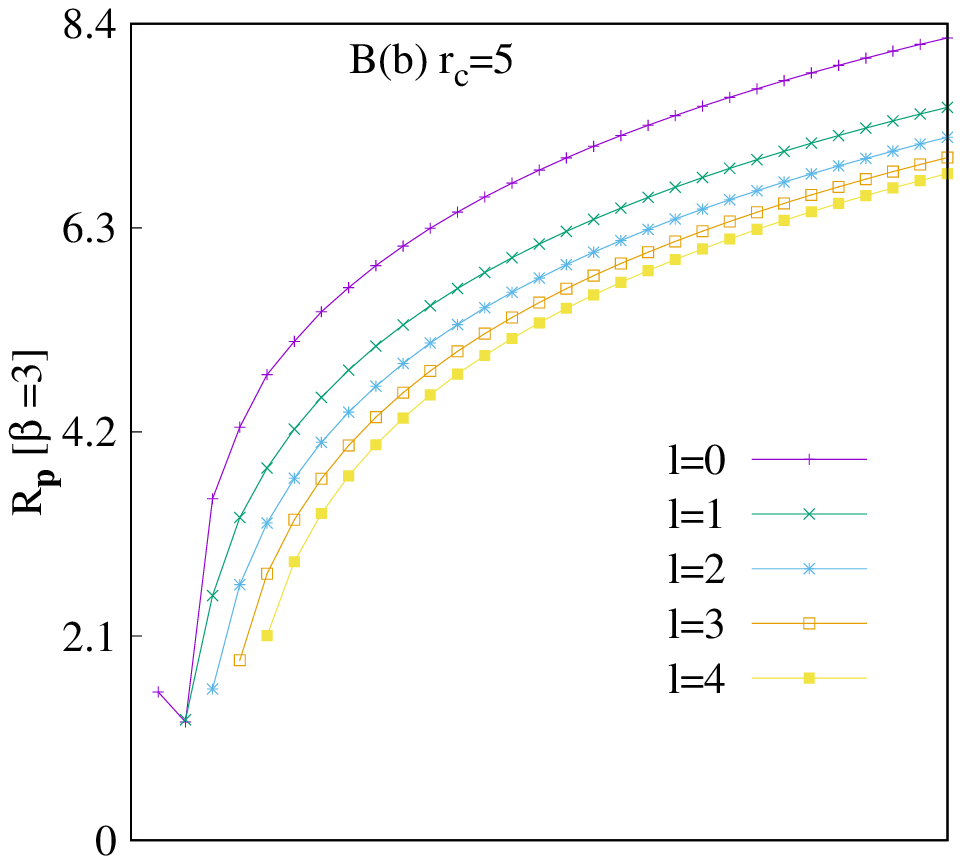}
\end{minipage}%
\begin{minipage}[c]{0.30\textwidth}\centering
\includegraphics[scale=0.40]{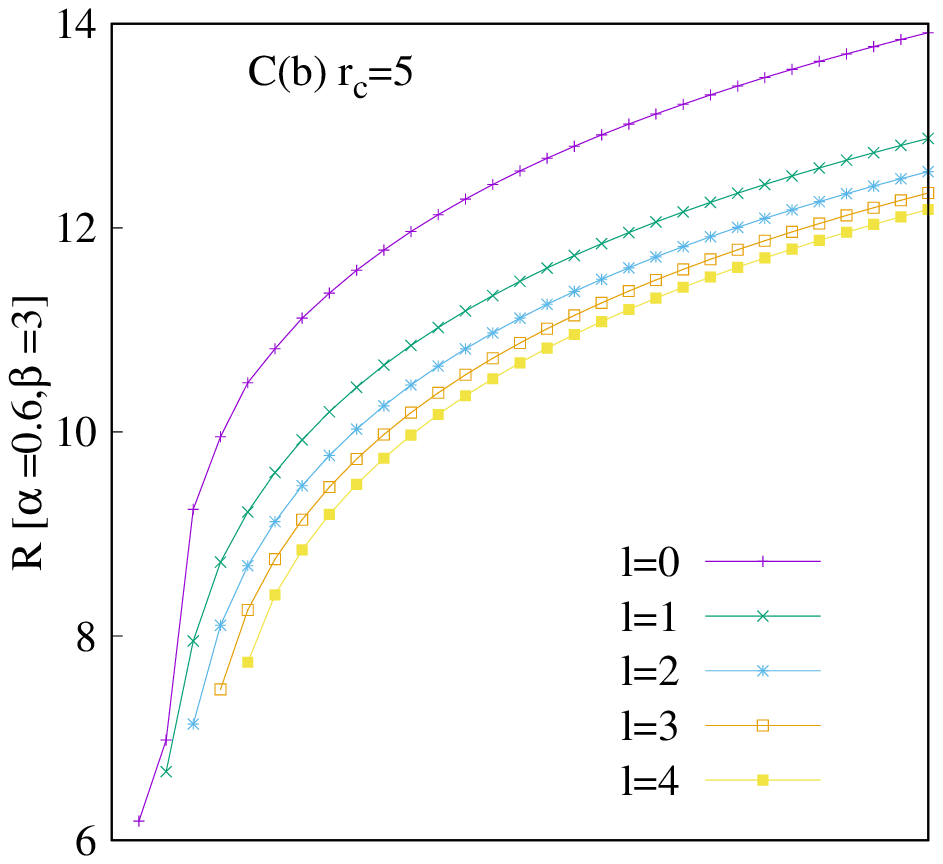}
\end{minipage}%
\hspace{0.2in}
\begin{minipage}[c]{0.30\textwidth}\centering
\includegraphics[scale=0.40]{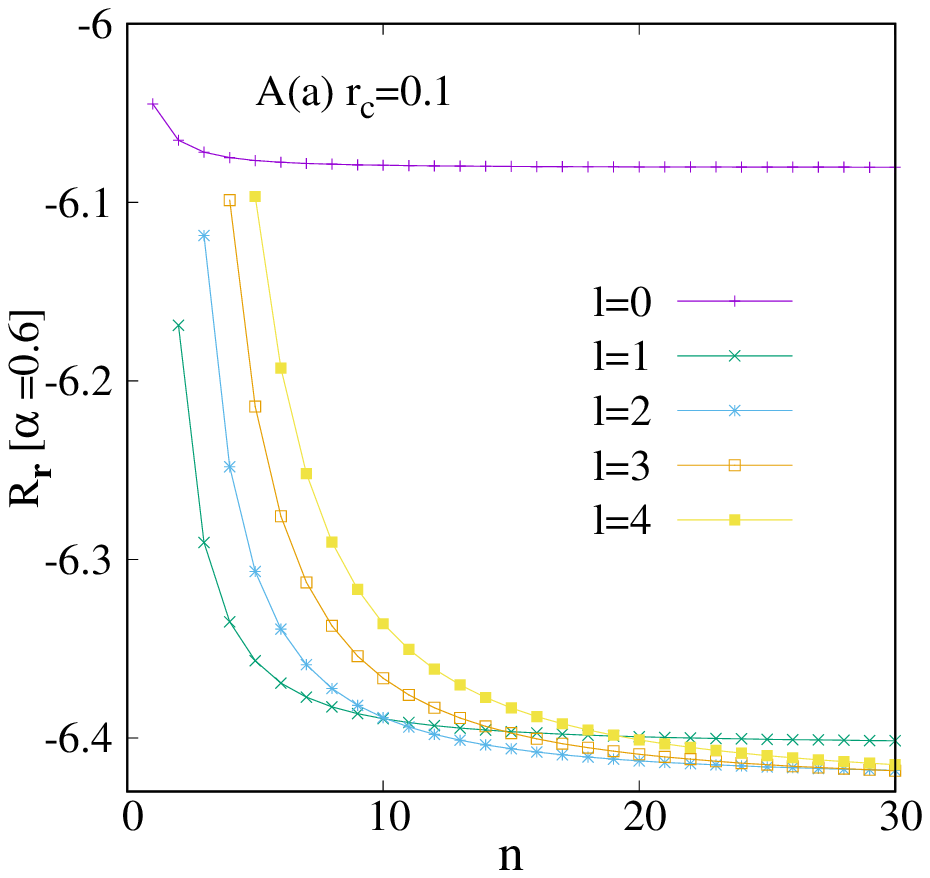}
\end{minipage}%
\begin{minipage}[c]{0.30\textwidth}\centering
\includegraphics[scale=0.40]{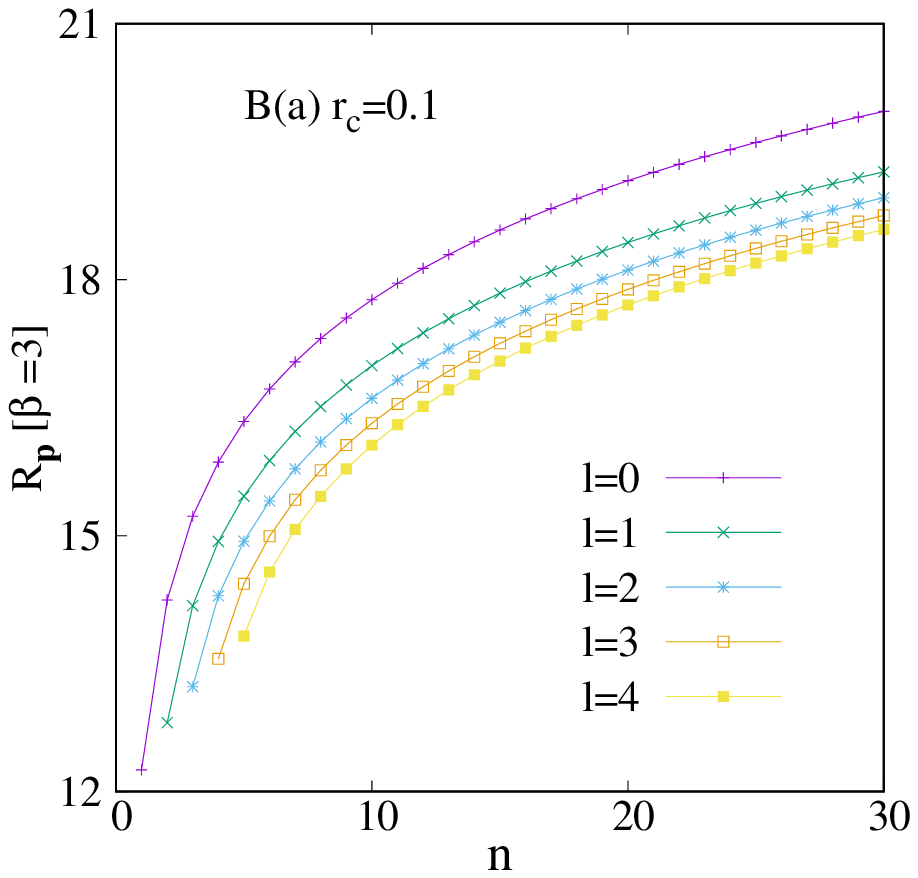}
\end{minipage}%
\begin{minipage}[c]{0.30\textwidth}\centering
\includegraphics[scale=0.40]{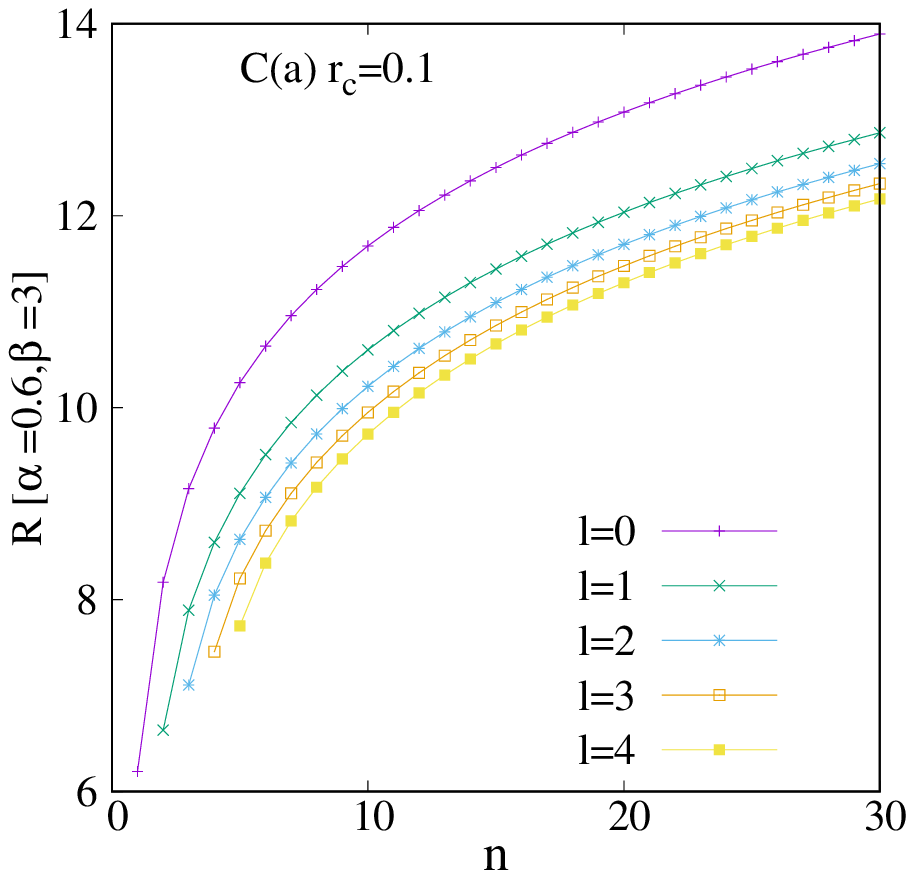}
\end{minipage}%
\caption{Plot of $R^{\alpha}_{\rvec}$ (A), $R^{\beta}_{\pvec}$ (B) and {\color{red} total PM R\'enyi entropy $R^{(\alpha, \beta)}$} (C) versus $n$ for $s,p,d,f,g$ 
orbitals at six particular $r_{c}$ values of CHA, namely, $0.1,5,10,15,25,75$ in panels (a)-(f).{\color{red} $R^{(\alpha, \beta)}$'s 
for all these states obey the lower bound given in Eq.~(19).} For more details, consult text.}
\end{figure}

To begin with, {\color{red}Table~VIII} displays calculated $R_{\rvec}^{\alpha},~R_{\pvec}^{\beta}$, {\color{red} total PM R\'enyi entropy $R^{(\alpha, \beta)}$} for $1s$, $2s$ 
states of CHA, at a selected set of $r_{c}$; which differ for the two. In this and all following tables of CHA, information 
quantities are furnished for these two states. In order to save space and volume of the length of this communication, higher 
states (especially having non-zero $l$) are omitted and may be presented elsewhere in future. This does not affect the main theme
of this work. However, to facilitate a clear understanding and presentation, these are well included in plots where required. 
In both occasions, $R_{\rvec}^{\alpha}$'s, starting from certain negative 
values at very small $r_c$, continuously progress, finally converging to the respective FHA behavior after some larger finite
$r_c$. In contrast, $R_{\pvec}^{\beta}$'s in both states generally tend to diminish with $r_{c}$, again merging to FHA in the end. 
In $2s$, eventually it goes to the ($-$)ve in large $r_c$. Consequently, the {\color{red} total PM R\'enyi entropy} for 
$1s$, depletes with $r_c$ and goes through a minimum before attaining FHA. For $2s$, however, it decays with $r_c$ to reach the 
borderline value. For two states under consideration, these convergences occur at roughly $r_c \approx 20$, 30 respectively. At 
very low $r_c$, $R_{\rvec}^{\alpha}(1s)> R_{\rvec}^{\alpha}(2s)$ but they cross each other at around $r_c \approx 0.4$. A similar exercise leads 
to $R_{\pvec}^{\beta}(2s)> R_{\pvec}^{\beta}(1s)$ at smaller $r_c$, with crossing occurring at nearly $r_c \approx 4-5$. There 
is no such crossover in $R^{(\alpha,\beta)}$ in any of these states. No literature results could be found for these quantities 
to make direct comparison. Above observations are vividly depicted in Fig.~7, where in three segments, (a)-(c), changes in 
$R_{\rvec}^{\alpha}$, $R_{\pvec}^{\beta}$, {\color{red} total PM R\'enyi entropy $R^{(\alpha, \beta)}$} of representative eight (covering 3 zero- and 5 non-zero-$l$) 
states, \emph{viz.,} $1s, 2s, 2p, 3s, 3p, 3d, 4f, 5g$ of CHA, with respect to $r_c$ are recorded. Panel (a) reveals that, for all 
of them, $R_{\rvec}^{\alpha}$'s quite smoothly advance initially with $r_{c}$, and finally coalesce to FHA. 
Likewise, from panel (b), clearly, $R_{\pvec}^{\beta}$ shows a reverse behavior with $r_c$, before reaching FHA-limit. Panel (c) 
portrays that for $1s$, {\color{red} $R^{(\alpha,~\beta)}$} in the beginning, shows a drop with $r_c$, then attains
a minimum, and decisively converges to limiting value of 6.213243 at nearly $r_{c} \approx 10$, whereas for $2s$, it continuously
falls off as $r_c$ extends, and thereafter reaching the FHA value of 6.2776 at around $r_c \approx 20$. For 
$3s$, $3p$ states $R^{(\alpha,\beta)}$'s first rise with $r_c$, then attain some maxima and decay until reaching FHA. For remaining 
node-less states, $R^{(\alpha,\beta)}$'s register growth with $r_c$, before permanently marching towards FHA. This characteristic $r_c$ 
generally shifts towards right, as $n$ moves upwards. It is important to note that, there appears multiple crossovers amongst various 
states in all these entropies, which occurs due to confinement. A more detailed, systematic study would be necessary to explain 
the pattern of this occurrence, which may be undertaken in future. 
           
\begingroup           %%Table 9
\squeezetable
\begin{table}
\caption{$T_{\rvec}^{\alpha}, T_{\pvec}^{\beta}$ and {\color{red} total PM Tsallis entropy $T^{(\alpha, \beta)} (=T_{\rvec}^{\alpha} T_{\pvec}^{\beta})$} for $1s$ and 
$2s$ states, at several particular $r_c$ in a CHA, for $\alpha=\frac{3}{5}, \beta=3$ respectively. Consult text for more 
details.}
\centering
\begin{ruledtabular}
\begin{tabular}{llllllll}
\multicolumn{4}{c}{$1s$}    &      \multicolumn{4}{c}{$2s$}    \\
\cline{1-4} \cline{5-8}
$r_c$  &    $T_{\rvec}^{\alpha}$     & $T_{\pvec}^{\beta}$  &  $T^{(\alpha, \beta)}$  &  
$r_c$  &    $T_{\rvec}^{\alpha}$     & $T_{\pvec}^{\beta}$  &  $T^{(\alpha, \beta)}$  \\
\hline 
0.1    &   $-$2.2772467171115   &   0.499999999988   &   $-$1.138623358529   &   0.1    &  $-$2.27905040787     
       &    0.4999999999997     &   $-$1.139525203938 \\
0.2    &   $-$1.9899960082576   &   0.499999999284   &   $-$0.994998002705   &   0.2    &  $-$1.99236352845   
       &    0.4999999999863     &   $-$0.996181764197 \\
0.5    &   $-$0.9853488252731   &   0.499999833837   &   $-$0.492674248908   &   0.5    &  $-$0.97510014910   
       &    0.4999999964918     &   $-$0.487550071130 \\
0.8    &      0.1298124908784   &   0.499997363767   &      0.064905903223   &   0.8    &     0.18219873857   
       &    0.4999999370189     &      0.091099357813 \\
1.0    &      0.9066489220414   &   0.499990349260   &      0.453315711188   &   1.0    &     1.00811263141   
       &    0.4999997455334     &      0.504056059175 \\
1.5    &      2.9023494698347   &   0.499902367380   &      1.450891370935   &   3.0    &    10.81989915020   
       &    0.4994369378358     &      5.403857299273 \\
2.5    &      6.8273006148635   &   0.498479920188   &      3.403272265597   &   5.0    &    22.59710251821   
       &    0.4560283384851     &      10.30491911595 \\
5.0    &     13.6370406555684   &   0.476304362681   &      6.495381958317   &   7.5    &    38.04973331016   
       &   $-$0.59440039327     &   $-$22.61677644352 \\
10.0   &     15.7718785921688   &   0.458021139475   &      7.223853804451   &   15.0   &    69.87499964359   
       &   $-$33.9829209887     &   $-$2374.556591979 \\
20.0   &     15.7955808066365   &   0.457903457550   &      7.232851065381   &   30.0   &    75.86611269904   
       &   $-$52.8750365859     &   $-$4011.423484596 \\
40.0   &     15.7955810228201   &   0.457903457462   &      7.232851162970   &   40.0   &    75.87378643375   
       &   $-$52.8898370490     &   $-$4012.952200777 \\
\end{tabular}
\end{ruledtabular}
\end{table}
\endgroup

To gain further insight, Fig.~8 portrays $R_{\rvec}^{\alpha}, R_{\pvec}^{\beta}$, {\color{red} total PM R\'enyi entropy $R^{(\alpha, \beta)}$} in left (A), 
middle (B), right (C) panels, for lowest five $l$ (0--4) as function of $n$, (maximum of 30). Six different $r_c$'s 
are chosen, i.e., 0.1, 5, 10, 15, 25, 75, in segments (a) through (f) from bottom to top. At $r_c=0.1$, for all 
$l$, $R_{\rvec}^{\alpha}$'s consistently go down with $n$, while $R_{\pvec}^{\beta}$, $R^{(\alpha, \beta)}$ show opposite 
trend. Furthermore, they follow similar pattern with $l$, for a given $n$. This indicates that, at small $r_c$, effect 
of confinement is more pronounced for high-$n,l$ states implying that, quantum nature gets amplified in this situation. 
Because, information content reduces, whereas {\color{red}} PM information (uncertainty) escalates. 
First two columns (A, B), interestingly show appearance of a maximum, minimum in $R_{\rvec}^{\alpha}$, $R_{\pvec}^{\beta}$ plots 
with regular advancement of $r_c$, as one moves up from bottom (a) to top (f) panel. Positions of these
maxima, minima move to right as $r_c$ intensifies. Apparently, there exists an interplay between two 
competing aspects: (i) radial confinement (localization) and (ii) accumulation in the nodes with $n$ (delocalization). With a
build-up in $r_c$, delocalization predominates for lower $n$, whereas extent of localization is more prominent for larger $n$. Hence 
with steady relaxation of confinement, states having greater $n$ get delocalized. In the limit of $r_c \rightarrow \infty$, where 
second effect prevails, system behaves as FHA, with the plots reducing to Figs.~3(a), 3(b) respectively. Third column 
portrays that, in all six $r_c$'s, $R^{(\alpha, \beta)}$'s strengthen with $n$. In all cases, $l=0$ plots remain rather
isolated from all higher $l$, which within themselves form a family. 

\begingroup           %%Table 10, S for 1s-2s in CHA, all checked 
\squeezetable
\begin{table}
\caption{$S_{\rvec}, S_{\pvec}$, {\color{red} total PM Shannon entropy $S$} for $1s$, $2s$ states in CHA at some chosen $r_c$. {\color{red} $S$'s 
for all these states obey the lower bound given in Eq.~(15).}  See text for details.}
\centering
\begin{ruledtabular}
\begin{tabular}{llllllll}
\multicolumn{4}{c}{$1s$}    &      \multicolumn{4}{c}{$2s$}    \\
\cline{1-4} \cline{5-8}
$r_c$  &    $S_{\rvec}$    & $S_{\pvec}$  &  $S=S_{\rvec}+S_{\pvec}$          &
$r_c$  &    $S_{\rvec}$    & $S_{\pvec}$  &  $S=S_{\rvec}+S_{\pvec}$          \\
\hline 
0.1$^{\S,\P}$  & $-$6.2445033842373  & 12.8535 &  6.6089   &   0.1    &  $-$6.4474579193881     &    14.638     &      8.1905    \\
0.2$^{\P}$   & $-$4.1778564051631  &  10.7787      &  6.6008	&   0.2$^{\ast}$    &  $-$4.3692335356773     &    12.5593    &      8.1900    \\
0.3   & $-$2.9747379859399  &   9.5675      &  6.5927	&   0.3    &  $-$3.1539053277870     &    11.343     &      8.189    \\
0.5$^{{\S},\dag,\P}$  & $-$1.4703406847180  &  8.0472  &  6.5768  &  0.5  &  $-$1.6230786943140  &  9.8112    &    8.1881    \\
0.6$^{\P}$   & $-$0.9382193800580  &   7.5073      &  6.5890    &   0.6$^{\ast}$    &  $-$1.0766799706228 &     9.2647    &      8.1880    \\
0.8   & $-$0.1065724371260  &   6.6609      &  6.5543    &   0.8    &  $-$0.2142040627489  &     8.4027    &      8.1884     \\
1.0$^{{\S},\dag,\ddag,\P}$   &    0.5290303076727  &   6.0114   &  6.5404  &   1.0 
$^{\ast}$  &  0.4554622941859  &  7.7347  &   8.1901    \\
1.5   &    1.6490560732453  &   4.8627      &  6.5117        &   3.0$^{\ast}$    &     3.8083926260850     &   4.454   &   8.262    \\
2.5   &    2.9291995226882  &   3.562952  &  6.492151  &   5.0$^{\ast}$  &   5.4641608279724   & 2.8173   &    8.2814    \\
3.0$^{\dag,\ddag,\P}$  &    3.3163654395398  &   3.1801450   &  6.496510   &   7.5  &  6.7230262418630  &  1.3022 &   8.025    \\
4.0   &    3.7942454904008  &   2.7241362   &  6.5183816   &   10.0$^{\ast}$   &     7.4461562639086     &   0.2765    &    7.7226    \\
5.0$^{\dag,\P}$  &    4.0174441862565  &   2.5243610   &  6.5418051   &  12.0   &   7.7816678917348  &  $-$0.23875  &  7.54291    \\
7.5   &    4.1393245365993  &  2.42550824  &  6.5648327    &  15.0  &   8.0218565650054   &  $-$0.6283   &  7.3935    \\
10.0$^{\P}$  &    4.1446014364987  &  2.42193665  &  6.56653808   &  20.0$^{\ast}$  &   8.1057256203059   &  $-$0.75320  &      7.35252    \\
20.0  &    4.1447298842431  &  2.42186233  &  6.56659221   &  30.0$^{\ast}$  &   8.1109253338427   &  $-$0.75758  &      7.35334    \\
40.0$^{\P}$  &    4.1447298842432  &  2.42186233  &  6.56659221   &  40.0  &   8.1109293629546   &  $-$0.75758  &      7.35334    \\
\end{tabular}
\end{ruledtabular}
\begin{tabbing}
$^\S$Literature results \cite{aquino13} of $(S_{\rvec}, S_{\pvec}, S)$ at $r_c=0.1$, 0.5, 1 in $1s$ state are: 
($-$6.2445, 12.8536, 6.6091), ($-$1.4702, 8.0473, 6.5771) \=    \\
and (0.5290, 6.0115, 6.5405) respectively.  \\
$^\dag$Literature results \cite{sen05} of $(S_{\rvec}, S_{\pvec}, S)$ at $r_c=0.5$, 1, 3, 5 in $1s$ state are: 
($-$1.47, 7.967, 6.497), (0.529, 5.991, 6.52), (3.316, 3.183,\\ 6.499) and (4.011, 2.533, 6.544) respectively. \\
$^\ddag$Literature results \cite{patil07} of $(S_{\rvec}, S_{\pvec}, S)$ at $r_c=1$, 3 in $1s$ state are: 
(0.52903, 6.011673, 6.54703) and (3.316365, 3.180236, 6.496602) \\
respectively. \\
[0pt \color{red}]$^{\P}$Literature results {\color{red}\cite{jiao17}} of $(S_{\rvec}, S_{\pvec}, S)$ at $r_c=0.1$, 0.2, 0.5, 0.6, 1, 3, 5, 10, 40 in $1s$ state are: \\
[0pt \color{red}]($-$6.24450338251, 12.85356277, 6.60905939),~($-$4.1778564034, 10.77871310, 6.60085670),~($-$1.47034068299, 8.04723315, 6.57689247), \\
[0pt \color{red}]~($-$0.9382193783, 7.50740813, 6.56918875),~(0.52903030941, 6.01144522, 6.54047553),~(3.31636544150, 3.18014501, 6.49651045), \\
[0pt \color{red}]~(4.01744418917, 2.52436106, 6.54180525),~(4.14460144459, 2.42193666, 6.56653810),~(4.14472988585, 2.42186234, 6.56659222) \\ 
[0pt \color{red}]respectively. \\
$^{\ast}$Literature results {\color{red}\cite{jiao17}} of $(S_{\rvec}, S_{\pvec}, S)$ at $r_c=0.2$, 0.6, 1, 3, 5, 10, 20, 30 in $2s$ state are:
($-$4.3692335342, 12.55940257, 8.19016903),\\
~($-$1.076679969, 9.26455393, 8.18787396),~(0.4554622961, 7.73472053, 8.19018283),~(3.8083926278, 4.45432335, 8.26271598), \\
~(5.4641608298, 2.8174878158, 8.28164864),~(7.4461562656, 0.27655185, 7.72270812),~(8.1057256268, $-$0.75319646, 7.35252916), \\
~(8.1109253319, $-$0.75758021, 7.35334511) respectively. 
\end{tabbing}
\end{table}
\endgroup

Next, {\color{red}Table~IX} reports numerical values of $T_{\rvec}^{\alpha},~T_{\pvec}^{\beta}$, {\color{red} total PM Tsallis entropy $T^{(\alpha, \beta)}$} for first two 
$s$ states of CHA at several distinct $r_{c}$.  These are selected so as to cover small, moderate and large cage radius. Once 
again, no reference work exists for these, which could be compared. Here, starting from a ($-$)ve value, 
$T_{\rvec}^{\alpha}$'s continually progress with $r_{c}$ for both states, and in the end, merge with FHA. The same for 
$T_{\pvec}^{\beta}$'s, from an initial value of 0.5, steadily fall off with $r_{c}$ before reaching the same fate of attaining 
FHA limit. While the change is rather mild for $1s$ for entire range, for $2s$, it is quite dramatic, especially around $r_c 
\approx 7.5$, from where it becomes ($-$)ve, and approach very large magnitude at the end. Like $R^{\alpha}_{\rvec}$, at very low 
$r_c$, $T_{\rvec}^{\alpha}(1s)> T_{\rvec}^{\alpha}(2s)$ but at nearly $r_{c} \approx 0.4$, they cross each other. Initially, 
in lower $r_c$, $T_{\pvec}^{\beta}(2s)> T_{\pvec}^{\beta}(1s)$; the ordering reverses in the $r_c$ range of 4-5. There is a 
fundamental difference in the nature of $T^{(\alpha, \beta)}$ of these states however; from a ($-$)ve value, ground state 
gradually progresses steadily, while in $2s$, it passes through a positive maximum, joining two terminal negatives. For 
further appreciation, panels (a)-(c) of {\color{red}Fig.~S1} reveal corresponding changes of Tsallis entropies with respect to $r_c$, 
in same eight states as in Fig.~7. In all occasions, $T_{\rvec}^{\alpha}$'s tend to enlarge in great extent with $r_{c}$ and 
at last coalesce to FHA. Panel (b), on the other hand gives an opposite effect for $T_{\pvec}^{\beta}$. 
Actually, for all states, they dip as $r_{c}$ grows up and eventually converge to FHA scenario. But the extent of downfall is 
not in same order; hence are not seemingly clear from the plot. Panel (c) suggests that, for ground state, {\color{red} total PM Tsallis entropy $T^{(\alpha,\beta)}$} 
grows with $r_c$ and finally merges to FHA limit. But for all other states, {\color{red} $T^{(\alpha,\beta)}$'s} slowly increase, then attain maxima
and lastly falls off prior to joining with FHA. Positions of these maxima shift to right as $n$ and number 
of nodes rise. Such attainment to FHA is not so conspicuous from panel (c), as they tend to approach much larger ($-$)ve 
values with $r_c$. But upon closer examination, they follow same trend as exemplified by $2p$. Behavior of $R^{\alpha}_{r}$ 
and $T^{\alpha}_{r}$ with change of $r_{c}$ are quite harmonious. But, variations of $T^{\beta}_{p}$ and $T^{(\alpha,\beta)}$ 
with development of $r_{c}$ are different from $R^{\beta}_{p}$ and $R^{(\alpha,\beta)}$ patterns, even though, one can draw analogous 
conclusion from study of $T$ and $R$. The graphs of $T$ versus $n$ (parallel to those in Fig.~8 for $R$), offer resembling 
motives in their nature. Hence they are not separately presented here. 

Now we move on to $S$ in {\color{red}Table~X}, where $S_{\rvec}, S_{\pvec}$ and {\color{red} total PM Shannon entropy $S$} are probed for lowest two $l=0$ states of CHA at same 
particular set of $r_c$ as in {\color{red}Table~VIII}. {\color{red} A handful of literature results are known for ground state; the reference values are  
duly quoted at $r_c$ of 0.1, 0.2, 0.5, 0.6, 1, 3, 5, 10 and 40 a.u., whereas, for $2s$ state these have been considered only in the recent 
work of \cite{jiao17} at $r_c$ values 0.2, 0.6, 1, 3, 5, 10, 20, 30 and 40 a.u.} Other than that we are not aware 
of any report on these quantities. Wherever applicable, present estimates are in decent agreement with reference data. $S_{\rvec}, S_{\pvec}, 
S$ imprint exactly analogous behavior of $R_{\rvec}^{\alpha},~R_{\pvec}^{\beta}$ and $R^{(\alpha,\beta)}$ respectively. Like 
$R_{\rvec}^{\alpha}$, $S_{\rvec}$'s also possess ($-$)ve values for $1s,~2s$ at very low $r_{c}$ and then continuously evolve, until reaching 
FHA-limit at some large $r_{c}$. However, like $R_{\pvec}^{\beta}$, $S_{\pvec}$ offers an opposite nature of $S_{\rvec}
(R_{\rvec}^{\alpha})$; from an initial ($+$)ve, consistently diminishes to reach FHA, which for $2s$, assumes a ($-$)ve 
($-$0.75758). One finds that, in smaller $r_c$ region, $S_{\rvec}(1s)>S_{\rvec}(2s)$; but at close to $r_c$ within 1-2, they 
cross each other. Likewise, in low to moderate $r_c$ area, $S_{\pvec} (1s) < S_{\pvec} (2s)$; near the region of 7-7.5, 
$S_{\pvec}(1s)$ overcomes $S_{\pvec}(2s)$. 
    
\begin{figure}                         %%%Fig. 9, CHA
\begin{minipage}[c]{0.35\textwidth}\centering
\includegraphics[scale=0.55]{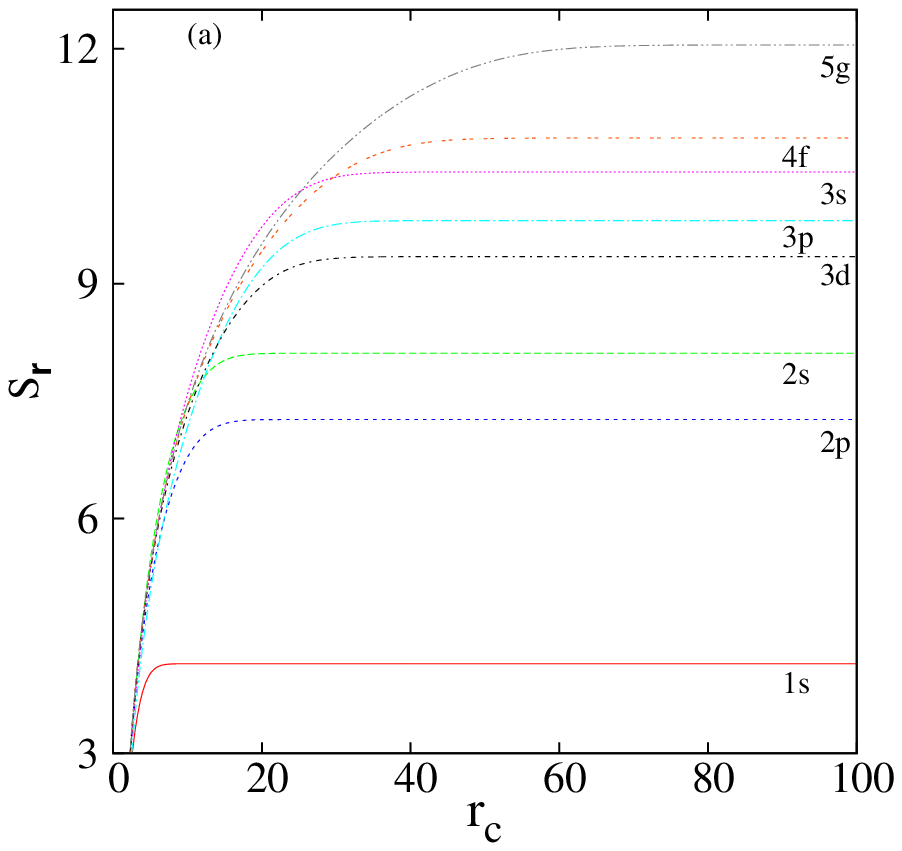}
\end{minipage}%
\begin{minipage}[c]{0.35\textwidth}\centering
\includegraphics[scale=0.55]{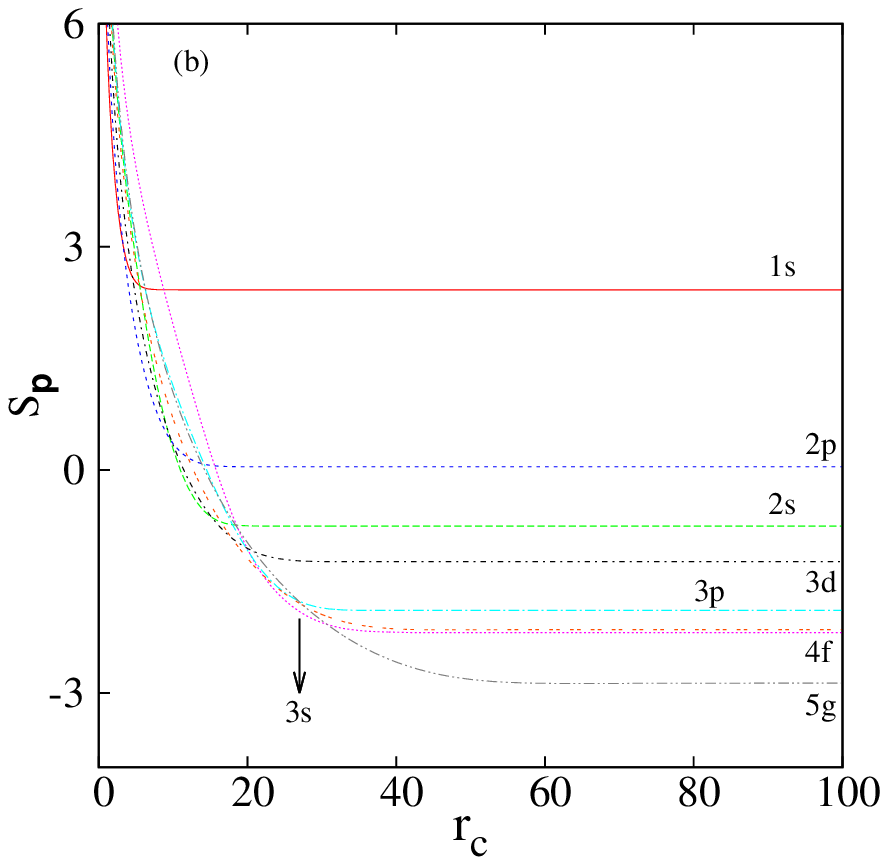}
\end{minipage}%
\begin{minipage}[c]{0.35\textwidth}\centering
\includegraphics[scale=0.55]{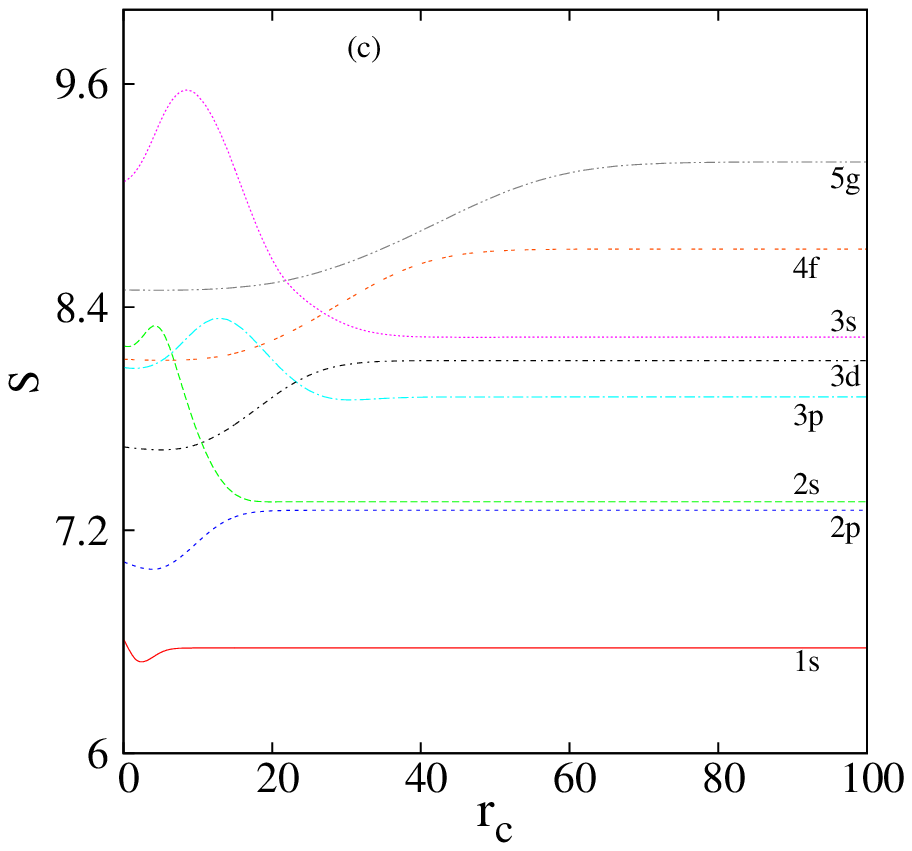}
\end{minipage}%
\caption{Plot of $S_{\rvec}$ (a), $S_{\pvec}$ (b), {\color{red} total PM Shannon entropy $S$} (c), in left, middle and right panels, against $r_c$ for a few selected 
low-lying states of CHA. {\color{red} $S$'s for all these states obey the lower bound condition given in Eq.~(15).} More details are available in text.}
\end{figure}

Next, {\color{red}Fig.~9} exhibits behavioral patterns of $S_{\rvec}, S_{\pvec}$, {\color{red} total PM Shannon entropy $S$} with $r_c$ in segments 
(a)-(c), for same eight states of Fig.~7. Note that, panels (a), (b), (c) of both {\color{red}Figs.~7, 9} show similar style.
For all of them, $S_{\rvec}$'s mount up with $r_c$ and finally converge to corresponding $r$-space FHA, while $S_{\pvec}$'s deplete 
before attaining that. In case of $S$'s in panel (c), for node-less states ($1s,2p,3d,4f,5g$), one finds an initial 
decay until arriving at some minima and then an expansion again. As usual, FHA-limit is retrieved after some large $r_c$. 

\begingroup           %%Table 11, I for 1s-2s in CHA, all checked 
\squeezetable
\begin{table}
\caption{$I_{\rvec}, I_{\pvec}$ for $1s$ and $2s$ states in CHA at some particular $r_c$. See text for details.}
\centering
\begin{ruledtabular}
\begin{tabular}{llllll}
\multicolumn{3}{c}{$1s^{\dag}$}    &      \multicolumn{3}{c}{$2s$}    \\
\cline{1-3} \cline{4-6}
$r_c$  &    $I_{\rvec}$  &   $I_{\pvec}$      &
$r_c$  &    $I_{\rvec}$  &   $I_{\pvec}$      \\ 
\hline
0.1\footnotemark[1]   &  3948.737092    &   0.01119745297   &  0.1  &  15791.82122     &    0.01284003608   \\
0.2\footnotemark[2]   &  987.8765878    &   0.04434444184   &  0.2  &  3948.29263      &    0.05141786856  \\
0.3\footnotemark[3]   &  439.586678     &   0.09875572074   &  0.3  &  1755.043513     &    0.11583040943  \\
0.5\footnotemark[4]   &  158.8961123    &   0.26851341481   &  0.5  &   632.0932498    &    0.32261837746  \\
0.6                   &  110.6681458    &  0.38237153819    &  0.6  &   439.0827084    &  0.46525991900  \\
0.8                   &  62.739860      &  0.66414501270    &  0.8  &   247.1624885    &    0.82981661465  \\
1.0\footnotemark[5]   &  40.58509174    &   1.01251135493   &  1.0  &   158.32289745   &    1.30128804642   \\
1.5                   &  18.79543801    & 2.12851618061     &  3.0  &   17.70794067    &   12.34353050970  \\
2.5                   &  7.90930147     &  4.99836645404    &  5.0  &    6.144128803   &  35.62201065982  \\
3.0                   &  6.17657298     &  6.49907451467    &  7.5  &    2.538369575   &   75.35119911871  \\
4.0                   &  4.67890854     &  9.08124532490    &  10.0 &    1.4882497628  &  114.09728048962  \\
5.0\footnotemark[6]   &  4.1962752      &  10.73988673564   &  12.0 &    1.1870576     &  138.20789171766  \\
7.5                   &  4.00555844     &  11.92721564499   &  15.0 &    1.037249102   &  158.95005505011  \\
10.0                  &  4.00009944     &  11.99783793184   &  20.0 &    1.001488032   &  167.39728283512  \\
20.0                  &  4.000000000    &  11.99999999999   &  30.0 &    1.0000006963  &  167.99942953967  \\
40.0\footnotemark[7]  &  4.000000000    &  12.00000000000   &  40.0 &    1.000000000   &  167.99999999101    \\
\end{tabular}
\end{ruledtabular}
\begin{tabbing}
$^{\mathrm{a}}${Reference result~\cite{aquino13}: $I_{\rvec}=3947.738178, I_{\pvec}= 0.011309. $} \hspace{30pt} \=    
$^{\mathrm{b}}${Reference result~\cite{aquino13}: $I_{\rvec}= 987.890146, I_{\pvec}= 0.043982. $} \\                  
$^{\mathrm{c}}${Reference result~\cite{aquino13}: $I_{\rvec}= 439.591750, I_{\pvec}= 0.099274. $}           \>        
$^{\mathrm{d}}${Reference result~\cite{aquino13}: $I_{\rvec}= 158.896729, I_{\pvec}= 0.269820. $}          \\         
$^{\mathrm{e}}${Reference result~\cite{aquino13}: $I_{\rvec}=  40.585607, I_{\pvec}= 1.012849. $}            \>       
$^{\mathrm{f}}${Reference result~\cite{aquino13}: $I_{\rvec}=  4.195911,  I_{\pvec}= 10.740746.$}  \\                 
$^{\mathrm{g}}${Reference result~\cite{aquino13}: $I_{\rvec}=  3.999875,  I_{\pvec}= 11.999627.$}     \\              
$^{\dag}${Reference values are multiplied with a 4$\pi$ factor}  
\end{tabbing}
\end{table}
\endgroup

In {\color{red}Fig.~S2}, $S_{\rvec}$ (A), $S_{\pvec}$ (B), {\color{red} total PM Shannon entropy $S$} (C) of $l=0-4$ states are plotted against $n$ at same six $r_c$ of Fig.~8,
in panels (a)-(f) from bottom to top. Again, the graphs in {\color{red}Fig.~S2} delineate similar shape and aptitude to that of Fig.~8. Thus, in
agreement with $R^{\alpha}_{\rvec}$ at $r_c=0.1$, for five $l$, $S_{\rvec}$'s get lowered in A(a), while $S_{\pvec}$'s and $S$'s 
progress with $n$ in B(a) 
and C(a) respectively. This augments our previous inference (as $R$ in Fig.~8) that, at very low $r_c$, effect of confinement is more
prevalent in high-lying states, signifying a magnification of quantum nature in such circumstances. As usual, like 
$R_{\rvec}^{\alpha}$ and $R_{\pvec}^{\beta}$ here also, the first two columns (A, B) of {\color{red}Fig.~S2} render the appearance of a maximum 
and minimum in $S_{\rvec}$ ((b) upwards), $S_{\pvec}$ ((c) upwards) plots with successive growth of $r_c$. Their positions get 
shifted to right as $r_c$ advances, which is indicative of the fact that, at $r_c \rightarrow \infty$ these plots merge to 
Figs.~5(a), 5(b) respectively (note that those graphs depicted radial quantities in $r$, $p$ spaces). Column C suggests that at 
all $r_c$, $S$ consistently broadens with $n$.      

Now we discuss $I$ in {\color{red}Table~XI}, by providing $I_{\rvec}$, $I_{\pvec}$ of lowest two $s$ states at selected $r_c$ used in {\color{red}Tables~VIII, X}. As a check, $I_{\rvec}$ was calculated in two ways: first one using direct expression of {\color{red}Eq.~(14)} needing $\nabla \rho (\rvec)$, and second one employing a 
simplified expression for central potentials requiring expectation values, namely {\color{red}Eq.~(23)}. They produce almost identical results, which 
are quoted in table for two states. Note that gradient 
of density and integrands in expectation values for CHA can be evaluated analytically ($m=0$ throughout). Only integrations 
needed to be performed numerically. Thus, it suffices to mention that, $I$'s can be accurately approached from a knowledge of 
$\langle p^2\rangle, \langle r^{-2}\rangle$, $\langle r^2\rangle, \langle p^{-2}\rangle$. Two possibilities may be envisaged: 
(i) first three evaluated in $r$, while $\langle p^{-2}\rangle$ in $p$ space (ii) $\langle r^2\rangle, \langle r^{-2}\rangle$ in $r$ 
space, while $\langle p^2\rangle, \langle p^{-2}\rangle$ in $p$ space. Here we adopted route (i) which obviates the necessity to 
do numerical differentiation in either space (all integrands are available analytically). Once again literature reports
are quite scanty; only for ground state some variational calculations were published in \cite{aquino13} for a few $r_c$. Present 
results show good agreement with these.     
  
\begin{figure}                         %%%Fig. 10, CHA
\begin{minipage}[c]{0.35\textwidth}\centering
\includegraphics[scale=0.55]{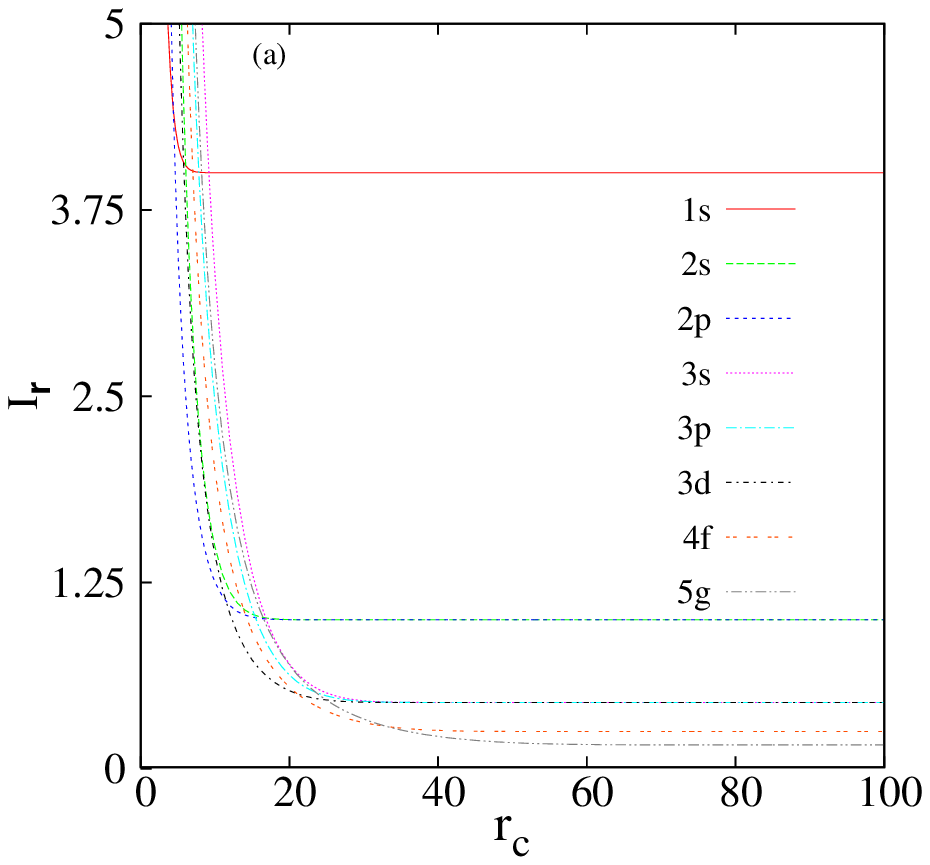}
\end{minipage}%
\begin{minipage}[c]{0.35\textwidth}\centering
\includegraphics[scale=0.55]{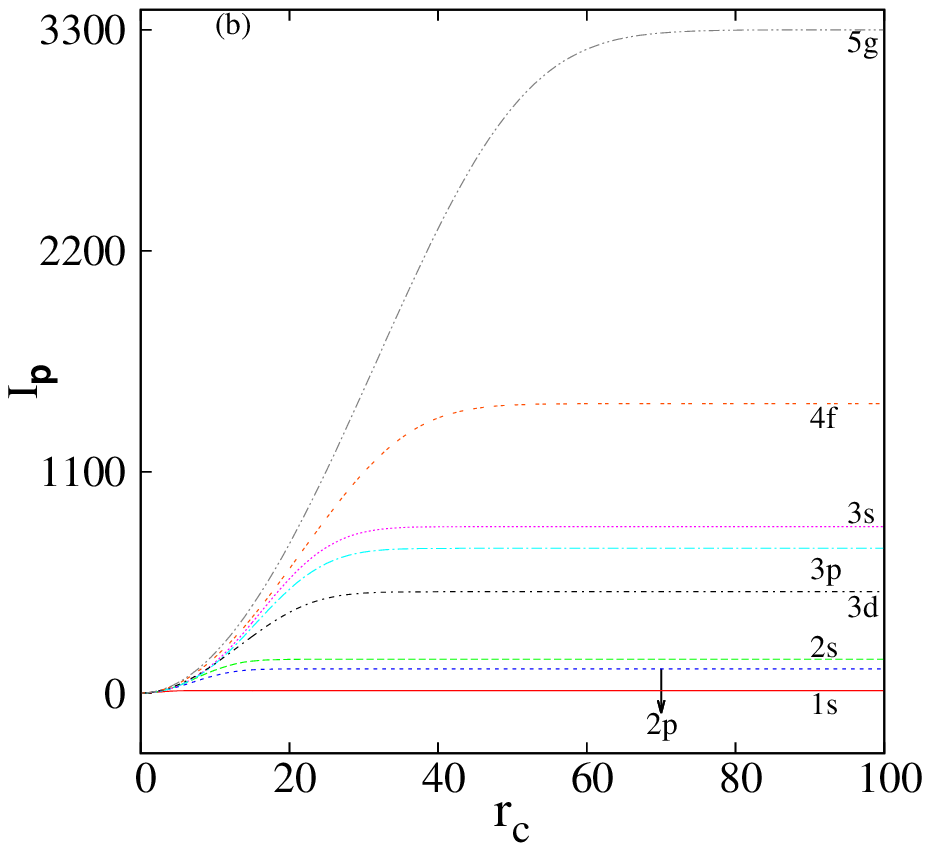}
\end{minipage}%
\begin{minipage}[c]{0.35\textwidth}\centering
\includegraphics[scale=0.55]{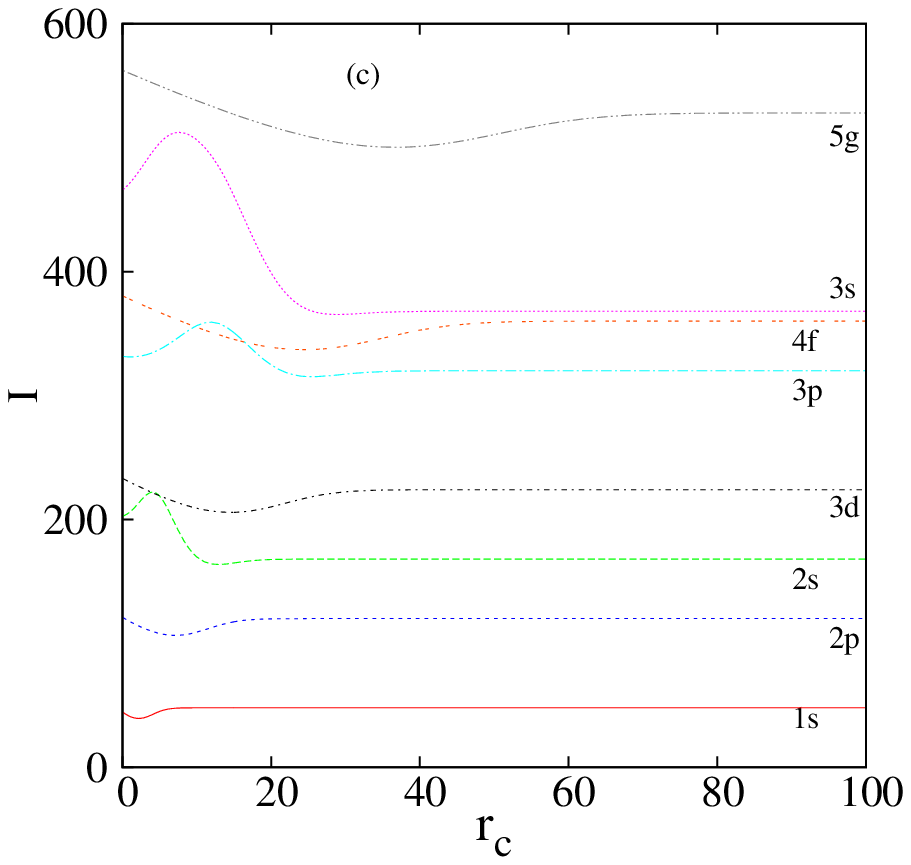}
\end{minipage}%
\caption{Plot of $I_{\rvec}$ (a), $I_{\pvec}$ (b), {\color{red} total PM Fisher information $I$} (c) in left, middle and right panels, against $r_c$ for a few selected 
low-lying states of CHA. $I$'s for all these states obey the lower and upper bound conditions given in {\color{red}Eq.~(14)}. More details can be found in text.}
\end{figure}

Now, {\color{red}Fig.~10} depicts the variation of $I_{\rvec}, I_{\pvec}$, {\color{red} total PM Fisher information $I=I_{\rvec} I_{\pvec}$,} in three columns labeled (a)-(c) from left, 
with of $r_c$. Keeping same presentation strategy as in Fig.~7, these are offered for same eight states. The trends of $I$ is
completely opposite to those observed in $R,~T, S$. One notices from (a), (b) that, with $r_c$, $I_{\rvec}$ and 
$I_{\pvec}$ behave in an opposite fashion--former grows while latter falls. Eventually they both approach FHA values after a 
certain $r_c$. For node-less states in (c), $I$'s mount up and thereafter merge to FHA. But, for non-circular states, $I$'s rise 
towards certain maxima, ultimately falling flat at FHA. 
 
Next, in {\color{red}Fig.~S3}, $I_{\rvec}, I_{\pvec}$, {\color{red} total PM Fisher information $I$} are displayed for several $l \! = \! 0-4$ states of CHA, as function of 
$n$ at same six (corresponding to panels (a)-(f)) $r_c$ of {\color{red}Fig.~S2}, in left (A), center (B), right (C) columns. 
At $r_c=0.1$ in (a), $I_{\rvec}$ and $I$ increase with $n$, while $I_{\pvec}$ behaves contrastingly. Thus, it follows that,  
at very low $r_c$, $\langle p^2 \rangle$ and $n$ go hand in hand. This is exactly reverse to FHA case, as delineated
before in {\color{red}Eq.~(23)}, where the same shows inverse relationship with $n$. Once again, the quantum nature of high-$n$ states 
intensifies at smaller $r_c$, possibly due to same reason as found in $R$, $S$. 
First two columns show that, a minimum and a maximum tends to develop as $r_c$ progresses from lower to
upper panels; their positions shift towards right on moving from (a) to (f). Here kinetic energy is gained with confinement
and lost with addition of radial nodes. This establishes that, as $r_c$ grows, effect of delocalization 
(number of nodes) predominates over localization, for lower $n$. Thus there appears minimum in $I_{\rvec}$ and maximum 
in $I_{\pvec}$. Eventually at $r_c \rightarrow \infty$, first effect is switched off; hence $\langle p^2 \rangle$ as well as 
$I_{\rvec}$ lower with $n$ and FHA situation is restored. In all $r_c$'s, $I$ progress with $n$, in last column. 

\begingroup           %%Table 12, E for 1s-2s in CHA, all checked 
\squeezetable
\begin{table}
\caption{$E_{\rvec}, E_{\pvec}$ for $1s$ and $2s$ states in CHA at some chosen $r_c$. See text for details.}
\centering
\begin{ruledtabular}
\begin{tabular}{llllll}
\multicolumn{3}{c}{$1s$}    &      \multicolumn{3}{c}{$2s$}    \\
\cline{1-3} \cline{4-6}
$r_c$  &    $E_{\rvec}$           & $E_{\pvec}$         & $r_c$    &  $E_{\rvec}$    &     $E_{\pvec}$   \\ 
\hline 
0.1     &  685.2442626946369   &    0.000003957597   &   0.1   &   1467.6825381961700     &  0.0000005701644   \\
0.2     &   87.4022739883438   &    0.000031421866   &   0.2   &   185.2798582651059      &  0.000004564133    \\
0.3     &   26.4463446487399   &    0.00010521164    &   0.3   &   55.4384452351512       &  0.000015417924    \\
0.5     &    5.9724213649058   &    0.0004788967     &   0.5   &   12.2085268184201       &  0.00007158083     \\
0.6     &    3.5387151037986   &    0.0008200589     &   0.6   &   7.1325907889262        &  0.00012393792     \\
0.8     &    1.5693288422636   &    0.0019064694     &   0.8   &   3.0655325000105        &  0.00029535230     \\
1.0     &    0.8479175599159   &    0.0036453711     &   1.0   &   1.5979206523341        &  0.0005811807      \\
1.5     &    0.2926831761804   &    0.011563379      &   3.0   &   0.0656052279197        &  0.02062639        \\
2.5     &    0.0931826682370   &    0.045113309      &   5.0   &   0.0126465348027        &  0.17568481        \\
3.0     &    0.0680640975474   &    0.069558611      &   7.5   &   0.0030330727129        &  1.0029980         \\
4.0     &    0.0481916949634   &    0.123248904      &   10.0  &   0.0013566103366        &  2.6889282         \\
5.0     &    0.0421759263287   &    0.167061238      &   12.0  &   0.0009842621480        &  4.3151579         \\
7.5     &    0.0398551249937   &    0.20591605       &   15.0  &   0.0008167874743        &  6.30370206        \\
10.0    &    0.0397899027431   &    0.208864145      &   20.0  &   0.0007786672679        &  7.5080031         \\
20.0    &    0.0397887357477   &    0.208974941      &   30.0  &   0.0007771237450        &  7.6497493         \\
40.0    &    0.0397887357477   &    0.208974941      &   40.0  &   0.0007771237450        &  7.6497493         \\ 
\end{tabular}
\end{ruledtabular}
\end{table}
\endgroup

At this stage, we move on to the last measure in this study, i.e., $E$ in {\color{red}Table~XII}. Here, the behavior compliments that of $I$ 
before. A cross-section of $E_{\rvec}, E_{\pvec}$ for $1s$, $2s$ states of CHA at same $r_c$ values introduced previously in 
{\color{red}Table~XI} are offered. Once again we observe that, at small $r_c$, $E_{\rvec}(1s)<E_{\rvec}(2s)$, which 
reverses after around $r_c \approx 3$. On the other hand, at moderately large (around 5) $r_c$, $E_{\pvec}(2s)$ exceeds   
$E_{\pvec}(1s)$. None of these have been reported before; hence cannot be compared. 

Above changes of $E_{\rvec}, E_{\pvec}$, {\color{red} total PM Onicescu energy $E$} with $r_c$ are graphically displayed in {\color{red}Fig.~11}, in left (a), middle (b), 
right (c) panels, for eight low-lying states. Like $I_{\rvec}$ in {\color{red}Fig.~10}, $E_{\rvec}$ falls off with $r_c$, with ground state
remaining well separated from others; all finally converging to FHA. Similarly, $E_{\pvec}$, like $I_{\pvec}$ of {\color{red}Fig.~10}
again, rises with $r_c$; then merges to FHA. At last, they converge to $E$ of FHA. In {\color{red}Fig.~(S4)}, $E_{\rvec}, E_{\pvec}$ 
and {\color{red} total PM Onicescu energy $E$} are depicted (in columns A, B, C) for $l \! = \! 0-4$ states as function of $n$ at six different $r_c$ (in segments 
(a)-(f)). At the lowest $r_c$ considered, these three behave qualitatively quite similarly as the respective $I$'s in {\color{red}Fig.~13}; 
$E_{\rvec}$ climb up while $E_{\pvec}$, $E$ record an opposite trend with $n$. This is in accordance with our earlier finding 
that, at very low $r_c$, 
confinement is more on higher states. First two columns suggest that, a minimum and maximum appears in 
$E_{\rvec}$, $E_{\pvec}$ graphs as $r_c$ gets extended. As in {\color{red}Fig.~(S3)}, positions of these extrema also shift towards right upon
proceeding from bottom to top panels in columns A, B. This supports that, at $r_c \rightarrow \infty$ CHA gets modified to FHA. 
Lastly the rightmost column records variation of {\color{red} total PM Onicescu energy $E$} against $n$. In all $r_c$'s, $E$ tend to grow with $n$.   

\begin{figure}                         %%%Fig. 11, CHA
\begin{minipage}[c]{0.35\textwidth}\centering
\includegraphics[scale=0.55]{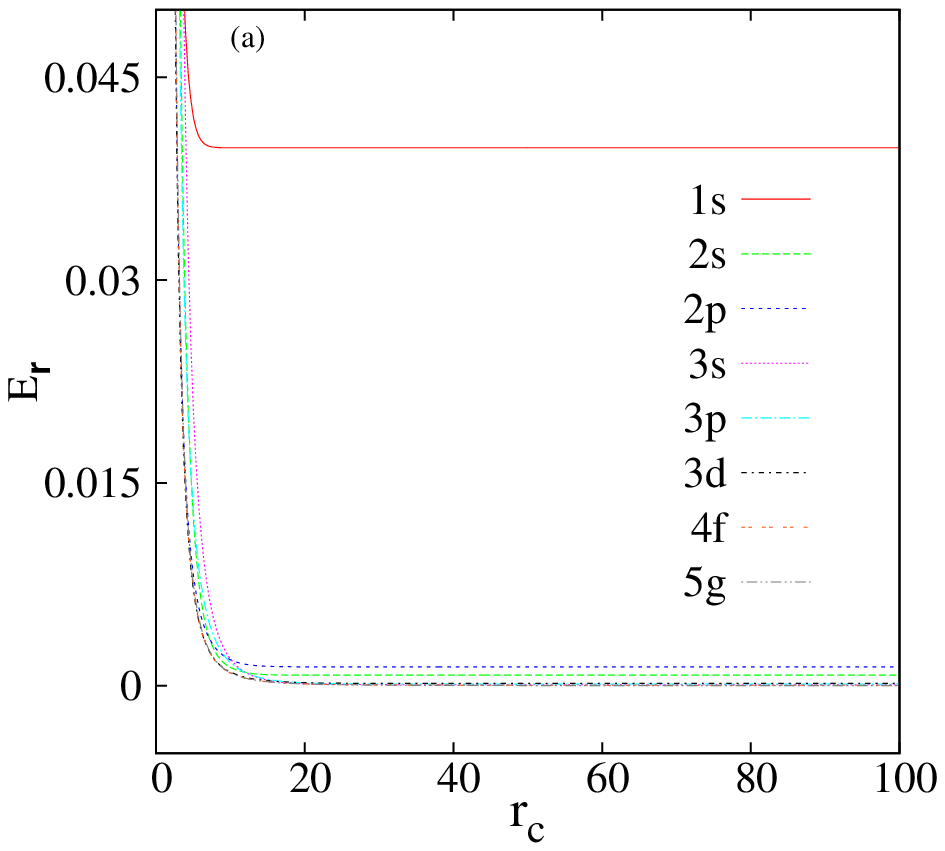}
\end{minipage}%
\begin{minipage}[c]{0.35\textwidth}\centering
\includegraphics[scale=0.55]{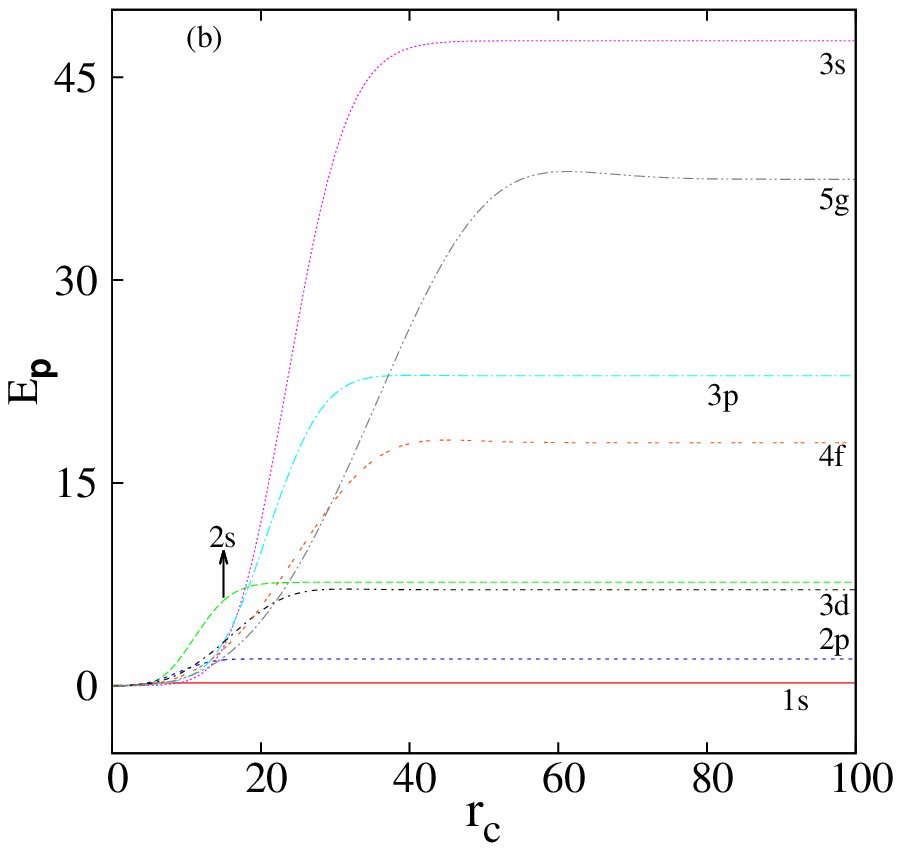}
\end{minipage}%
\begin{minipage}[c]{0.35\textwidth}\centering
\includegraphics[scale=0.55]{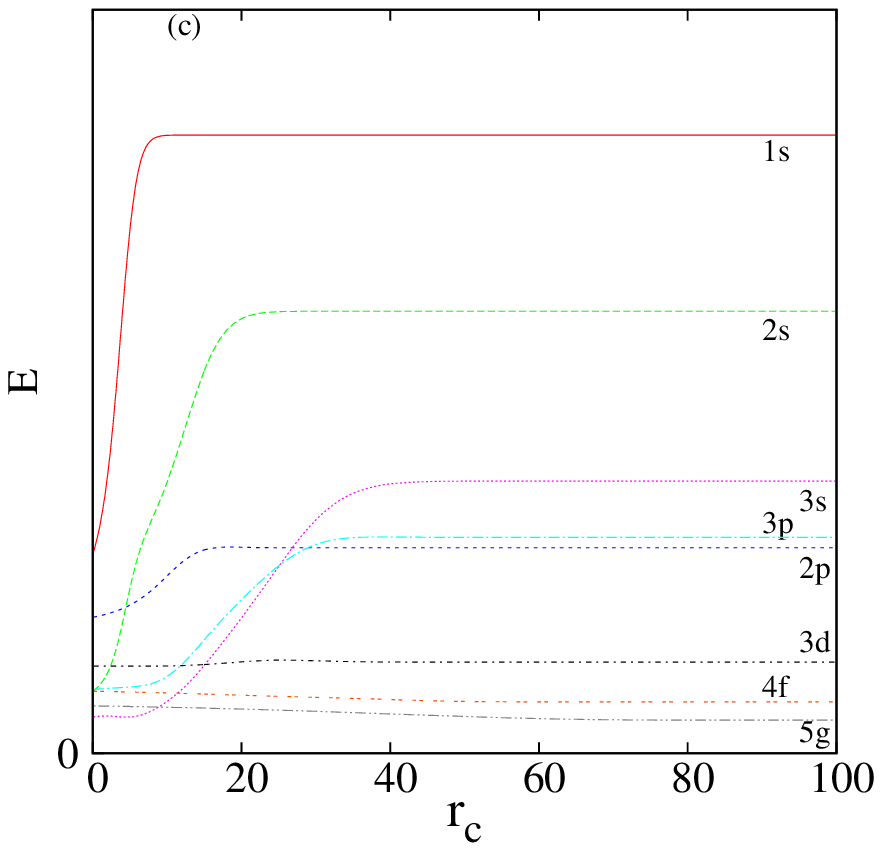}
\end{minipage}%
\caption{Variation of $E_{\rvec}, E_{\pvec}$ and $E$ with $r_{c}$ for CHA. For more details, see text.}
\end{figure}

\section{Future and Outlook}
Information theoretic measures like $R,~T,~S,~I,~E$ are pursued for a FHA and CHA in both $r$, $p$ spaces, along with their {\color{red} total
PM measures}. Accurate results of angular contributions to $R,~T,~S,~E$ are reported for $l\leq 9$ states ($m=0$), besides 
their radial counterparts. For CHA, \emph{combined} or total information measures (radial \emph{plus} angular) are provided.

In FHA, for \emph{node-less states}, while \emph{exact} analytical expressions for $I$, $S$ have been published, the same for $R,~T,~E$ 
are as yet unknown and derived here, again in both $r$, $p$ space. Illustrative calculations were made for both $s$ and non-zero $l$ 
states ($1s-4f$, $10s-10m$), out of which $R,~T$ and $E$ are completely new. It is found that, with growth of $n$, $R_{r},~T_{r}$, 
$S_{r}$ increase and $E_{r}$ decreases, which effectively points to the addition of radial node as well as spread of wave function. Thus, 
these quantities may be exploited to understand the diffuse nature of orbitals, especially
for high-lying states. Like FHA, $R,T$, $E$ for all states are given first time in a CHA. For $S$, $I$, even, excepting 
the lowest state, all results are new here. Among many interesting features, one notices that, at very low 
$r_{c}$, kinetic energy rises, while $R^{\alpha}_{\rvec},~T^{\beta}_{\rvec},~S_{\rvec}$ fall, as $n$ advances, which is 
in sharp contrast to that found in FHA. 

Overall, we have presented an elaborate account of the nature of a multitude of information measures under hard confinement. 
Further, it establishes the validity and utility of $R,T,S,I,E$ in the context of confinement in a CHA. These may
be useful to explore the so-called \emph{complexity} measures in a CHA, in future. There are several open questions that may lead to 
important conclusions and requires further scrutiny, such as, the effect of non-zero $m$ and a penetrable cavity. It may also be 
worthwhile to examine these quantities in the realm of Rydberg states under certain boundary conditions. A parallel inspection 
on many-electron systems would be highly desirable. 

\section{Acknowledgement}
Financial support from DST SERB, New Delhi, India (sanction order: EMR/2014/000838) is gratefully acknowledged. NM thanks DST SERB, 
New Delhi, India, for a National-post-doctoral fellowship (sanction order: PDF/2016/000014/CS). {\color{red} Critical constructive comments from 
two anonymous referees is greatly appreciated.}

\end{document}